\documentclass[preprint,amsmath,amssymb,aps,a4]{revtex4-2}
\usepackage{graphicx}
\usepackage{lineno}
\usepackage{bm}% bold math
\usepackage{txfonts}
\usepackage{caption}
\usepackage{amsmath}
\usepackage{amssymb}
\usepackage[section]{placeins}
\usepackage{tabularx}
\usepackage{braket}
\usepackage{multirow}
\usepackage{subcaption}
\usepackage{makecell}
\usepackage{enumitem}
\usepackage{times}
\usepackage{setspace}
\usepackage[justification=centering, skip=3pt]{caption}
\usepackage{booktabs}
\usepackage{rotating}
\usepackage{cancel}
\usepackage{braket}
\usepackage{tikz}
\usepackage{amsfonts}
\usepackage{xcolor}
\usepackage{hyperref}
\date{\today}
\newcommand{\be}{\begin{eqnarray}}
\newcommand{\ee}{\end{eqnarray}}

\begin{document}
\title{\(D\) and \(D^*\) mesons in isospin asymmetric nuclear medium }

\author{Anshu Gautam}
\email{gautamanshu681@gmail.com}
\author{Dhananjay Singh}
\email{snaks16aug@gmail.com}
\author{Navpreet Kaur}
\email{knavpreet.hep@gmail.com}
\author{Satyajit Puhan}
\email{puhansatyajit@gmail.com}
\author{Suneel Dutt}
\email{dutts@nitj.ac.in}
\author{Harleen Dahiya}
\email{dahiyah@nitj.ac.in}
\author{Arvind Kumar}
\email{kumara@nitj.ac.in}

\affiliation{ Department of Physics, Dr. B.R. Ambedkar National
	Institute of Technology, Jalandhar, 144008, India}

\date{\today}% 
\begin{abstract}
We investigate the properties of pseudoscalar \(D\) and vector \(D^*\) mesons in an isospin asymmetric nuclear medium using a hybrid approach that integrates the light-front quark model with the chiral SU(3) quark mean field model. The influence of isospin asymmetric nuclear medium is examined by utilizing the in-medium quark masses derived from the chiral SU(3) quark mean field model as an input in the light-front quark model to study the medium modification of \(D\) mesons. We examine the impact of isospin asymmetry and baryon density at zero and finite temperature on the effective masses, weak decay constants, and distribution amplitudes of the pseudoscalar mesons \(D^0\), \(D^+\), \(D_s\), and the vector mesons \(D^{0*}\), \(D^{+*}\), and \(D_s^*\). Our results indicate significant medium-induced changes for pseudoscalar \(D\) and vector \(D^*\) mesons having \(u/d\) as one of their constituent quarks, while a comparatively reduced effect is observed for mesons containing a strange quark. In contrast to temperature and isospin asymmetry, changes in the baryon density of the nuclear medium have a larger effect on different properties of \(D\) and \(D^*\) mesons.

 \vspace{0.1cm}
\end{abstract}
%====================================================
%
\maketitle
%
%\tableofcontents
%

\section{INTRODUCTION}
Investigating the properties and internal structure of hadrons in the nuclear medium is among the most active areas of research in hadronic physics, essential for understanding strongly interacting matter under extreme conditions.
%It provides insights into phenomena like chiral symmetry restoration and modifications of hadron spectral functions. These studies are crucial for interpreting experimental signals from heavy-ion collisions, where quark-gluon plasma effects are significant, and for understanding the physics of compact stars in astrophysics.
% The European Muon Collaboration (EMC)\cite{EuropeanMuon:1983wih} was the first to observe that the ratio of nucleon structure functions for iron and deuterium in deep inelastic muon scattering showed considerable deviations from well-established theoretical predictions. 
Through the deep inelastic scattering (DIS) experiments, the European Muon Collaboration (EMC) observed that the structure function of nucleons bound inside the nuclei is different compared to free space \cite{EuropeanMuon:1983wih}. This finding suggests that the interaction between  nucleons and their surrounding medium alters the internal structure of the nucleus. Moreover, the medium modification of hadron properties may relate to the partial restoration of chiral symmetry. Experimental evidence supporting this phenomenon has been observed in deeply bound pionic atoms \cite{Suzuki:2002ae}, low-energy pion-nucleus scattering \cite{Friedman:2004jh}, and di-pion production in hadron and photon-nucleus interactions \cite{CHAOS:1996nql, CHAOS:2004rhl}, all highlighting substantial modifications to pion properties in the nuclear environment. Specifically, the studies of deeply bound pionic atoms \cite{Suzuki:2002ae} and pion-nucleus scattering \cite{Friedman:2004jh} have shown a reduction in the pion decay constant, indicating that pions behave differently when bound within a nuclear medium.
\par Investigating how hadron properties are affected in nuclear medium at various temperatures and densities is crucial in high-energy physics as it provides insight into the behavior of strongly interacting matter under extreme conditions. Different experimental facilities, such as the proton-antiproton annihilation (PANDA) of GSI, Germany, focus on investigations related to hadron spectroscopy and the modifications of mass and width within the charm region \cite{PANDA:2009yku}. The Japan proton accelerator research complex (J-PARC) in Japan \cite{Ohnishi:2019cif, Sawada:2007gy} is designed to undertake measurements of hadrons, fluctuations in conserved quantities, dileptons, and multi-strange hypernuclei by utilizing the world's most intense heavy-ion beam to address the physics of high-density matter \cite{Sakaguchi:2019xjv}. In parallel, the compressed baryonic matter experiment (CBM) at GSI, Germany \cite{ Senger:2012wr} and the nuclotron-based ion collider facility (NICA) in Dubna, Russia \cite{Kekelidze:2017ghu, Kekelidze:2017tgp} aim to probe the nonperturbative aspects of quantum chromodynamics (QCD) at  finite baryon density \cite{Rapp:2008tf}. 

\par Motivated by these experimental developments, various theoretical approaches have been employed in the past to study the in-medium properties of hadrons such as the two-part model for the EMC effect based on light-front holographic QCD (LFHQCD) \cite{Kim:2022lng}, Nambu–Jona-Lasinio (NJL) model combined with the quark meson coupling (QMC) model \cite{Hutauruk:2018qku}, the Dyson-Schwinger equation (DSE) based approach \cite{Roberts:2000aa} and the hybrid light front-quark meson coupling (LF-QMC) \cite{deMelo:2016uwj,deMelo:2018hfw}. Additionally, the impact of nuclear medium on the properties of pions and kaons is studied within the light cone quark model (LCQM) and chiral SU(3) quark mean field (CQMF) model \cite{Kaur:2024wze,Puhan:2025ibn,Puhan:2024xdq, Singh:2024lra}.
The properties of $B$ mesons have also been investigated recently using the 
combined approach of LCQM and CQMF models \cite{Tanisha:2025glu}.
 In Refs.~\cite {Tsushima:2016sep, Hutauruk:2018qku, Zschiesche:2003qq}, studies demonstrate how hadrons' internal structure and properties are modified in a nuclear medium due to changes in the underlying quark condensates and meson-baryon interactions. Nonetheless, extensive research has been conducted on the medium modification of mesons in both symmetric and asymmetric nuclear medium. In the present work, we are primarily concerned with how the properties of pseudoscalar \(D\) mesons \(\left( D^0, D^+, D_s\right)\) and vector \(D^*\) mesons \(\left(D^{0*}, D^{+*}, D_s^*\right)\), such as their masses, weak decay constants, and distribution amplitudes (DAs) are affected in a nuclear medium. 

% The modification of meson mass in a medium is primarily influenced by the interaction of light quarks with the nuclear medium, as they are responsive to variations in the quark condensates linked to chiral symmetry breaking. In contrast, heavy quarks are only slightly influenced by the medium because of their significant mass. 
% By conducting these experiments, we can gain a deeper understanding of QCD matter in strongly interacting scenarios  
% % providing additional insights into the dynamics of heavy-ion collisions and the characteristics of dense stellar environments.These studies are 
% particularly relevant for interpreting experimental observables from heavy-ion collisions \cite{Aoki:2023qgl, CBELSATAPS:2005iwc, KEK-PS-E325:2005wbm} and understanding compact-star phenomena, where extreme conditions of high density and low temperature nuclear matter exist. Thus, the modifications of hadron properties in the nuclear medium are key to understanding the intricate physics of neutron stars and other dense astrophysical entities.

\par The interest in the properties of both open and hidden charmed mesons was sparked over two decades ago, mainly due to the investigation of relativistic nucleus-nucleus collisions. A key reason for this was charmonium suppression \cite{Matsui:1986dk}, proposed as an indicator of quark gluon plasma (QGP) formation. Following the observation of \(J/\psi\) suppression at super proton synchroton (SPS) energies by the NA50 collaboration \cite{NA50:1996lag}, the medium modifications of \(D\) mesons have become particularly significant, as it provides crucial insights into the behavior of heavy quarks within dense nuclear matter, which is vital for understanding the dynamics of QGP and the hadronization process in high-energy collisions. One of the consequences of in-medium interactions is that the attractive potential experienced by \(D\) mesons in the nuclear medium may also lead to the formation of D-mesic  \cite{Garcia-Recio:2010fiq} and \(J/\psi\)-mesic nuclei \cite{Tsushima:2011kh, Tolos:2013gta}. To probe these medium effects, the spectral density of \(D\) mesons has been analyzed using a self-consistent coupled channel approach \cite{Tolos:2005ft, Tolos:2004yg, Lutz:2005vx}, indicating that in-medium modifications predominantly affect the meson width, contrary to earlier expectations about the mass shift, particularly in the context of nucleus-nucleus collisions at FAIR (GSI) \cite{Cassing:2000vx}. A negative mass shift of \(\sim 50\) MeV for the \(D\) meson   was observed in the symmetric  nuclear matter by employing the QCD sum rule \cite{Hayashigaki:2000es}. Moreover, a comparable reduction in the mass of around \(60\) MeV at nuclear saturation density was obtained in the calculations using the QMC model  \cite{Tsushima:1998ru}.
The chiral SU(3) hadronic mean field model has been used to explore the
in-medium masses of $D$ and $\bar{D}$ mesons in nuclear and strange medium \cite{Mishra:2008cd,Kumar:2011ff}.
 The interaction of \(D\) mesons with the nucleons and the scalar mesons in a magnetized asymmetric nuclear medium has been studied using an effective hadronic model \cite{ReddyP:2017tgo}. The integration of the chiral SU(3) hadronic framework with QCD sum rules has been used to explore the mass shifts and decay constants of both pseudoscalar and scalar $D$ mesons in a hot magnetized asymmetric nuclear medium \cite{Kumar:2019axp}. Furthermore, this combined approach has also been applied to evaluate the in-medium masses and decay constants of pseudoscalar, scalar, vector, and axial-vector $D$ mesons in strange hadronic environments \cite{Chhabra:2017emy, Chhabra:2016vhp, Kumar:2013tna, Kumar:2015fca}.

% Recent research has examined the interaction of \(D\) mesons with the
% nucleons and the scalar mesons in the effective hadronic model \cite{Mishra:2008cd}. The integration of the chiral hadronic SU(3) framework with QCD sum rules has been used to explore the mass shifts and decay constants of both pseudoscalar and scalar $D$ mesons within a hot, magnetized, asymmetric nuclear medium \cite{Kumar:2019axp}. Furthermore, this combined approach has also been applied to evaluate the in-medium masses and decay constants of pseudoscalar, scalar, vector, and axial-vector $D$ mesons in strange hadronic environments \cite{Chhabra:2017emy, Chhabra:2016vhp, Kumar:2013tna, Kumar:2015fca}. The spectral density of \(D\) mesons has been analyzed through a self-consistent coupled-channel approach \cite{Tolos:2005ft, Tolos:2004yg, Lutz:2005vx}, indicating that the medium modification for \(D\) meson predominantly affect the width, contrary to earlier assumptions about the mass, during nucleus-nucleus collisions at FAIR (GSI) \cite{Cassing:2000vx}. A mass shift of \(\sim 50\) MeV for the \(D\) meson compared to the \(J/\psi\) was observed in isospin-averaged nuclear matter by employing the QCD sum rule \cite{Hayashigaki:2000es}. Moreover, a comparable reduction in the mass of the \(D\) meson, around \(60\) MeV at a density of \(\rho = \rho_0\), was investigated through the quark meson coupling model in Ref.~\cite{Tsushima:1998ru}.

\par In the present manuscript, we investigate the properties of pseudoscalar \(D\) and vector \(D^*\) mesons under varying conditions of isospin asymmetry and temperature of the dense nuclear medium. The modification of meson masses in a medium is primarily governed by the interaction of light quarks with the nuclear medium, as they respond sensitively to changes in quark condensates associated with chiral symmetry breaking \cite{Hutauruk:2018qku, Hosaka:2016ypm}. In contrast, heavy quarks are only slightly influenced due to their large masses. The weak decay constant represents the meson's decay amplitude via weak interactions and serves as a key indicator of chiral symmetry breaking in dense matter \cite{Chhabra:2017rxz, Chhabra:2018oyl}. Furthermore, the DAs of mesons are crucial for analyzing SU(3) symmetry breaking effects and play an important role in understanding hard exclusive QCD processes involving large momentum transfers \cite{Arifi:2024mff, Arifi:2023jfe}. In this study, we employ a hybrid framework wherein meson properties are evaluated using the light-front quark model (LFQM), while in-medium quark masses are supplied by the chiral SU(3) quark mean field (CQMF) model. The LFQM framework provides an effective nonperturbative method that uses quantization at equal light-front time \((x^+ = t + z)\) and evolves the system via the light-front Hamiltonian \(P^-\) \cite{Brodsky:1997de, Wilson:1994fk}. The model uses a Gaussian wave function as the trial state for applying the variational principle to the QCD motivated Hamiltonian that incorporates the quark potential, which effectively captures the interaction between quark and antiquark within the meson. While the nuclear medium notably impacts lighter quarks more than heavier ones, isospin asymmetric matter relevant to both neutron stars and heavy-ion collisions introduces additional complexity. We include the effects of isospin asymmetry via in-medium quark masses computed using the CQMF model. In the CQMF model, quarks are confined within baryons by means of a confining potential and engage in interactions through the exchange of scalar  ($\sigma$, $\zeta$, $\delta$) and vector  ($\omega$, $\rho$) fields. 
%In earlier applications of this model to isospin asymmetric matter, only the vector-isovector field $\rho$ was included. However, the significant role of the scalar-isovector $\delta$ meson has led to its inclusion in the current analysis \cite{Liu:2001iz, Thakur:2022dxb, Li:2022okx}.
 %A similar methodology was previously employed to study the weak decay constant, DAs, and parton distribution functions (PDFs) of the valence quarks in kaons \cite{Singh:2024lra}, and in pions under isospin asymmetric nuclear environment \cite{Puhan:2024xdq}.
% \par The present work focuses on the modifications in the masses of the pseudoscalar mesons \(D^0\), \(D^+\), \(D_s\), and the vector mesons \(D^{0*}\), \(D^{+*}\), \(D_s^*\). We study the weak decay constants and distribution amplitudes under varying conditions of isospin asymmetry and temperature. We compare pseudoscalar and vector mesons across isospin asymmetries and temperatures to highlight the influence of the nuclear medium on mesonic properties.

\par The paper is organized as follows: In Sec.~\ref {sec:1},  we present the CQMF model employed for computing the in-medium quark masses. Sec.~\ref{sec:2} introduces the effective Hamiltonian and light front wave function (LFWF) adopted in the present approach.
% Additionally, we describe how the model parameters are fitted using the variational principle. 
Results of the present work on the in-medium properties of pseudoscalar and vector \(D\) mesons are presented in Sec.~\ref {sec:3}, and lastly in Sec.~\ref {sec:4}, we summarize our findings.

 \section{METHODOLOGY}
 \subsection{Chiral SU(3) quark mean field model}
 \label{sec:1}
The CQMF model is used to evaluate effective quark masses within an isospin asymmetric nuclear medium, which serve as input to investigate the medium-modified properties of \(D\) mesons. The model treats quarks and mesons as fundamental degrees of freedom and incorporates important low-energy features of QCD, particularly the spontaneous and explicit breaking of chiral symmetry \cite{Wang:2001jw}. As discussed earlier, quarks are confined within baryons through an effective potential and acquire their masses via interactions with the scalar-isoscalar fields $\sigma$ and $\zeta$, the scalar-isovector field $\delta$, and the vector-isoscalar $\omega$ along with the vector-isovector field $\rho$. 
% The isospin-asymmetry of the medium is incorporated by the introduction of fields $\delta$ and $\rho$ \cite{Kumar:2010hs}, while the scalar dilation field $\chi$ is added to preserve the trace anomaly property of QCD \cite{Papazoglou:1998vr, Kumar:2018ujk}. 
The trace anomaly property of QCD is accounted for by the inclusion of scalar dilation field \(\chi\) \cite{Papazoglou:1998vr, Kumar:2018ujk}, while the \(\delta\) and \(\rho\) fields introduce isospin asymmetry into the medium \cite{Kumar:2010hs}.

\par The thermodynamic potential for isospin asymmetric nuclear matter at finite temperature and density is expressed as 
\begin{eqnarray}
\Omega & = & - \frac{k_B T}{(2\pi)^3} \sum_i \gamma_i \int_0^\infty d^3k \left[ \ln \left( 1 + e^{-\left(\frac{E^*_i(k) - \nu^*_i}{k_B T}\right)} \right) + \ln \left( 1 + e^{-\left(\frac{E^*_i(k) + \nu^*_i}{k_B T}\right)} \right) \right] \nonumber\\&-& \mathcal{L}_M - \mathcal{V}_{\text{vac}},
\end{eqnarray}
where the summation over nucleons in the medium is represented by $i = p/n$ and the degeneracy factor $\gamma_i = 2$ accounts for two spin states of each nucleon. The effective  energy $E_i^*(k)$  and the effective chemical potential  $\nu_i^*$ of the baryon  are expressed as  
\begin{eqnarray}
E_i^* &=& \sqrt{{M_i^*}^2 + k^2}, \\
\nu^*_i &=& \nu_i - g_\omega^i \, \omega - g_\rho^i \, I^{3i} \rho.
\end{eqnarray}
In above, $M_i^*$ is the effective mass of the baryon and $\nu_i$ is the free chemical potential. 
To obtain the density and temperature dependent values of scalar, vector, and dilaton fields, the thermodynamic potential is minimized as
\begin{eqnarray}
\frac{\partial \Omega}{\partial \sigma}=
\frac{\partial \Omega}{\partial \zeta}=
\frac{\partial \Omega}{\partial \delta} =
\frac{\partial \Omega}{\partial \chi} =
\frac{\partial \Omega}{\partial \omega} = 
\frac{\partial \Omega}{\partial \rho} = 0.
%\frac{\partial \Omega}{\partial \phi} & 
\end{eqnarray}
This yields a set of nonlinear equations, which are solved for different values of baryon density, temperature, and isospin asymmetry. The effect of isospin asymmetry is quantified by the parameter $\eta=-\frac{\sum_i I_{3i} \rho_i}{\rho_B}$, where $I_{3i}$ represents the $3^{rd}$ component of isospin quantum number and $\rho_B$ denotes the total baryon density of the medium. 
The effective Lagrangian density which encapsulates the interactions within the CQMF model is given by \cite{Wang:2001jw}
\begin{eqnarray}
\label{eq:9}
\mathcal{L}_{\text{eff}} &=& \mathcal{L}_{q0} + \mathcal{L}_{qm} + \mathcal{L}_{\Sigma \Sigma} + \mathcal{L}_{VV} + \mathcal{L}_{\chi SB} + \mathcal{L}_{\Delta m} + \mathcal{L}_{c}.
\end{eqnarray}
In the above, $ \mathcal{L}_{q0}$ denotes the kinetic term and $\mathcal{L}_{qm}$ represents the quark-meson interaction term, including quark interactions with both spin-0 and spin-1 mesons \cite{Wang:2001jw, Singh:2016hiw, Wang:2001hw}. The expressions for \(\mathcal{L}_{q0}\) and \(\mathcal{L}_{qm}\) are written as
\begin{eqnarray}
\mathcal{L}_{q0} &=& \bar{q} \, i \gamma^\mu \partial_\mu q,
\end{eqnarray}
and 
\begin{eqnarray}
\mathcal{L}_{qm} &=& g_s \left(\bar{\Psi}_L M \Psi_R + \bar{\Psi}_R M^\dagger \Psi_L\right) 
- g_v \left(\bar{\Psi}_L \gamma^\mu l_\mu \Psi_L + \bar{\Psi}_R \gamma^\mu r_\mu \Psi_R\right) \nonumber \\
&=& \frac{g_s}{\sqrt{2}} \, \bar{\Psi} \left( \sum_{a=0}^8 s_a \lambda_a + i \gamma^5 \sum_{a=0}^8 p_a \lambda_a \right) \Psi \nonumber \\
&& - \frac{g_v}{2\sqrt{2}} \, \bar{\Psi} \left( \gamma^\mu \sum_{a=0}^8 v_a^\mu \lambda_a 
- \gamma^\mu \gamma^5 \sum_{a=0}^8 a^\mu_a \lambda_a \right) \Psi,
\end{eqnarray}
respectively.
Here, \( q = \begin{pmatrix} u \\ d \end{pmatrix} \)
and parameters $g_s$ and $g_v$ describe couplings of quark with scalar and vector meson fields. From Eq.~\eqref{eq:9}, the chiral invariant self-interaction term for scalar mesons \(\mathcal{L}_{\Sigma \Sigma}\) is expressed as 
\begin{equation}
\begin{aligned}
\label{eq:5}
\mathcal{L}_{\Sigma\Sigma} = & -\frac{1}{2} k_0 \chi^2 \left(\sigma^2 + \zeta^2 + \delta^2\right) 
+ k_1 \left(\sigma^2 + \zeta^2 + \delta^2\right)^2 
+ k_2 \left(\frac{\sigma^4}{2} + \frac{\delta^4}{2} + 3\sigma^2\delta^2 + \zeta^4\right) \\
& + k_3 \chi \left(\sigma^2 - \delta^2\right) \zeta 
- k_4 \chi^4 - \frac{1}{4} \chi^4 \ln\frac{\chi^4}{\chi_0^4} 
+ \frac{\xi}{3} \chi^4 \ln \left(\left(\frac{(\sigma^2 - \delta^2)\zeta}{\sigma_0^2\zeta_0}\right) 
\left(\frac{\chi^3}{\chi_{0}^3}\right)\right).
\end{aligned}
\end{equation}
In the above, the last two logarithmic terms are included to ensure the trace of the energy-momentum tensor is proportional to the fourth power of the dilaton field $\chi$. For vector mesons, the self-interaction term \(\mathcal{L}_{V V}\) is given by 
\begin{eqnarray}
\mathcal{L}_{VV} & = & \frac{1}{2} \frac{\chi^2}{\chi_0^2} \left( m^2_\omega \omega^2 + m^2_\rho \rho^2  \right) + g_4 \left( \omega^4 + 6\omega^2\rho^2 + \rho^4  \right).
\end{eqnarray}
% Here to express the trace of energy-momentum tensor proportional to the fourth power of the dilaton field $\chi$, the last two logarithmic terms are included in Eqn~\ref{eq:5}.  
To incorporate the non-zero masses of pseudoscalar mesons, the explicit symmetry-breaking term $\mathcal{L}_{\chi SB}$ is introduced in Eq~\eqref{eq:9} as \cite{Papazoglou:1998vr, Wang:2002aq}
\begin{eqnarray}
\mathcal{L}_{\chi SB} & = &-  \frac{\chi^2}{\chi_0^2} \left[ m_\pi^2 f_\pi \sigma + \left(\sqrt{2} m_K^2 f_K -\frac{ m_\pi^2} {\sqrt{2}} f_\pi \right)\zeta \right].
\end{eqnarray}
% At zero baryon density and temperature, the constituent masses of light $u$ and $d$ quarks, in terms of the vacuum expectation values $\sigma_0$ and $\zeta_0$ corresponding to the scalar fields $\sigma$ and $\zeta$, are given as
% \begin{eqnarray}
% m_u=m_d=-g^q_\sigma \sigma_0=-\frac{g_s}{\sqrt{2}}\sigma_0,
% \end{eqnarray}
% The value of $g_s$ is determined such that $m_u=253$ MeV. 
% An additional mass term $\Delta m$ is incorporated through an extra explicit symmetry-breaking term in order to achieve an accurate value for the strange quark mass $m_s$, which is defined as \cite{Wang:2001jw}, \cite{Wang:2002aq}
% \begin{eqnarray}
% \mathcal{L}_{\Delta m}=-(\Delta m)\bar\psi S_1 \psi,
% \end{eqnarray}
% where $S_1$ is the strange quark matrix operator
% % and is given as 
% % \begin{eqnarray}
% % S_1=\frac{1}{3}\left(I-\lambda_8 \sqrt{3}\right)=diag(0,0,1).
% % \end{eqnarray}
% which results in the equation for the vacuum mass of the strange quark
% % \begin{eqnarray}
% $m_s=-g_\zeta^s \zeta_0+\Delta m.$
% % \end{eqnarray}
Finally, to represent quark confinement within baryons, the confining potential term $\mathcal{L}_c$ is written as 
\begin{eqnarray}
\mathcal{L}_c=-\bar\psi \chi_c \psi.
\end{eqnarray}
% \par To analyze the properties of isospin asymmetric nuclear matter at finite temperature and density, the mean-field approximation is employed \cite{Wang:2001jw}. 
In the presence of meson mean fields, the Dirac equation for a quark field $\Psi_{qi}$ becomes
\begin{eqnarray}
\left[- i \boldsymbol{\alpha} \cdot \nabla + \chi_c(r) + \beta m^*_q \right]\Psi_{qi} = e^*_q \, \Psi_{qi},
\end{eqnarray}
where the subscript $q$ represents the quarks within a baryon of type $i$ (where $i$=$n$, $p$) and $\boldsymbol\alpha$, $\beta$ denote the standard Dirac matrices. The effective quark mass $m_q^*$ and energy $e_q^*$ are expressed in terms of relevant coupling constants, and the scalar and vector meson fields as \cite{Kumar:2023owb}
\begin{eqnarray}
m^*_q & = & - g^q_\sigma \sigma - g^q_\zeta \zeta - g_\delta^q I^{3q} \delta + \Delta m,
\label{eq:4}
\end{eqnarray}
and
\begin{eqnarray}
e^*_q & = & e_q - g^q_\omega \omega - g_\rho^q I^{3q} \rho.
\end{eqnarray}
% The value of various parameters and coupling constants used in our work are given in Ref. \cite {Kumar:2023owb}. 
The Lagrangian density \(\mathcal{L}_{\Delta m}=-(\Delta m)\bar\psi S_1 \psi\) is incorporated through an explicit symmetry breaking term to attain a realistic value for the strange quark mass $m_s$ 
% The additional mass term $\Delta m$ is incorporated through an extra explicit symmetry breaking term to attain a reasonable value for the strange quark mass $m_s$ and is defined as \cite{Wang:2001jw}, \cite{Wang:2002aq} \(\mathcal{L}_{\Delta m}=-(\Delta m)\bar\psi S_1 \psi\), 
, where $S_1$ is the strange quark matrix operator and is defined as 
$S_1=\frac{1}{3}\left(I-\lambda_8 \sqrt{3}\right)=diag(0,0,1)$ \cite{Wang:2001jw, Wang:2002aq}.
% which results in the equation for the vacuum mass of the strange quark
% \begin{eqnarray}
% $m_s=-g_\zeta^s \zeta_0+\Delta m.$
% \end{eqnarray}
The value of $\Delta m$ for the $u$ and $d$ quarks is zero, while for the $s$ quark $\Delta m=77$ MeV. The various parameters and coupling constants used in our calculations are listed in Table~\ref{tab:1}.
\par The in-medium spurious center-of-mass momentum $\langle p^{*2}_{i\:\text{cm}} \rangle$ \cite{Barik:1985rm, Barik:2013lna} and the effective energy of constituent quarks $e^*_q$ are related to the effective mass of the $i^{th}$ baryon through the relation
\begin{eqnarray}
M^*_i & = & \sqrt{\left( \sum_q n_{qi} e^*_q + E_{i\:\text{spin}}\right)^2 - \langle p^{*2}_{i\:\text{cm}} \rangle}.
\end{eqnarray}

Here, $n_{qi}$ denotes the number of quarks of type $q$ within the $i^{th}$ baryon. Additionally, the in-medium spurious center-of-mass correction $\langle p^{*2}_{i\:\text{cm}} \rangle$  is given by
\begin{eqnarray}
\langle p^{*2}_{i\:\text{cm}}\rangle & = & \sum_q\frac{\left( 11 e^*_q + m^*_q \right)}{ 6\left( 3 e^*_q + m^*_q \right)} \left( e^{*2}_q - m^{*2}_q \right).
\end{eqnarray}

\subsection{Light-front quark model}
\label{sec:2}
This section describes the fundamental idea of LFQM, which includes the effective Hamiltonian and the light front wave function (LFWF).
% In the LFQM, a meson state is defined as a bound state of a constituent quark-antiquark pair within a non-interacting framework, following the Bakamjian-Thomas construction \cite{Bakamjian:1953kh}. 
To preserve compliance with the group structure satisfying the Poincaré algebraic commutation relations, the interactions are incorporated into the meson mass operator \cite{ Keister:1991sb}. 
% the interaction is incorporated into the meson mass operator to preserve the structure of the Poincaré group \cite{ Keister:1991sb}, thereby encoding the interaction dynamics within the mass eigenfunction. The core idea of the LFQM is to use 
The model is based on the idea of using a radial wave function as a trial state for the QCD motivated effective Hamiltonian with a linear confinement potential \cite{Choi:1997iq}, which determines the mass eigenvalue by employing the variational principle framework.
% , and the interactions are incorporated into the meson mass operator to preserve compliance with the group structure satisfying the Poincar´e algebraic commutation relations \cite{ Keister:1991sb}.
\par The meson bound system at rest satisfying the eigenvalue equation of the QCD inspired effective Hamiltonian is described as \cite{Arifi:2024mff, Choi:2009ai, Choi:1999nu},
\begin{eqnarray}
\left( H_0 + V_{q\bar{q}} \right) \left| \mathbf \Psi_{q\bar{q}} \right\rangle = M_{q\bar{q}}^* \left| \mathbf \Psi_{q\bar{q}} \right\rangle,
\end{eqnarray}
where  $\mathbf \Psi_{q\bar{q}}$ and $M_{q\bar{q}}$ are the eigenfunction and mass eigenvalue of the meson, respectively. The relativistic kinetic energy of the quark and antiquark is given by
\begin{eqnarray}
H_0 &=& \sqrt{m_q^{*2} + \textbf{p}_q^2} + \sqrt{m_{\bar{q}}^{*2} + \textbf{p}_{\bar{q}}^2}.
\end{eqnarray}
The effective potential $V_{q\bar{q}} $ between the quark and antiquark in the rest frame of meson is  comprised of  linear confining potential $ V_{\text{Conf}}$, the one-gluon-exchange Coulomb potential $V_{\text{Coul}}$, and the hyperfine potential $ V_{\text{Hyp}}$ (essential to distinguish between the pseudoscalar $(0^{-+})$ and vector $(1^{--})$ mesons) and is defined as 
%\begin{eqnarray}
%V_{q\bar{q}} &=& V_{\text{Conf}} + V_{\text{Coul}} + V_{\text{Hyp}},
%\end{eqnarray}
\begin{eqnarray}
\label{eq:6}
V_{q\bar {q}}=
\underbrace{a + b r}_{\text{conf}} 
- \underbrace{\frac{4 \alpha_s}{3r}}_{\text{coul}} 
+ \underbrace{\frac{32\pi \alpha_s \, \langle\mathbf{S}_q \cdot \mathbf{S}_{\bar{q}}\rangle}{9 m^*_q m^*_{\bar{q}}} \delta^{(3)}(r)}_{\text{hyp}},
\end{eqnarray}
% \begin{eqnarray}
% V_{\text{Conf}} &=& a + br,\\
% V_{\text{Coul}} &=& -\frac{4\alpha_s}{3r},\\
% V_{\text{Hyp}} &=& \frac{32\pi \alpha_s \, \langle\mathbf{S}_q \cdot \mathbf{S}_{\bar{q}}\rangle}{9 m_q m_{\bar{q}}} \delta^{(3)}(r),
% \end{eqnarray}
here $a$ and $b$ parameterize the linear confining
potential, $\alpha_s$ is the strong running coupling, and the spin term
$\langle\mathbf{S}_q \cdot \mathbf{S}_{\bar{q}}\rangle$
yields the values of $1/4$ and $-3/4$ for the vector and pseudoscalar mesons, respectively.
% \par The meson state $\left| \mathcal M(\textit{P}, \textit{J},\textit{J}_z) \right\rangle$, as a bound state of the
% constituent quark $q$ and antiquark $\bar q$ with meson momentum $\textit{P}$ and total angular momentum $(\textit{J},\textit{J}_z)$ , can be written as \cite{Arifi:2023jfe}
% \begin{eqnarray}
% |\mathcal M(\textit{P}, \textit{J}, \textit{J}_z)\rangle=\int \left[d^3 \mathbf{p}_q\right] \left[d^3 \mathbf{p}_{\bar{q}}\right] {2(2\pi)^3} \, \delta^3(\mathbf{P} - \mathbf{p}_q - \mathbf{p}_{\bar{q}}) \nonumber \\
% \times\sum_{\lambda_q, \lambda_{\bar{q}}} \mathbf \Psi_{\lambda_q \lambda_{\bar{q}}}^{\textit{J} \textit{J}_z}(x, \mathbf{k}_\perp) |q_{\lambda_q}(p_q) \bar{q}_{\lambda_{\bar{q}}}(p_{\bar{q}})\rangle,
% \end{eqnarray}
% where $\mathbf{p}_i = \left(p_i^+, \mathbf{p}_{i\perp}\right)$ and $\left[d^3 \mathbf{p}_i\right]$ =$ d p_i^+ d^2 \mathbf{p}_{i\perp}/[2(2\pi)^3]$. 
% The pair $\left(p_i, \lambda_i\right)$ represents the on-mass shell light-front momentum and helicity of the quark ($i = q$) and antiquark ($i = \bar{q}$), respectively. 
\par The LFWF is characterized by Lorentz-invariant internal variables, such as the longitudinal momentum fraction \( x_j = \frac{p_j^+}{P^+} \), intrinsic transverse momentum \( \mathbf{k}_{\perp j} = \mathbf{p}_{\perp j} - x_j \mathbf{P}_\perp \), along with the pair \((p_j, \lambda_j)\) which denotes the on-mass-shell light-front momentum and the helicity of the quark or antiquark (\(j = q/\bar{q}\)), respectively. In the momentum space, the LFWF is given by \cite{Arifi:2024mff, Arifi:2022pal}
\begin{eqnarray}
\mathbf \Psi_{\lambda_q \lambda_{\bar{q}}}^{\textit{J} \textit{J}_z}(x, \mathbf{k}_\perp) &=& \mathbf \Phi(x, \mathbf{k}_\perp) \mathcal R_{\lambda_q \lambda_{\bar{q}}}^{\textit{J} \textit{J}_z}(x, \mathbf{k}_\perp),
\end{eqnarray}
where $\mathbf \Phi(x, \mathbf{k}_\perp)$ and $ \mathcal R_{\lambda_q \lambda_{\bar{q}}}^{\textit{J} \textit{J}_z}(x, \mathbf{k}_\perp)$ denote the vacuum radial wave function and interaction independent spin-orbit wave function, respectively, the latter distinguishing between vector $D^{*}$ and pseudoscalar $D$  mesons. 
As discussed before, the in-medium effects on the properties of $D$ mesons are
simulated through the effective quark masses. 
The radial wave function $\mathbf \Phi(x, \mathbf{k}_\perp)$ and the
spin-orbit wave function $ \mathcal R_{\lambda_q \lambda_{\bar{q}}}^{\textit{J} \textit{J}_z}(x, \mathbf{k}_\perp)$ depend upon the quark masses and hence, will be modified at finite density of the medium, compared to free space.
 The explicit form of \( \mathcal{R}_{\lambda_q \lambda_{\bar{q}}}^{*\textit{J} \textit{J}_z}(x, \mathbf{k}_\perp) \)   is expressed as (arising from the  Melosh transformation) \cite{Arifi:2024mff}

\begin{eqnarray}
\mathcal{R}^{*00}_{\lambda_q \lambda_{\bar{q}}} &=& 
\frac{1}{\sqrt{2} \sqrt{\mathcal{A^*}^2 + \mathbf{k}_\perp^2}} 
\begin{pmatrix}
-k^L & \mathcal{A^*} \\
-\mathcal{A^*} & -k^R
\end{pmatrix}
\end{eqnarray}
and
\begin{eqnarray}
\mathcal{R}^{*11}_{\lambda_q \lambda_{\bar{q}}} &=& 
\frac{1}{\sqrt{\mathcal{A}^{*2} + \mathbf{k}_\perp^2}} 
\begin{pmatrix}
\mathcal{A^*} + \frac{\mathbf{k}_\perp^2}{\mathcal{M^*}} & k^R \frac{\mathcal{M}^*_1}{\mathcal{M^*}} \\
-k^R \frac{\mathcal{M}^*_2}{\mathcal{M^*}} &  -  \frac{\left( k^R \right)^2}{\mathcal{M^*}}
\end{pmatrix}, \\[10pt]
\mathcal{R}^{*10}_{\lambda_q \lambda_{\bar{q}}} &=& 
\frac{1}{\sqrt{2} \sqrt{\mathcal{A}^{*2} + \mathbf{k}_\perp^2}} 
\begin{pmatrix}
k^L \frac{\mathcal{M}^*_3}{\mathcal{M}^*} & \mathcal{A^*} + \frac{2\mathbf{k}_\perp^2}{\mathcal{M^*}} \\
\mathcal{A}^* + \frac{2\mathbf{k}_\perp^2}{\mathcal{M}^*} & -k^R \frac{\mathcal{M}^*_3}{\mathcal{M}^*}
\end{pmatrix}, \\[10pt]
\mathcal{R}^{*1-1}_{\lambda_q \lambda_{\bar{q}}} &=& 
\frac{1}{\sqrt{\mathcal{A}^{*2} + \mathbf{k}_\perp^2}} 
\begin{pmatrix}
- \frac{\left(k^L\right)^2}{\mathcal{M}^*}  & k^L \frac{\mathcal{M}^*_2}{\mathcal{M}^*} \\
-k^L \frac{\mathcal{M}^*_1}{\mathcal{M}^*} & \mathcal{A}^* + \frac{\mathbf{k}_\perp^2}{\mathcal{M}^*}
\end{pmatrix},
\end{eqnarray}
% \begin{eqnarray}
% \mathcal{R}^{\textit{J} \textit{J}_z}_{\lambda_q \lambda_{\bar{q}}}(x, \mathbf{k}_\perp) 
% &=& \sum_{\lambda'_q, \lambda'_{\bar{q}}} 
% \braket{\lambda_q | \mathcal{R}_M^\dagger(x, \mathbf{k}_\perp, m_q) | \lambda'_q} \times \braket{\lambda_{\bar{q}} | \mathcal{R}_M^\dagger(1-x, -\mathbf{k}_\perp, m_{\bar{q}}) | \lambda'_{\bar{q}}} 
% \braket{\tfrac{1}{2} \lambda'_q \tfrac{1}{2} \lambda'_{\bar{q}} | \textit{J} \textit{J}_z},
% \label{eq:1.7}
% \end{eqnarray}
% where the Melosh transformation operator with unit vector $\hat{\mathbf{n}} = (0, 0, 1)$ in the $z$-direction is given by
% \begin{eqnarray}
% \mathcal{R}_M(x, \mathbf{k}_\perp, m) 
% &=& \frac{m + x M_0 - i \sigma \cdot (\mathbf{\hat{n}} \times \mathbf{\hat{k}})}
% {\sqrt{(m + x M_0)^2 + \mathbf{k}_\perp^2}}.
% \label{eq:1.8}
% \end{eqnarray}
for pseudoscalar and vector mesons, respectively. Here, \(k^{R(L)}= k_x+ik_y, \mathcal{A^*}=(1-x)m^*_q+xm^*_{\bar q}, \mathcal{M}^*=M^*_0+m^*_q+m^*_{\bar q}, \mathcal{M}^*_1=xM^*_0+m^*_q, \mathcal{M}^*_2=(1-x)M^*_0+m^*_{\bar q}\) and \(\mathcal{M}^*_3=\mathcal{M}^*_2-\mathcal{M}^*_1\). Asterik $(*)$ sign symbolize that the quantities are medium dependent. 
The boost invariant mass of meson squared, \(M^{*2}_0\) is defined as 
 % and the invariant meson mass $M_0^2$ is defined as
\begin{eqnarray}
M_0^{*2} &=& \frac{\mathbf{k}_\perp^2 + {m_q^*}^2}{x} + \frac{\mathbf{k}_\perp^2 + {m_{\bar{q}}^*}^2}{1 - x}.
\end{eqnarray}
Additonally, the spin-orbit wave functions $\mathcal R^{*\textit{J} \textit{J}_z}_{\lambda_q \, \lambda_{\bar{q}}}$ automatically satisfy the unitary condition $\Large\langle \mathcal R_{\lambda_q \lambda_{\bar{q}}}^{*\textit{J} \textit{J}_z}|\mathcal R_{\lambda_q \lambda_{\bar{q}}}^{*\textit{J} \textit{J}_z}\Large\rangle$ =1. To use a variational principle, we employ the lowest-order harmonic oscillator wave function as the in-medium trial radial wave function $\mathbf\Phi^*(x, \mathbf{k}_\perp)$ for both pseudoscalar and vector mesons, 
\begin{eqnarray}
\mathbf \Phi^*_{1S}(x, \mathbf{k}_\perp) &=& \frac{4\pi^{3/4}} {\beta^{3/2}} \sqrt{\frac{\partial
{k_z}^*}{\partial x}} e^{-\frac{\mathbf{k}^2}{2\beta^2}}.
\end{eqnarray}
where $\beta$ is the variational parameter controlling the wavefunction scale in our mass spectroscopic analysis, and is determined by applying the variational principle, under the condition that $\alpha_s$ is uniform across all the mesons. The variable transformation $\{k_z, \mathbf{\textbf{k}_\perp}\} \rightarrow \{x, \mathbf{\textbf{k}_\perp}\}$ involves the Jacobian,
\begin{eqnarray}
\frac{\partial k_z^*}{\partial x} & = & \frac{M^*_0}{4x(1 - x)} \left[ 1 - \frac{(m_q^{*2} - m_{\bar{q}}^{*2})^2}{M_0^{*4}} \right],
\end{eqnarray}
where longitudinal component $k_z^* = \left(x - 1/2\right) M^*_0 + (m_{\bar{q}}^{*2} - m_q^{*2})/2M^*_0$. It is important to mention that the normalized LFWF will be utilized for additional computations \cite{Arifi:2023jfe}.

\par We perform the variational analysis to evaluate the expectation value of the QCD motivated Hamiltonian $H_{q\bar{q}}$ with the trial function $\mathbf \Phi^*_{1S}(x, \mathbf{k}_\perp)$ which has variational parameter $\beta$ dependence. Once the parameter fixing is done by minimizing the expectation value, we can determine the mass eigenvalue of the meson as $M^*_{q\bar{q}} = \langle \mathbf \Psi_{q\bar{q}} | H_{q\bar{q}} | \mathbf \Psi_{q\bar{q}} \rangle = \langle \mathbf \Phi^*_{1S} | H_{q\bar{q}} | \mathbf \Phi^*_{1S} \rangle$. In order to avoid negative infinity, a Gaussian smearing function is used to weaken the singularity such that $\delta^3(\mathbf{r})\rightarrow(\kappa^3/\pi^{3/2})e^{-\kappa^2\mathbf{r}^2}$ \cite{Choi:2015ywa}. Finally, the in-medium masses of $D$ and $D^{*}$ mesons
are evaluated using the expression \cite{Arifi:2023jfe,Choi:2009ai} 
\begin{eqnarray}
M^*_{q\bar{q}} & = & \frac{\beta}{\sqrt{\pi}} \sum_{i=q, \bar{q}} z_i e^{z_i/2} \textbf{\textit{K}}_1\left(\frac{z_i}{2}\right) + a + \frac{2b} {\beta \sqrt{\pi}} -\frac{ 8 \alpha_s \beta}{3 \sqrt{\pi}} +\frac{ 32 \alpha_s \beta^3\langle \mathbf S_q \cdot \mathbf S_{\bar{q}}\rangle}{9 \sqrt{\pi} m^*_q m^*_{\bar{q}}}.
\label{eq:1}
\end{eqnarray}
where ${z_i} = m_i^{*2}/{\beta^2}$ and $\textbf{\textit{K}}_1$ is the modified Bessel function of the second kind. 

\section{RESULTS AND DISCUSSION}
\label{sec:3}
In this section, we present our results for the medium-modified masses, weak decay constants, and DAs for the pseudoscalar mesons \(D^0\), \(D^+\), \(D_s\) and vector mesons \(D^{0*}\), \(D^{+*}\), \(D_s^*\). We analyze the influence of finite baryon density $\rho_B$, isospin asymmetry $\eta$, and temperature \(T\) on the in-medium properties of mesons by incorporating effective quark masses derived using the CQMF model. The parameters utilized to solve the equations of motion to derive the density and temperature dependence of the scalar fields ($\sigma$, $\zeta$, $\delta$) and vector fields ($\omega$, $\rho$) are detailed in Table~\ref{tab:1}. The vacuum value of \(\chi_0\) and the coupling constant \(g_4\) are fitted to the effective nucleon mass. The parameter \(\xi\) is determined from the QCD \(\beta\)-function at one loop level for three colors and three flavours \cite{Papazoglou:1998vr}. The value of parameters \(k_0, k_1, k_2, k_3\) and \(k_4\) used in Eq.~\eqref{eq:5}, are determined utilizing the \(\pi\) meson mass, \(m_\pi\), \(K\) meson mass, \(m_K\) and the averaged mass of \(\eta\) and \(\eta'\) mesons \cite{Kumar:2023owb}. The vacuum expectation values of the scalar meson fields \(\sigma\) and \(\zeta\) represented by \(\sigma_0\) and \(\zeta_0\), are defined using the decay constants of the pion (\(f_\pi\)) and the kaon (\(f_K\)). These are given by the relations \(\sigma_0=-f_\pi\) and \(\zeta_0=\frac{1}{2}(f_\pi-2f_K)\) \cite{Kumar:2023owb}. The values of constituent quark masses in the free space (in units of GeV) and other potential parameters obtained by fitting the ground state mass spectra of \(D^0\) and \(D^{0*}\) mesons are summarized as $m_{u/d}=0.256$ GeV, $m_s=0.457$ GeV, $m_c=1.27$ GeV \cite{ParticleDataGroup:2024cfk}, $\alpha_s=0.565$ GeV, $a=-0.043$ GeV, $b=0.18$ GeV$^2$ \cite{Arifi:2023jfe} and $\kappa=0.451$ \cite{Dhiman:2019ddr}. The $\beta$ parameters for the pseudoscalar \(D\) and vector \(D^*\) mesons computed by the variational principle (in the units of GeV) are given in Table~\ref{tab:2}.

\par The masses of the ground state pseudoscalar \(D^0, D^+, D_s\) and vector \(D^{0*}, D^{+*}, D_s^*\) mesons calculated using the above discussed model parameters are given in Table~\ref{tab:3}. We compared our findings with experimental data \cite{ParticleDataGroup:2024cfk} and previously obtained results via LFQM \cite{Choi:2015ywa} and QCD sum rules \cite{Wang:2015mxa}. Our predicted meson masses are in good agreement with those obtained from LFQM and QCD sum rules, and are consistent with experimental data \cite{ParticleDataGroup:2024cfk}. The calculated weak decay constants in vacuum are presented in Table~\ref{tab:4}, alongside results from lattice QCD \cite{Becirevic:1998ua}, QCD sum rules \cite{Wang:2015mxa}, the Bethe-Salpeter (BS) model \cite{Cvetic:2004qg, Wang:2005qx}, in addition to the experimental data \cite{ParticleDataGroup:2024cfk}. For pseudoscalar \(D\) mesons, our predictions are in reasonable agreement with those from the BS model, whereas for the case of vector \(D^*\) mesons, our results show closer agreement with lattice QCD and QCD sum rules. 
%In both cases, our predicted decay constants are comparable to the available experimental values.

\begin{table}[h]
    \centering
    \renewcommand{\arraystretch}{1.8} % Adjusted for better readability
    \resizebox{0.6\textwidth}{!}{
    \begin{tabular}{|c|c|c|c|c|}
        \hline
        $k_0$ & $k_1$ & $k_2$ & $k_3$ & $k_4$ \\ 
        \hline
        4.94 & 2.12 & -10.16 & -5.38 & -0.06 \\ 
        \hline
        $\sigma_0$ (MeV) & $\zeta_0$ (MeV) & $\chi_0$ (MeV) &  $\xi$ & $\rho_0 (\text{fm}^{-3})$ \\ 
        \hline
        -92.8 & -96.5 & 254.6 & 6/33 & 0.16 \\ 
        \hline
        $g_{\sigma}^u = g_{\sigma}^d$ & $g_{\sigma}^s = g_{\zeta}^u = g_{\zeta}^d$ & $g_{\delta}^u$ & $g_{\zeta}^s = g_s$ & $g_4$ \\ 
        \hline
        2.72 & 0 & 2.72 & 3.847 & 37.4 \\ 
        \hline
        $g_{\delta}^p = g_{\delta}^u$ & $g_{\omega}^N = 3g_{\omega}^u$ & $g_{\rho}^p$ & $m_\pi$(MeV)& $m_K$(MeV) \\ 
        \hline
        2.72 & 9.69 & 8.886 & 139 & 494 \\ 
        \hline
        $f_\pi$(MeV) & $f_K$(MeV) & $g_{\delta}^s$ & &  \\
        \hline
        92.8&115& 0 & &\\
        \hline
    \end{tabular}}
    \vspace{0.5cm}
    \caption{List of model parameters used in the present work \cite{Wang:2001jw,Kumar:2023owb}.}
    \label{tab:1}
\end{table}
\begin{table}[h]
    \centering
    \resizebox{0.34\textwidth}{!}{
    \renewcommand{\arraystretch}{1.8} % Increases row height
    \begin{tabular}{|c|c|c|c|}
        \hline
        $J^{PC}$ & $\beta_{c\bar{u}}$ & $\beta_{c\bar{d}}$ & $\beta_{c\bar{s}}$ \\
        \hline
        $0^{-+}$ & 0.5331 & 0.5189 & 0.6066 \\
        $1^{--}$ & 0.4896 & 0.4681 & 0.4742 \\
        \hline
    \end{tabular}}
    \vspace{0.5cm}
    \caption{Gaussian parameter $\beta$ for pseudoscalar mesons (\(D^0\), \(D^+\), \(D_s\)) and vector mesons (\(D^{0*}\), \(D^{+*}\), \(D_s^*\)), determined using the variational principle.}
    \label{tab:2}
\end{table}

% \begin{table}[h]
%     \centering
%     \resizebox{1\textwidth}{!}{
%     \renewcommand{\arraystretch}{1.8} % Adjust row height
%     % \setlength{\tabcolsep}{2.9pt} % Adjust column spacing
% \begin{tabular}{| l | c c c c c c | c c c c c c |}
%         \hline
%         \multirow{2}{*}{} & \multicolumn{6}{c|}{\textbf{Mass of Meson (GeV)}} & \multicolumn{6}{c|}{\textbf{Weak Decay Constant (GeV)}} \\
%         \hline
%         & $M_{D^0}$ & $M_{D^+}$ & $M_{D_s}$ & $M_{D^{0*}}$ & $M_{D^{+*}}$ & $M_{D_s^*}$ & $f_{D^0}$ & $f_{D^+}$ & $f_{D_s}$ & $f_{D^{0*}}$ & $f_{D^{+*}}$ & $f_{D_s^*}$\\
%         \hline
%         Present Work & 1.865 & 1.868 & 2.008 & 2.007 & 2.009 & 2.111 & 0.234 & 0.228 & 0.284 & 0.282 & 0.268 & 0.276 \\
%         Exp.\cite{ParticleDataGroup:2024cfk}         & 1.865 & 1.869 & 2.007 & 2.010 & 1.968 & 2.112 & 0.206 & {-} & 0.257 & 0.239 & {-} & {-} \\
%         \hline
%     \end{tabular}}

%     \vspace{0.5cm}
    
%     \caption{The predicted ground state mass spectra and weak decay constant of pseudoscalar mesons (\(D^0\),\(D^+\),\(D_s\)) and vector mesons (\(D^{0*}\),\(D^{+*}\),\(D_s^*\)) and their comparison with the experimental data \cite{ParticleDataGroup:2024cfk}.}
%     \label{tab:3}
% \end{table}

\begin{table}[h]
    \centering
    \renewcommand{\arraystretch}{1.8} % Adjust row height
    \resizebox{0.9\textwidth}{!}{
    \begin{tabular}{| l | c c c c c c |}
        \hline
        \multirow{2}{*}{} & \multicolumn{6}{c|}{\textbf{Mass of Meson (GeV)}} \\
        \cline{2-7}
        & $M_{D^0}$ & $M_{D^+}$ & $M_{D_s}$ & $M_{D^{0*}}$ & $M_{D^{+*}}$ & $M_{D_s^*}$ \\
        \hline
        Present Work & 1.865 & 1.868 & 2.010 & 2.007 & 2.009 & 2.112 \\
        Exp.\cite{ParticleDataGroup:2024cfk} & 1.865 & 1.869 & 1.968 & 2.007 & 2.010 & 2.112 \\
        LFQM\ \makecell[l]{\cite{Choi:2015ywa}} & 1.875 &{-}& 1.981 &1.962 &{-}&2.031 \\
        QCD sum rules\ \makecell[l]{\cite{Wang:2015mxa}} &$1.87 \pm 0.10$  &{-}& \(1.97 \pm 0.10\)&$2.01 \pm 0.08$ &{-}&$2.11 \pm 0.07$ \\
        \hline
    \end{tabular}}

    \vspace{0.5cm}
    \caption{Predicted ground-state mass spectra (in GeV) of pseudoscalar  ($D^0$, $D^+$, $D_s$) and vector  ($D^{0*}$, $D^{+*}$, $D_s^*$) mesons, compared with experimental data from Ref. \cite{ParticleDataGroup:2024cfk} and other theoretical model predictions.}
    \label{tab:3}
\end{table}

\begin{table}[h]
    \centering
    \renewcommand{\arraystretch}{1.8} % Adjust row height
    \resizebox{0.9\textwidth}{!}{
    \begin{tabular}{| l | c c c c c c |}
        \hline
        \multirow{4}{*}{} & \multicolumn{6}{c|}{\textbf{Weak Decay Constant (MeV)}} \\
        \cline{2-7}
        & $f_{D^0}$ & $f_{D^+}$ & $f_{D_s}$ & $f_{D^{0*}}$ & $f_{D^{+*}}$ & $f_{D_s^*}$ \\
        \hline
        Present Work & 234 & 228 & 285 & 282 & 268 & 280 \\
        Exp.\cite{ParticleDataGroup:2024cfk} & 206 & {-} & 257 & 239 & {-} & {-} \\
        Lattice QCD\ \makecell[l]{\cite{Becirevic:1998ua}} & $211 \pm 14^{+0}_{-12}$ & {-} & $231 \pm 12^{+6}_{-1}$ & $245 \pm 20^{+0}_{-2}$ & {-} & $272 \pm 16^{+0}_{-20}$ \\
        QCD sum rules\ \makecell[l]{\cite{Wang:2015mxa}} &$208 \pm 10$  &{-}& \(240 \pm 10\)&$263 \pm 21$ &{-}&$308 \pm 21$ \\
        BS\ \makecell[l]{\cite{Cvetic:2004qg, Wang:2005qx}} &$230 \pm 25$  &$230 \pm 25$ & \(248 \pm 27\)&$340 \pm 23$ &$340 \pm 23$ &$375 \pm 24$ \\
        \hline
    \end{tabular}}
    \vspace{0.5cm}
    \caption{Predicted weak decay constants (in MeV) for pseudoscalar  ($D^0$, $D^+$, $D_s$) and vector  ($D^{0*}$, $D^{+*}$, $D_s^*$) mesons in the free space, compared with experimental data from Ref. \cite{ParticleDataGroup:2024cfk} and other theoretical model predictions.}
    \label{tab:4}
\end{table}

\subsection{In-medium masses of \(D\) and \(D^*\) meson}

\begin{figure}[h]
    \centering
    % First Row (T = 0)
    \begin{minipage}{0.33\textwidth}
        \centering
        \includegraphics[width=\textwidth,height=1.05\textwidth]{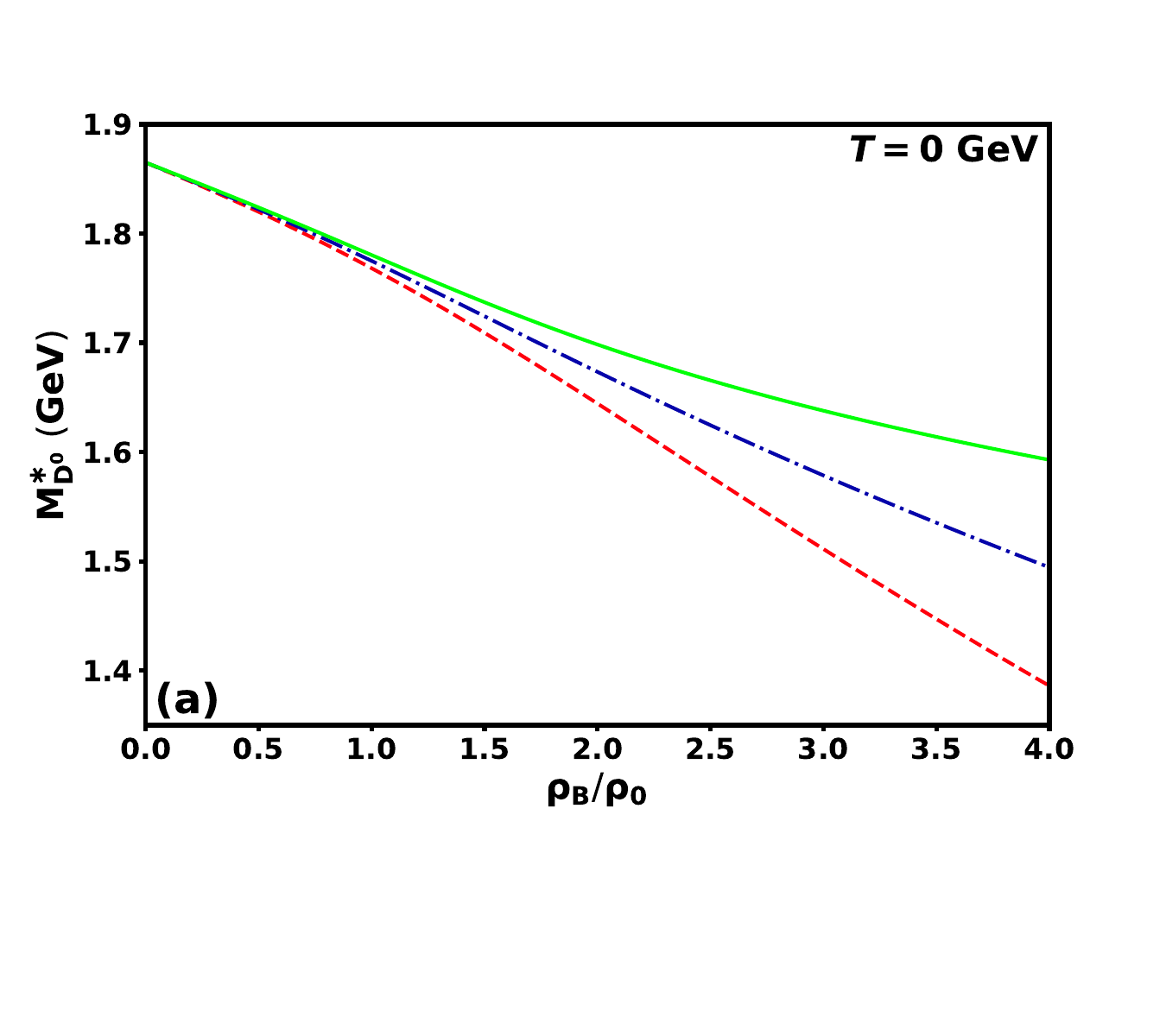}
    \end{minipage}\hspace{-1mm}
    \begin{minipage}{0.33\textwidth}
        \centering
        \includegraphics[width=\textwidth,height=1.05\textwidth]{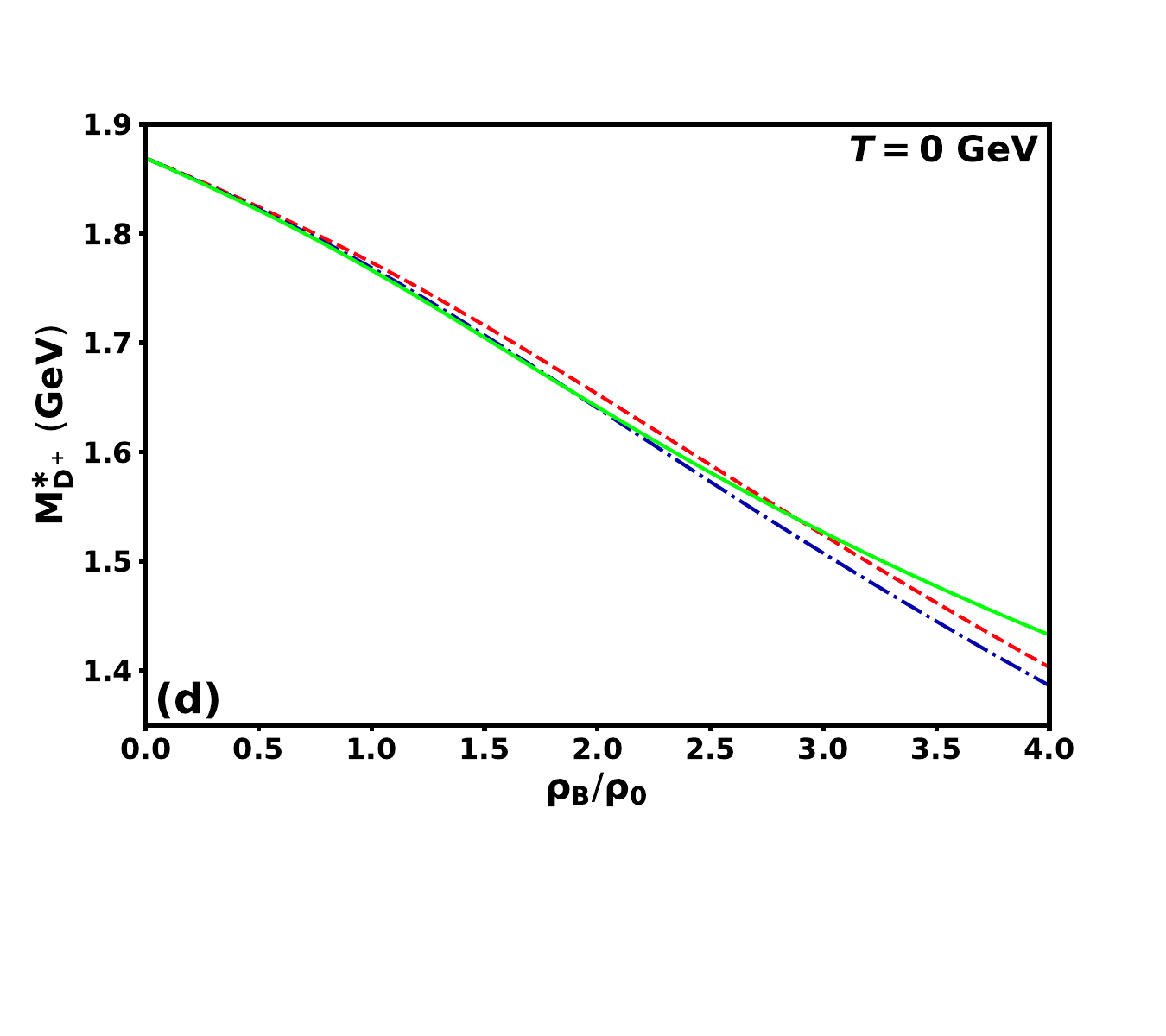}
    \end{minipage}\hspace{-1mm}
    \begin{minipage}{0.33\textwidth}
        \centering
        \includegraphics[width=\textwidth,height=1.05\textwidth]{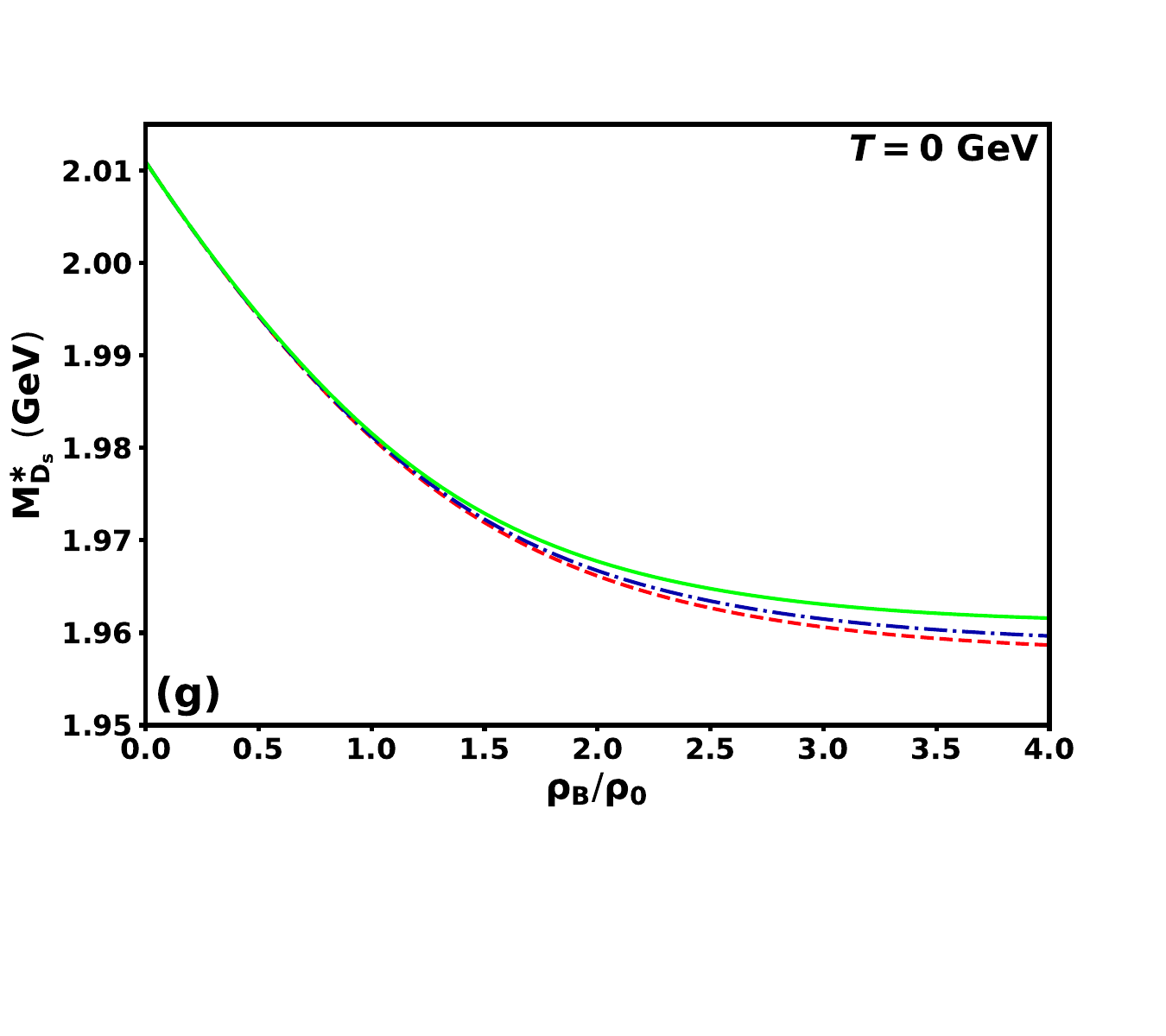}
    \end{minipage}
    
    \vspace{-10mm} % Reduce vertical spacing

    % Second Row (T = 100)
    \begin{minipage}{0.33\textwidth}
        \centering
        \includegraphics[width=\textwidth,height=1.05\textwidth]{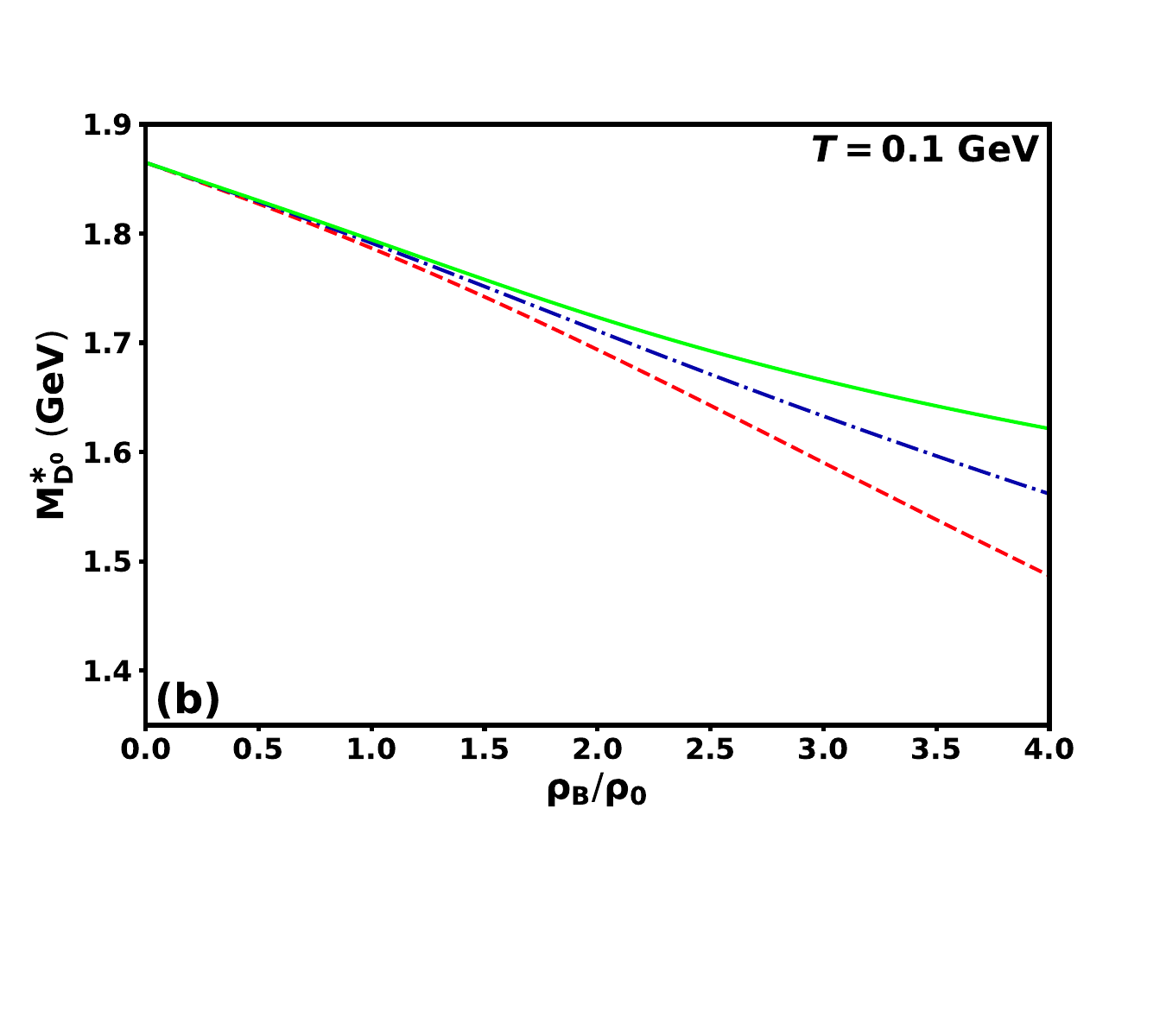}
    \end{minipage}\hspace{-1mm}
    \begin{minipage}{0.33\textwidth}
        \centering
        \includegraphics[width=\textwidth,height=1.05\textwidth]{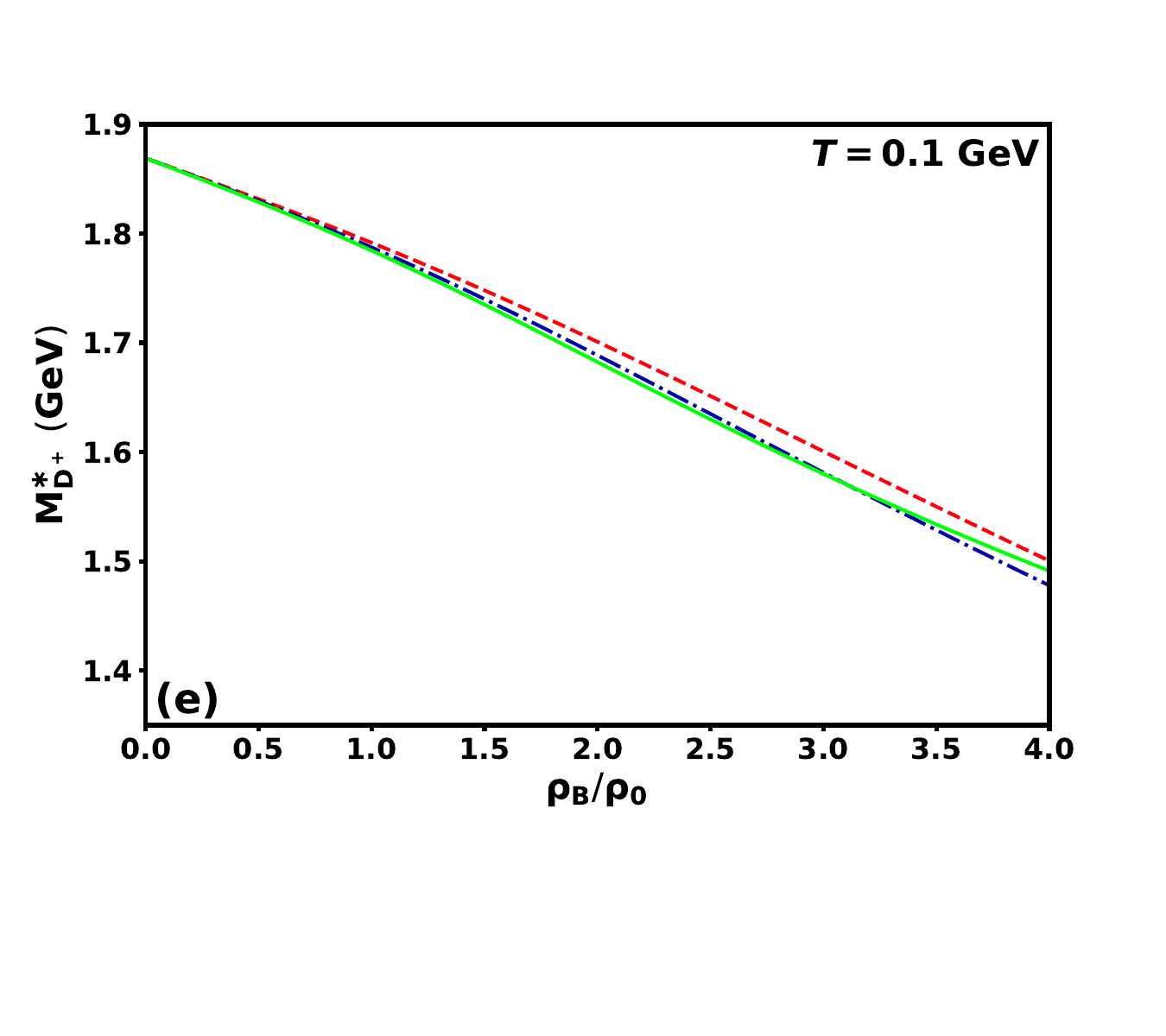}
    \end{minipage}\hspace{-1mm}
    \begin{minipage}{0.33\textwidth}
        \centering
        \includegraphics[width=\textwidth,height=1.05\textwidth]{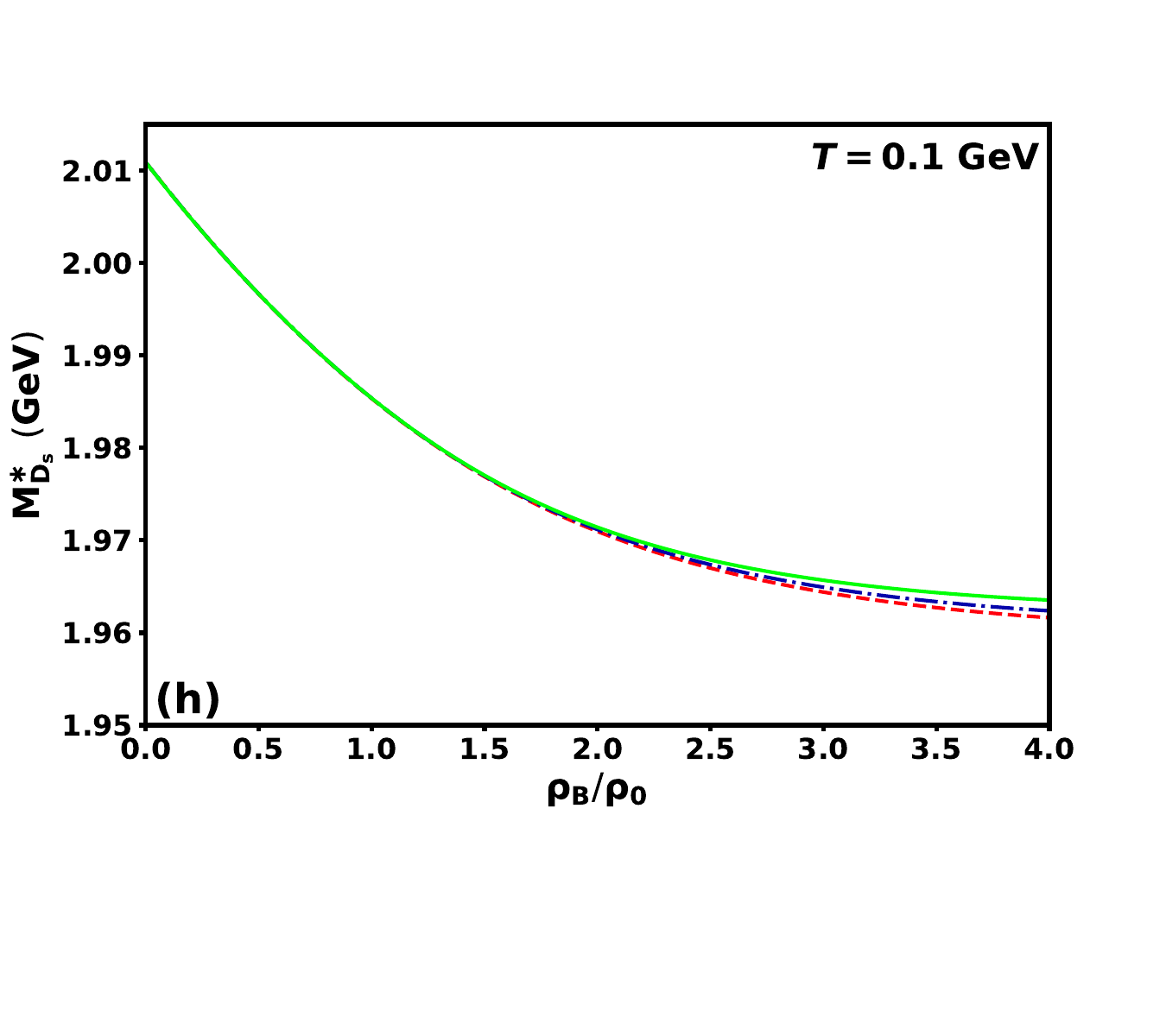}
    \end{minipage}

    \vspace{-10mm} % Reduce vertical spacing

    % Third Row (T = 150)
    \begin{minipage}{0.33\textwidth}
        \centering
        \includegraphics[width=\textwidth,height=1.05\textwidth]{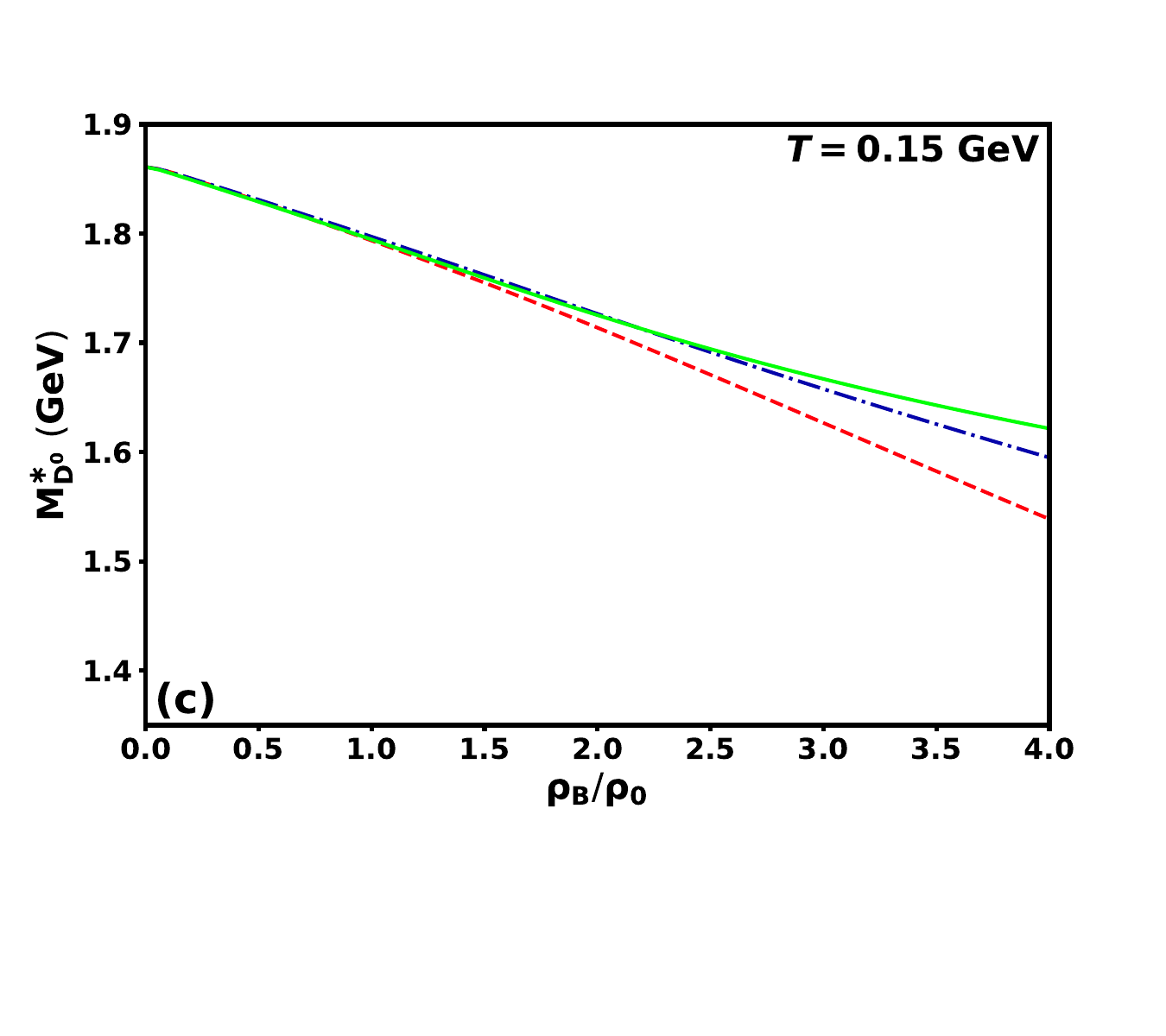}
    \end{minipage}\hspace{-1mm}
    \begin{minipage}{0.33\textwidth}
        \centering
        \includegraphics[width=\textwidth,height=1.05\textwidth]{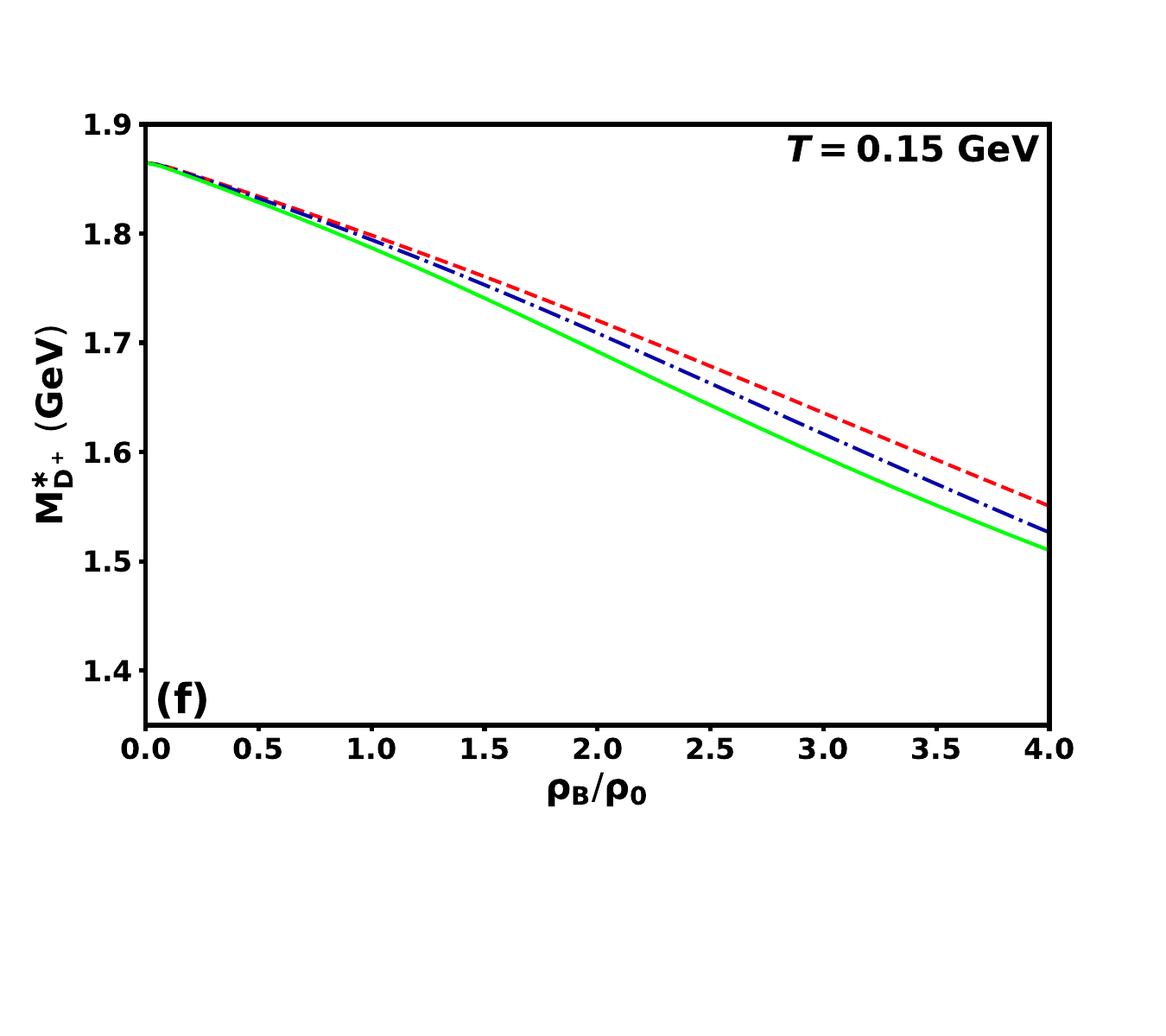}
    \end{minipage}\hspace{-1mm}
    \begin{minipage}{0.33\textwidth}
        \centering
        \includegraphics[width=\textwidth,height=1.05\textwidth]{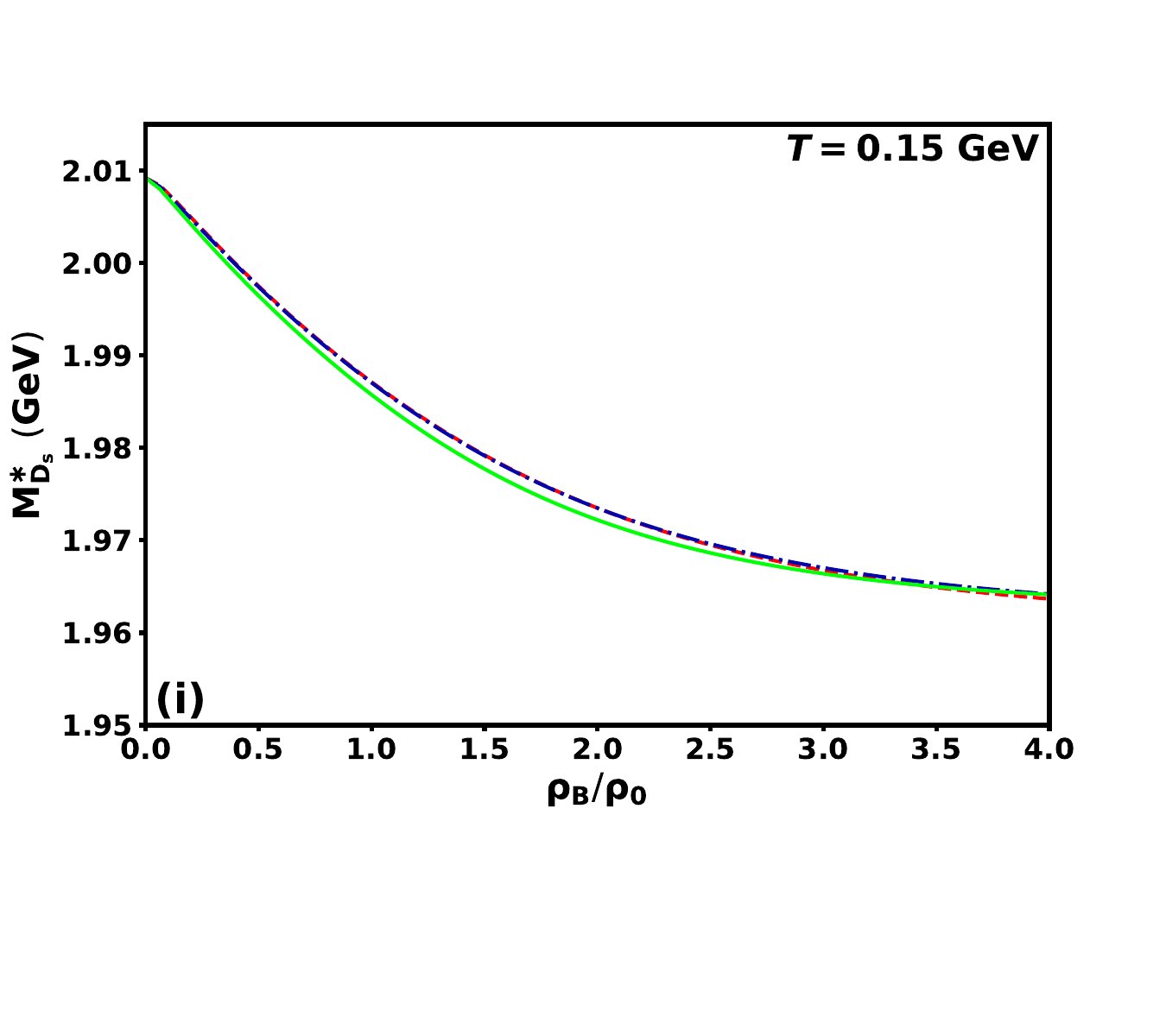}
    \end{minipage}
    
    \vspace{-10mm}
    
    \begin{minipage}{0.67\textwidth} 
        \centering
        \includegraphics[width=\textwidth,height=0.15\textwidth]{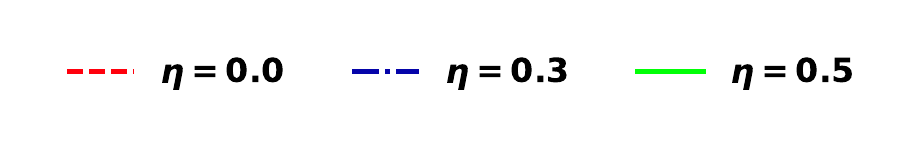} % Extra image that spans across
    \end{minipage}
    \caption{In-medium masses of pseudoscalar \(D^0\), \(D^+\) and \(D_s\) mesons as a function of baryon density $\rho_B$ (in units of nuclear saturation density $\rho_0$). Results are shown for temperatures \(T=0\) GeV [subplots (a), (d) and (g)], \(T=0.1\) GeV [subplots (b), (e) and (h)] and \(T=0.15\) GeV [subplots (c), (f) and (i)] GeV, at isospin asymmetry values $\eta= 0, 0.3$ and \(0.5\).}
    \label{fig:1}
\end{figure}

Using the effective quark masses \(m_q^*\) obtained from the CQMF model in  Eq.~\eqref{eq:1}, the in-medium masses of pseudoscalar \(D\) and vector \(D^*\) mesons are calculated in an isospin asymmetric nuclear medium . The variation of the medium-modified masses of pseudoscalar \(D^0\), \(D^+\) and \(D_s\) mesons as a function of baryon density $\rho_B$ (in units of nuclear saturation density $\rho_0$) for different values of temperatures \(T=0, 0.1,\) and \(0.15\) GeV is shown in Fig.~\ref{fig:1}. In each subplot, the results are plotted for the isospin asymmetry values $\eta = 0, 0.3$, and $0.5$. For all three pseudoscalar mesons, the effective mass decreases with increasing $\rho_B$. The reduction is more pronounced for \(D^0\) and \(D^+\) mesons compared to the \(D_s\) meson. This is due to the presence of lighter \(u\) and \(d\) quarks in the former, which are more sensitive to medium effects. As shown in Figs.~\ref{fig:1}(a) and \ref{fig:1}(d), isospin asymmetry significantly impacts the behavior of the members of isospin doublet \(D=\begin{pmatrix} D^+ \\ D^0 \end{pmatrix}\) in the nuclear matter. The mass of \(D^0\) meson increases notably as \(\eta\) rises from 0 to \(0.5\). For the \(D^+\) meson, the in-medium  mass increases slightly  as $\eta$ changes from $0$ to  $0.3$. However, at low baryon densities, effective mass of $D^{+}$ mesons is observed to be less  at \(\eta=0.5\) compared to $0.3$.
%,  it initially decreases at low $\rho_B$ and then increases at higher d
%ensities. 
This splitting in the effective masses of \(D^0\) and \(D^+\) mesons originates from the scalar-isoscalar field $\delta$, which becomes non-zero in an isospin asymmetric medium. Because \(I^{3u}=1/2=-I^{3d}\), the resulting difference between the effective masses of \(u\) and \(d\) quarks leads to a difference in the in-medium masses of \(D^0\) and \(D^+\) mesons. Additionally, as the temperature increases, the mass difference attributed to isospin asymmetry for the \(D^0\) meson becomes less pronounced, as shown in Figs.~\ref{fig:1}(b) and \ref{fig:1}(c). In contrast, for the \(D^+\) meson, increasing the temperature from \(T=0\) to \(0.15\) GeV, gradually reverses the trend, with its effective mass decreasing under higher isospin asymmetry, as illustrated in Fig.~\ref{fig:1}(f). For a symmetric nuclear medium (\(\eta=0\)) at \(\rho_B=\rho_0(3\rho_0)\), the mass shifts of \(D^0\) meson are found to be \(-0.097(-0.354)\), \(-0.078(-0.275)\), and \(-0.067(-0.234)\) GeV at temperatures $T=0,0.1,$ and 0.15 GeV, respectively. When isospin asymmetry is introduced with \(\eta=0.5\), the corresponding shifts reduce to \(-0.085(-0.227)\), \(-0.07(-0.199)\), and \(-0.066(-0.194)\) GeV from its vacuum mass of \(1.865\) GeV. 
%In contrast, the \(D^+\) meson shows an opposite trend.
 For the \(D^+\) meson, at $\eta=0$, the mass shifts are observed to be \(-0.095(-0.345)\), \(-0.077(-0.268)\), and \(-0.067(-0.229)\) GeV at the same temperatures and densities. Under isospin asymmetry \(\eta=0.5\), values of mass shifts change to \(-0.103(-0.342)\), \(-0.085(-0.289)\), and \(-0.078(-0.269)\) GeV, indicating that the effective mass of the \(D^+\) meson decreases with higher isospin asymmetry (except for  higher baryon densities at $T = 0$ GeV).

\par
\begin{figure}[h]
    \centering
    % First Row (T = 0)
    \begin{minipage}{0.33\textwidth}
        \centering
        \includegraphics[width=\textwidth,height=1.2\textwidth]{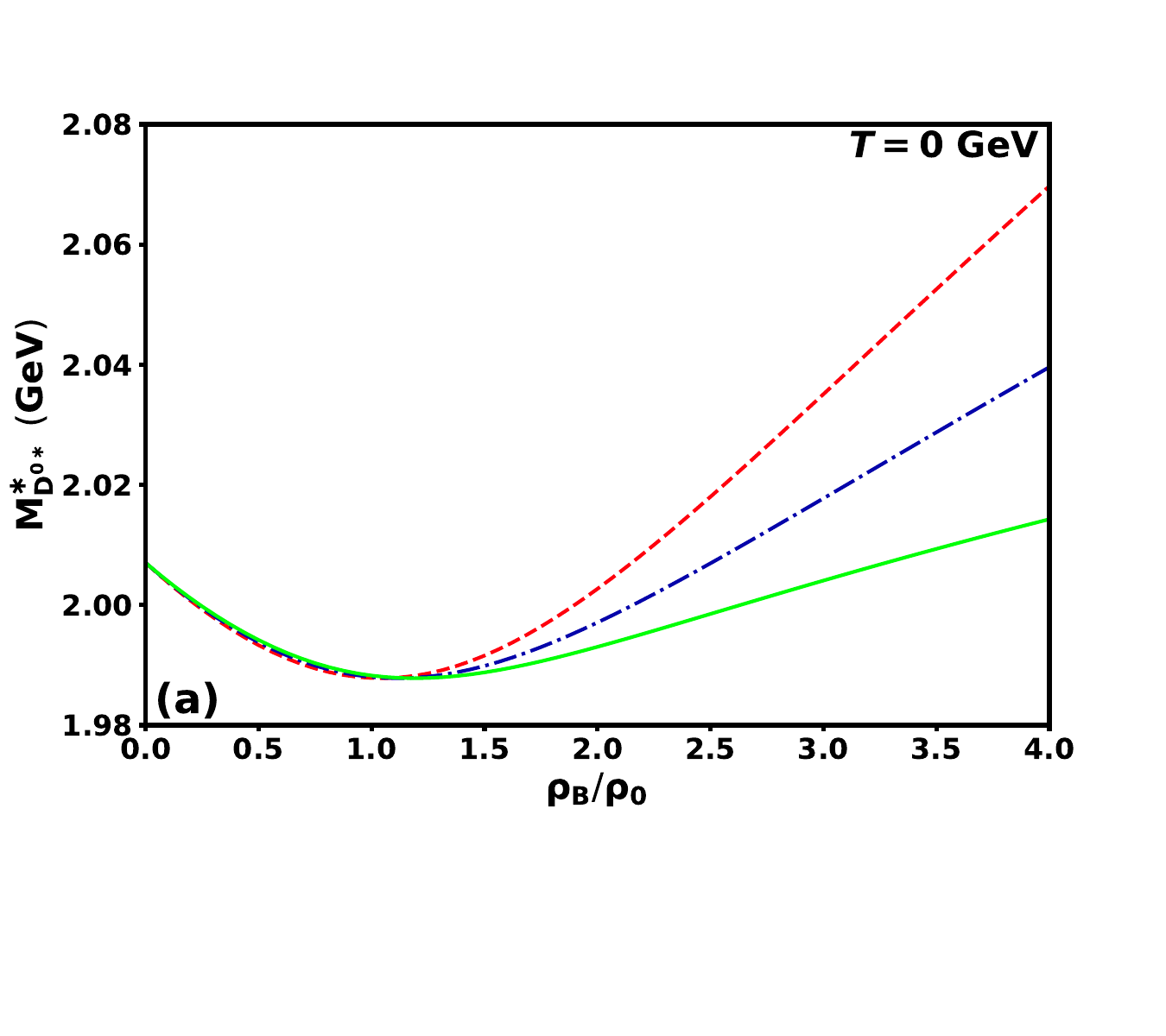}
    \end{minipage}\hspace{-1mm}
    \begin{minipage}{0.33\textwidth}
        \centering
        \includegraphics[width=\textwidth,height=1.2\textwidth]{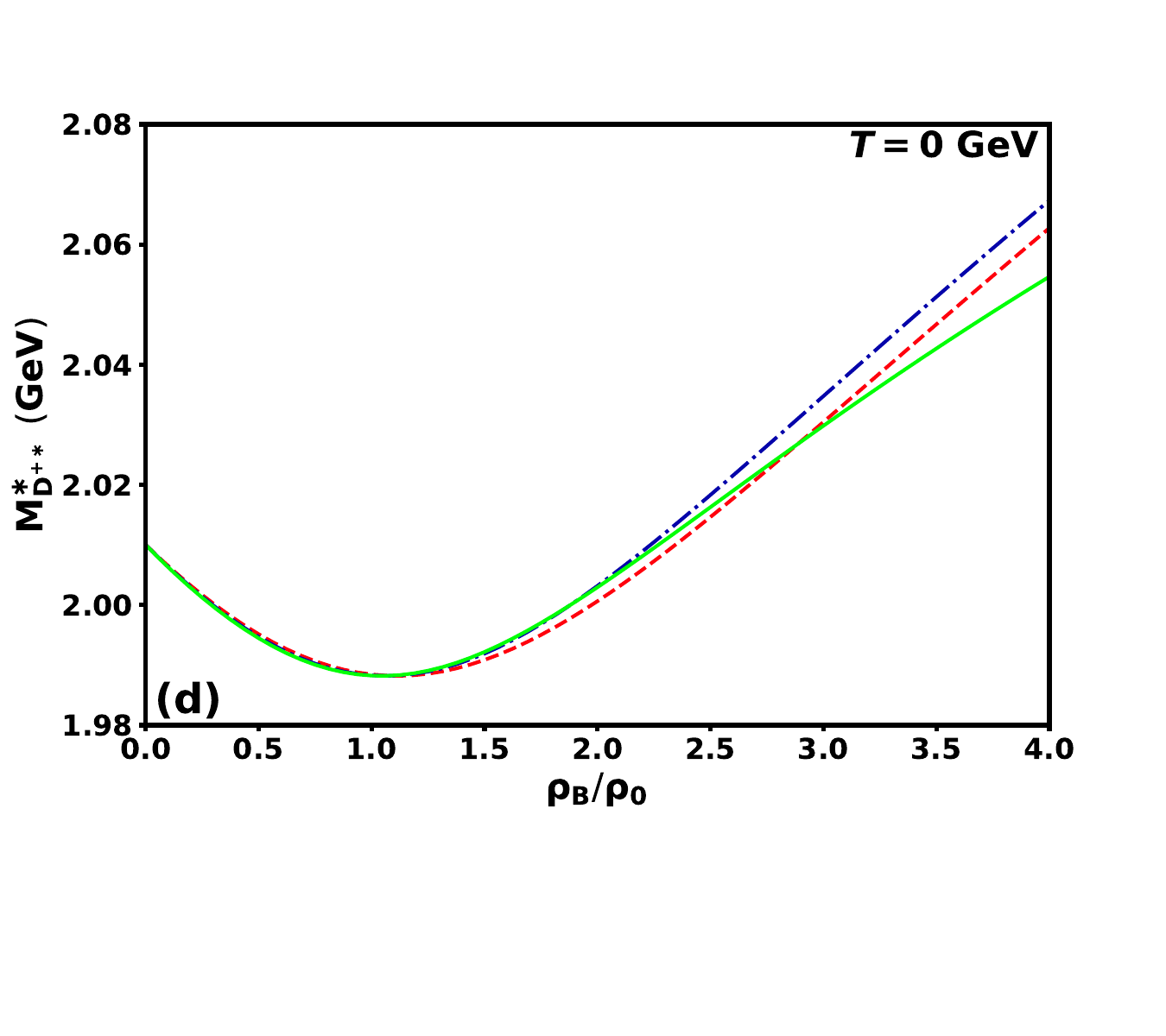}
    \end{minipage}\hspace{-1mm}
    \begin{minipage}{0.33\textwidth}
        \centering
        \includegraphics[width=\textwidth,height=1.2\textwidth]{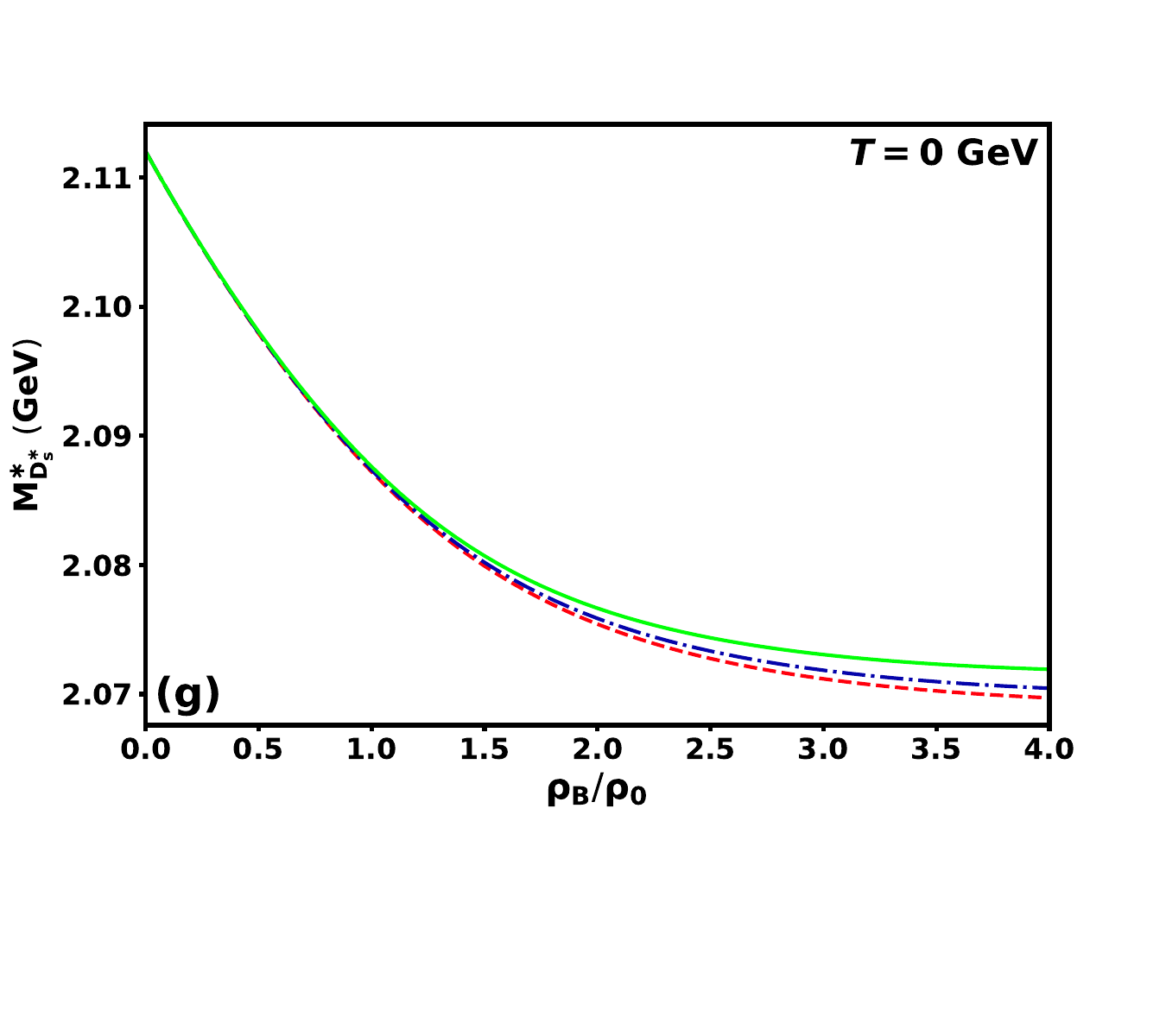}
    \end{minipage}
    
    \vspace{-13mm} % Reduce vertical spacing

    % Second Row (T = 100)
    \begin{minipage}{0.33\textwidth}
        \centering
        \includegraphics[width=\textwidth,height=1.2\textwidth]{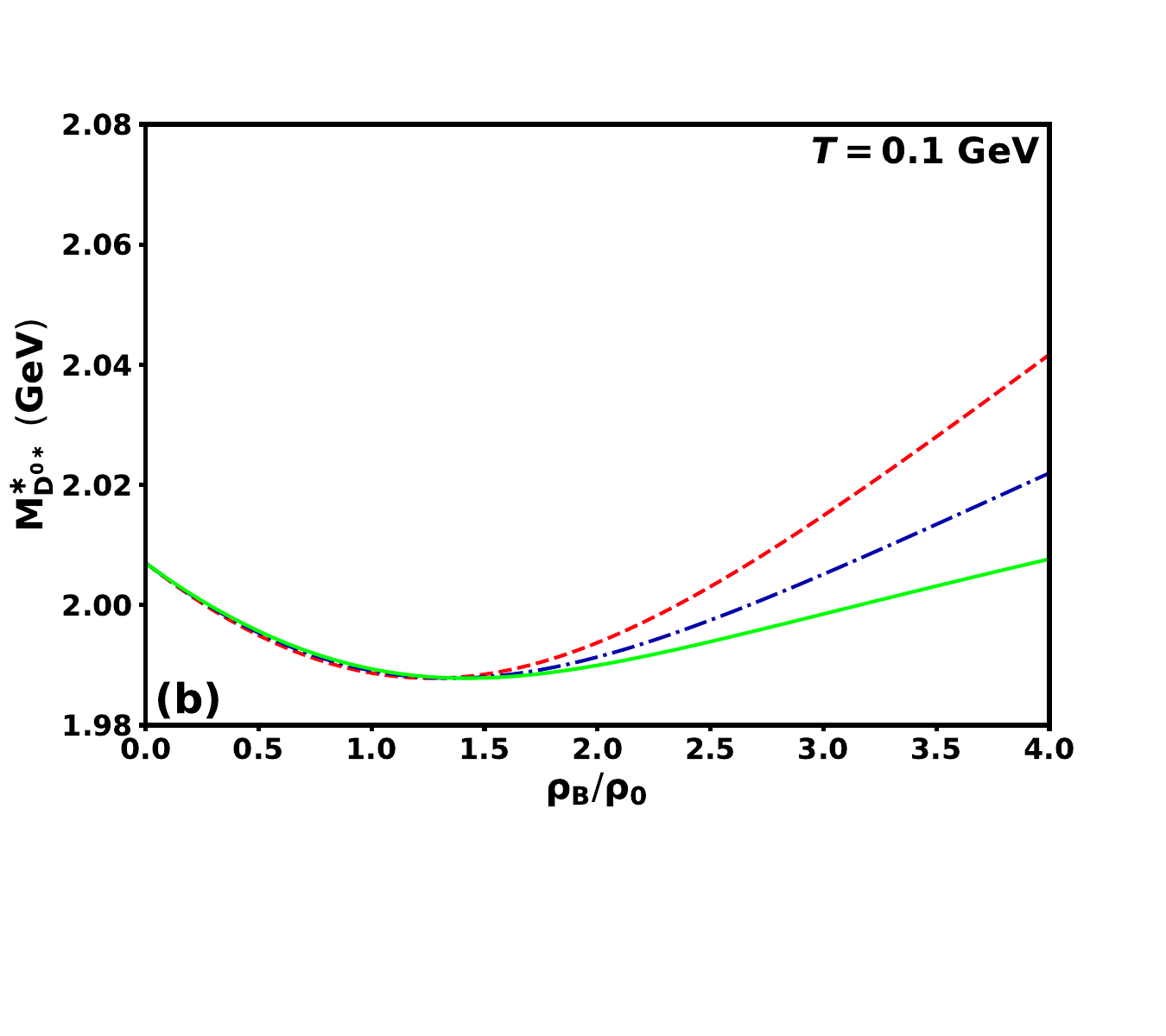}
    \end{minipage}\hspace{-1mm}
    \begin{minipage}{0.33\textwidth}
        \centering
        \includegraphics[width=\textwidth,height=1.2\textwidth]{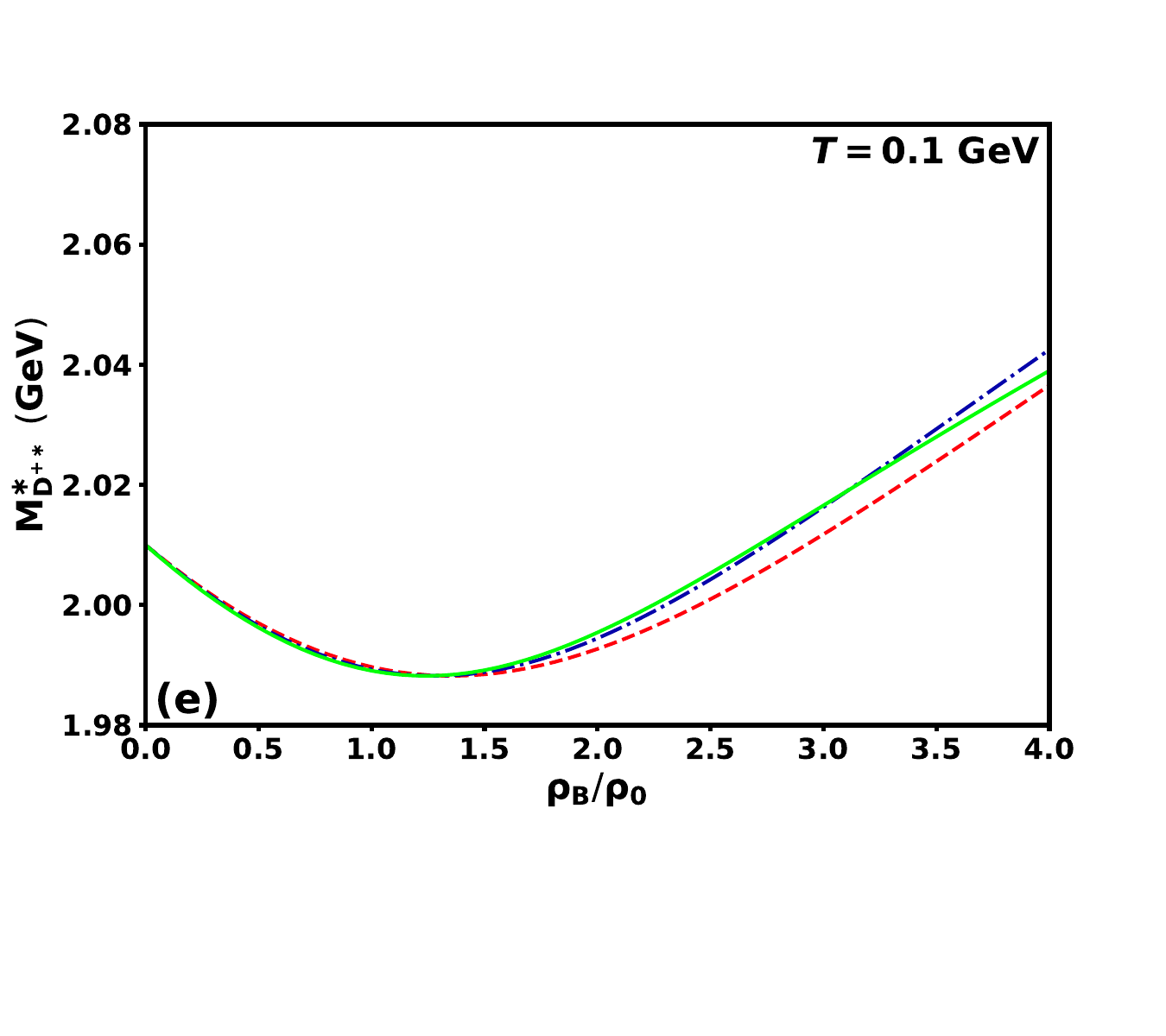}
    \end{minipage}\hspace{-1mm}
    \begin{minipage}{0.33\textwidth}
        \centering
        \includegraphics[width=\textwidth,height=1.2\textwidth]{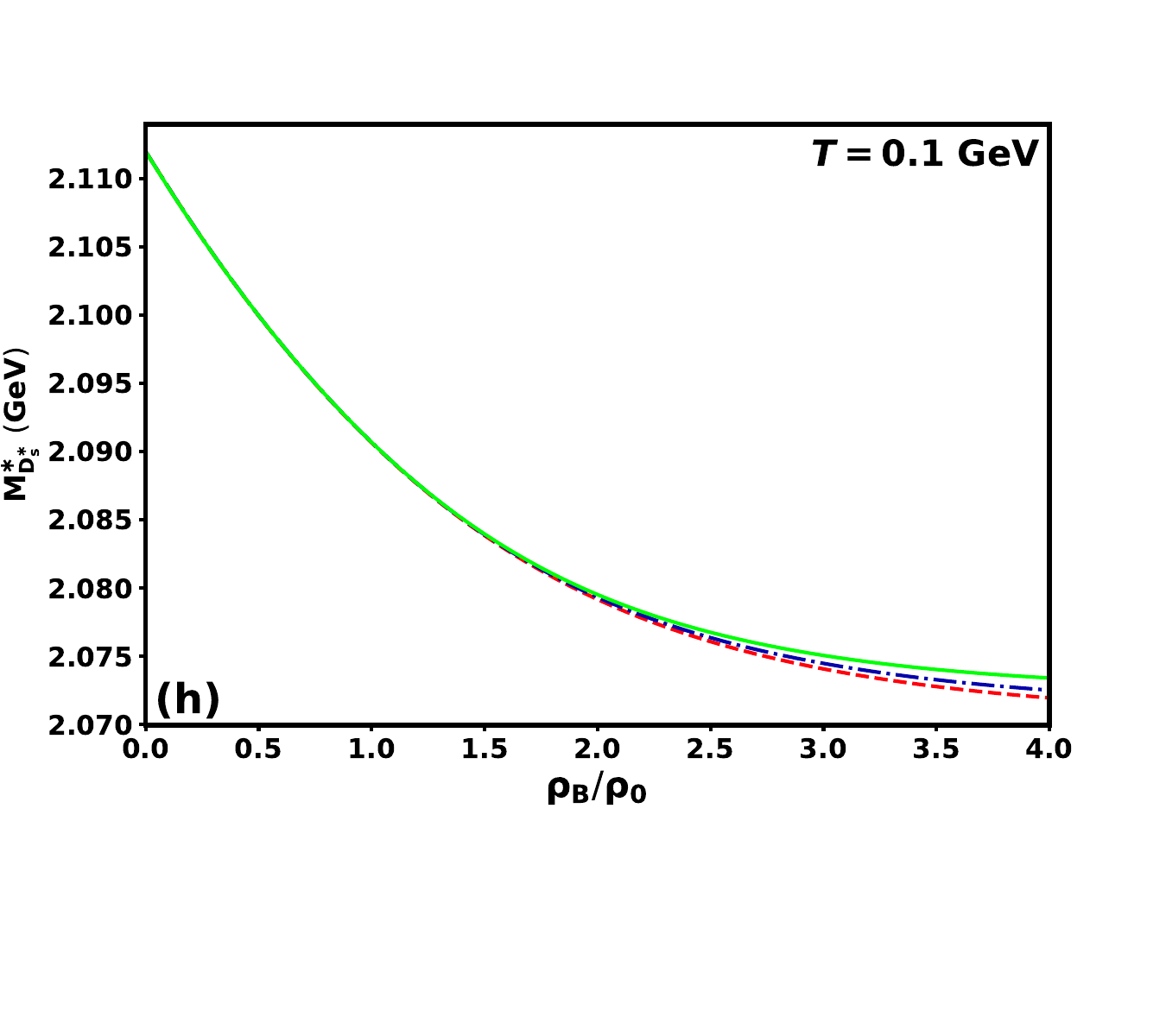}
    \end{minipage}

    \vspace{-13mm} % Reduce vertical spacing

    % Third Row (T = 150)
    \begin{minipage}{0.33\textwidth}
        \centering
        \includegraphics[width=\textwidth,height=1.2\textwidth]{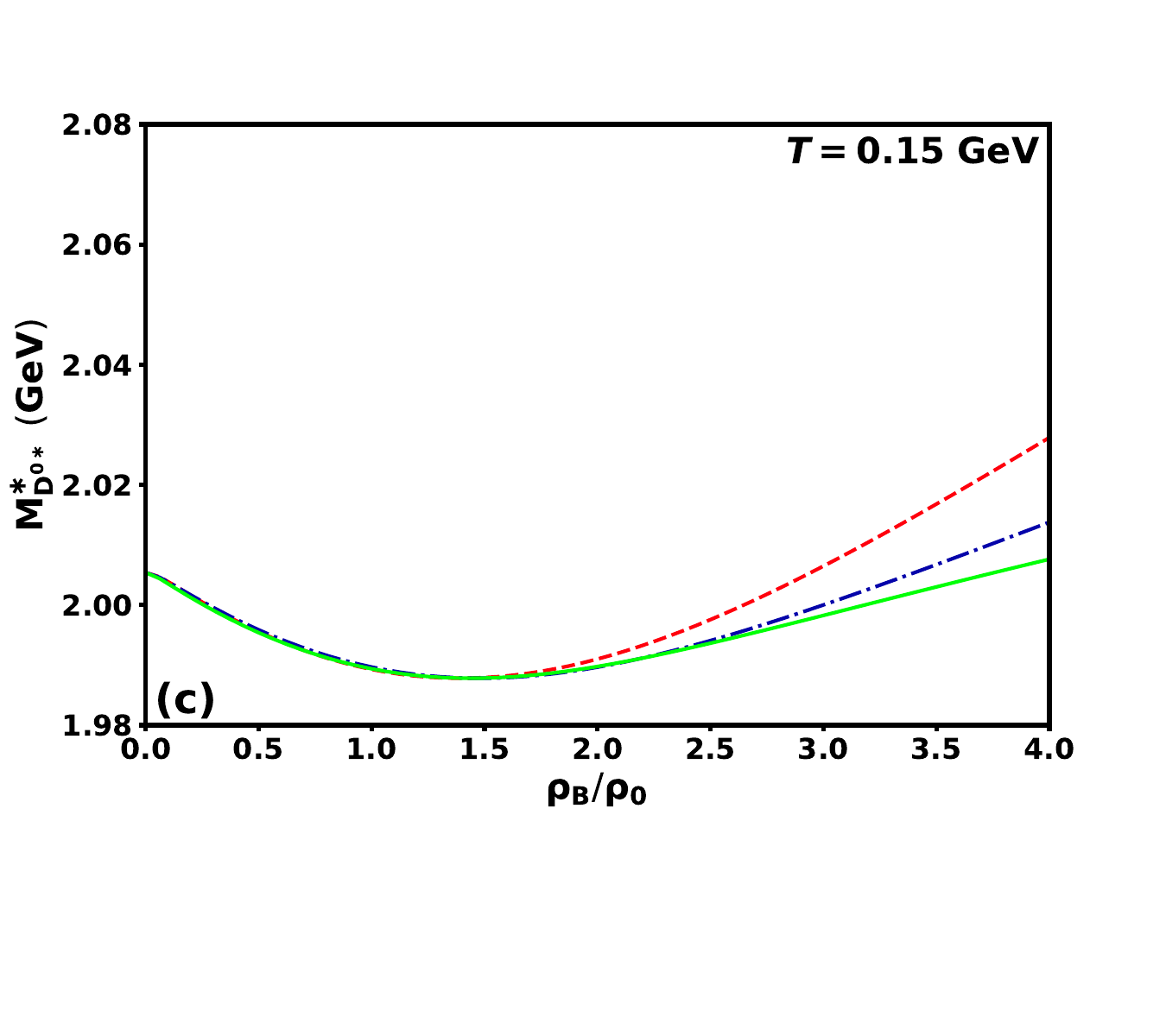}
    \end{minipage}\hspace{-0.8mm}
    \begin{minipage}{0.33\textwidth}
        \centering
        \includegraphics[width=\textwidth,height=1.2\textwidth]{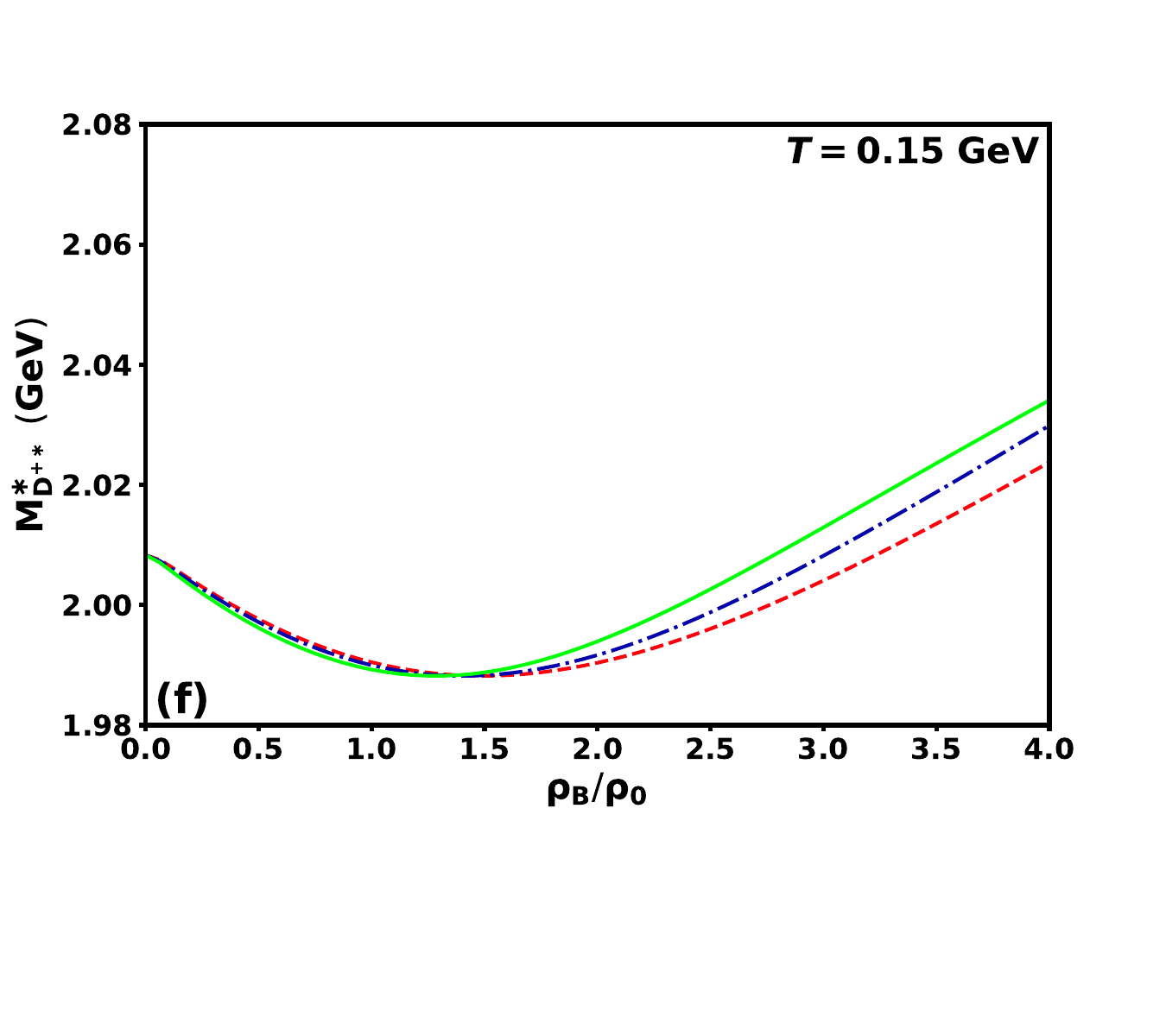}
    \end{minipage}\hspace{-0.8mm}
    \begin{minipage}{0.33\textwidth}
        \centering
        \includegraphics[width=\textwidth,height=1.2\textwidth]{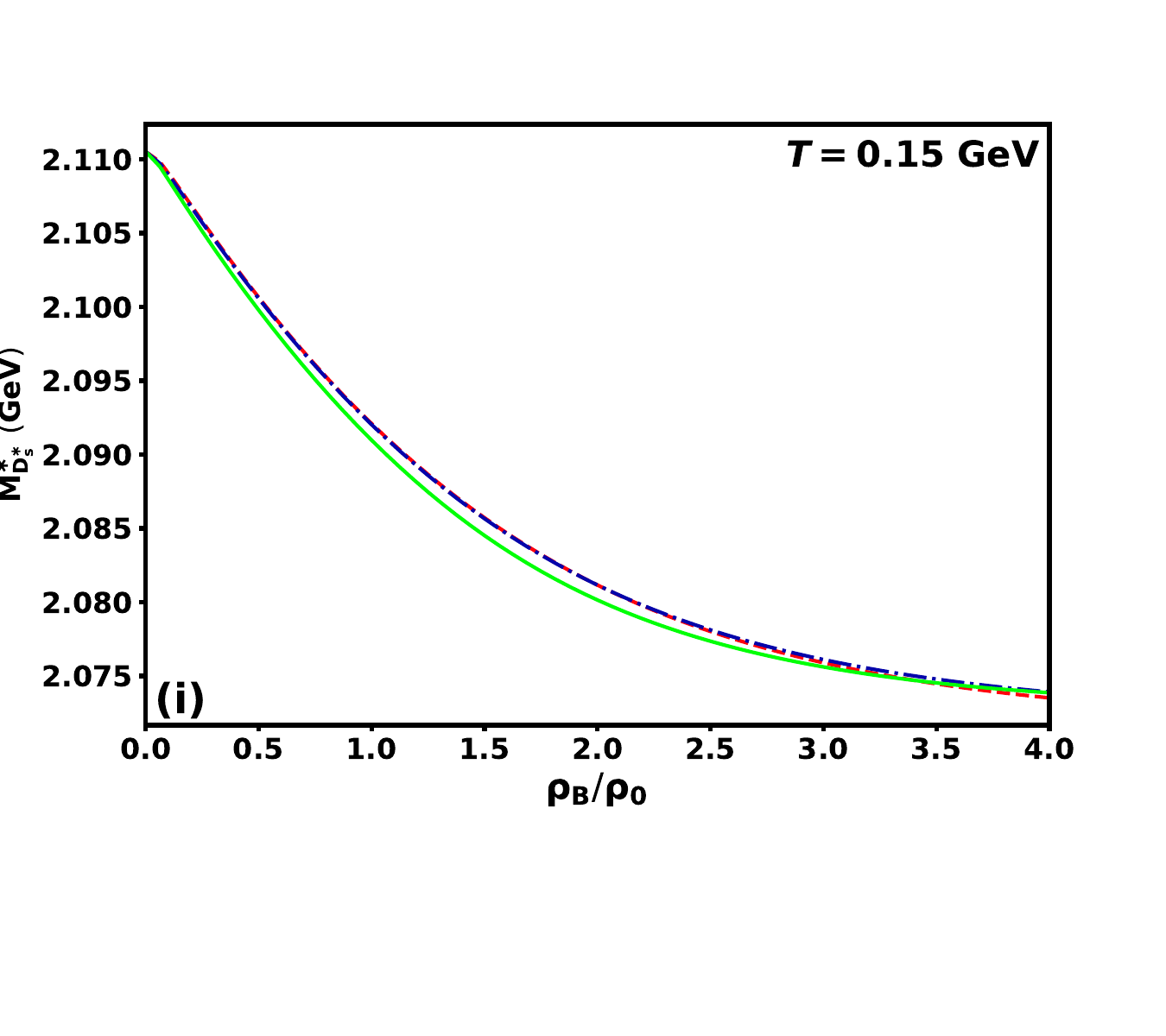}
    \end{minipage}
    
    \vspace{-13mm}
    
    \begin{minipage}{0.67\textwidth} 
        \centering
        \includegraphics[width=\textwidth,height=0.15\textwidth]{Masses/T0/Mass_D0_T0_diff_eta_legend.pdf} % Extra image that spans across
    \end{minipage}
    \caption{In-medium masses of vector \(D^{0*}\), \(D^{+*}\) and \(D_s^*\) mesons plotted as a function of baryon density $\rho_B$ (in units of $\rho_0$). Results are shown for temperatures \(T=0\) GeV [subplots (a), (d) and (g)], \(T=0.1\) GeV[subplots (b), (e) and (h)] and \(T=0.15\) GeV[subplots (c), (f) and (i)] GeV, at isospin asymmetry values $\eta= 0, 0.3$ and \(0.5\).}
    \label{fig:2}
\end{figure}

\begin{figure}[htbp]
    \centering
    % 2x2 Grid of Figures
    \begin{minipage}{0.48\linewidth}
        \centering
        \includegraphics[width=\linewidth]{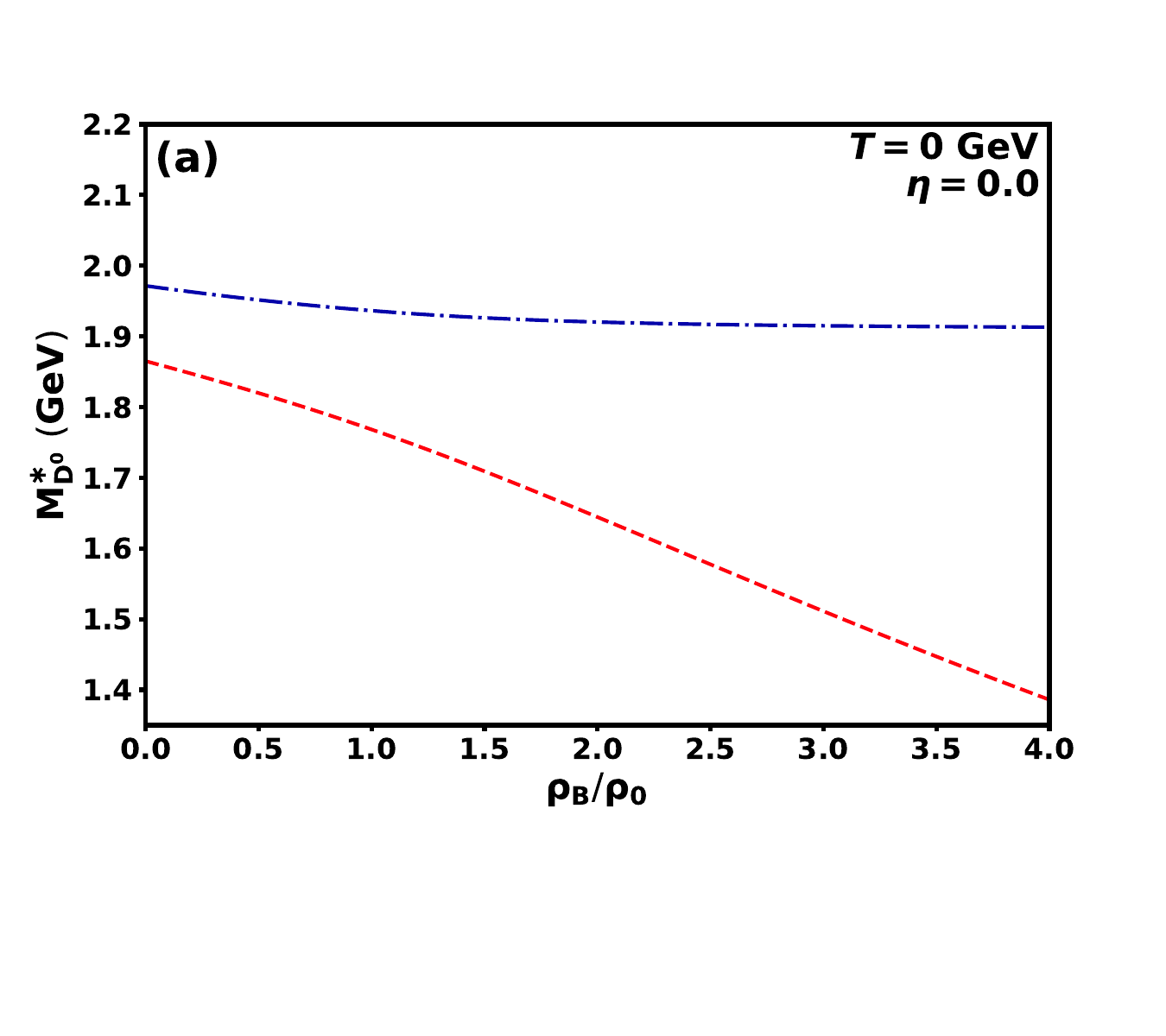}
    \end{minipage} \hspace{-6pt}
    \begin{minipage}{0.48\linewidth}
        \centering
        \includegraphics[width=\linewidth]{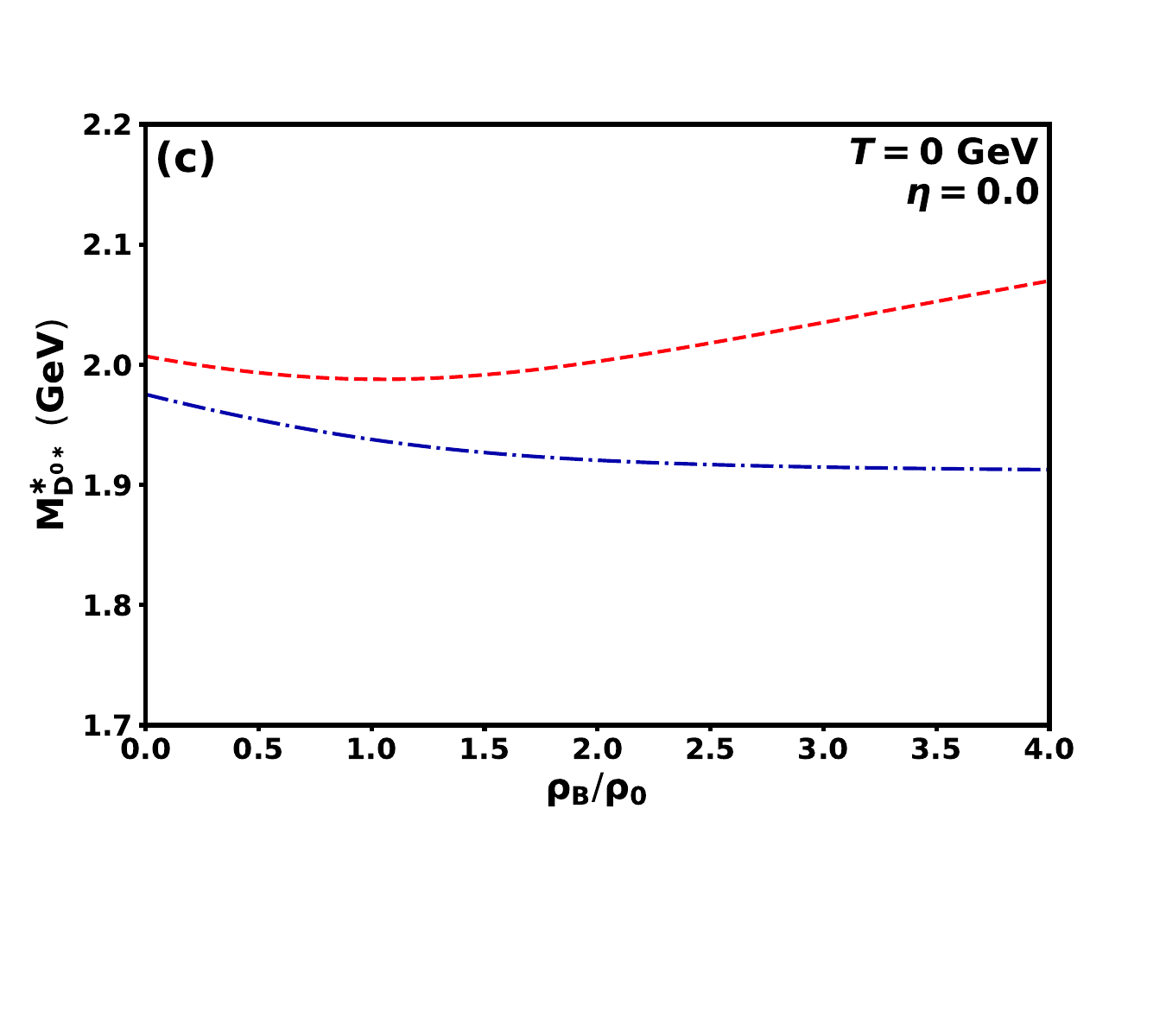}
    \end{minipage}

    \vspace{-15mm} % Reduce vertical space

    \begin{minipage}{0.48\linewidth}
        \centering
        \includegraphics[width=\linewidth]{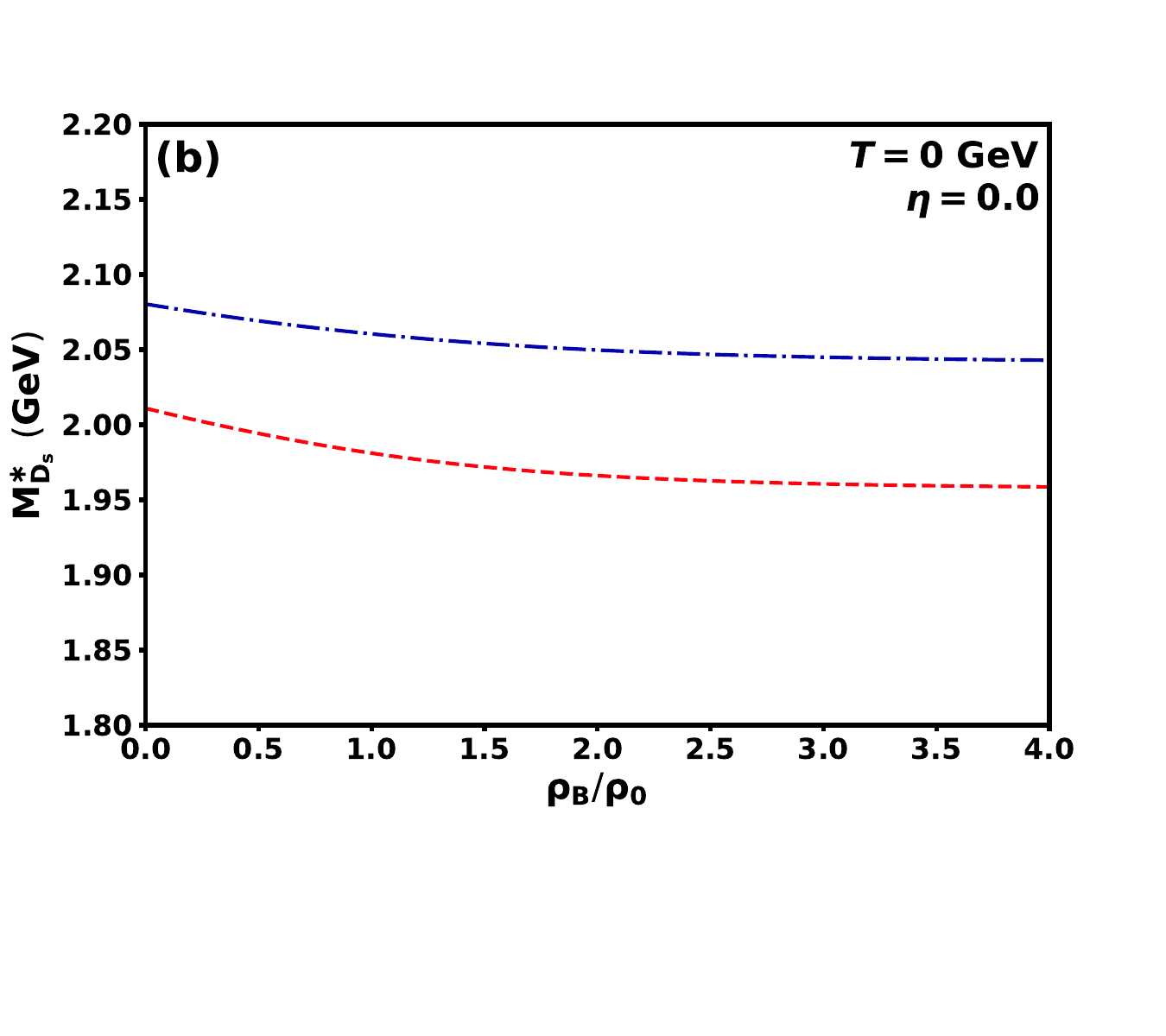}
    \end{minipage} \hspace{-6pt}
    \begin{minipage}{0.48\linewidth}
        \centering
        \includegraphics[width=\linewidth]{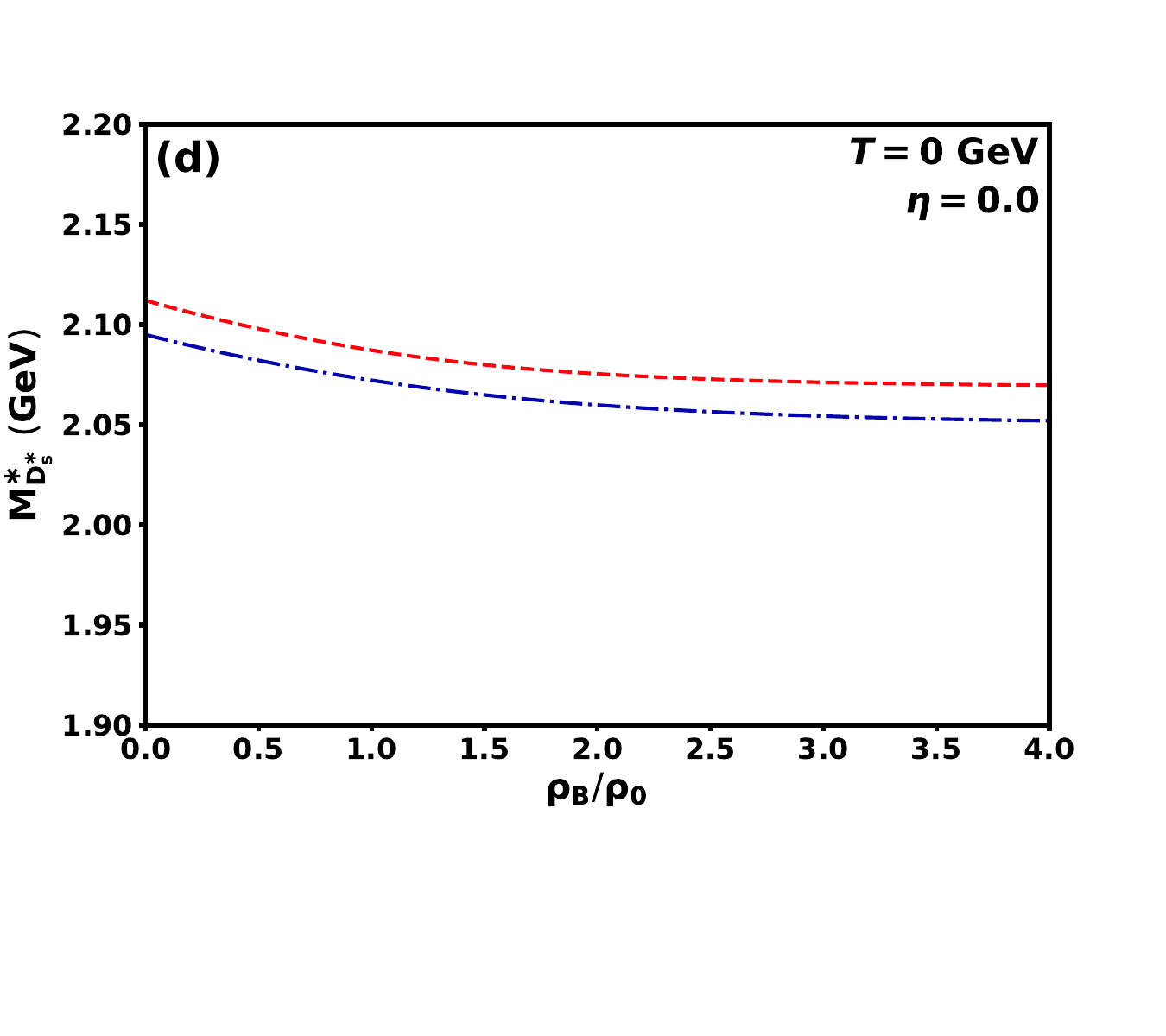}
    \end{minipage}

    \vspace{-15mm} % Reduce space before the last figure

    % Last Figure (Centered Below)
    \begin{minipage}{0.48\linewidth}
        \centering
        \includegraphics[width=\linewidth]{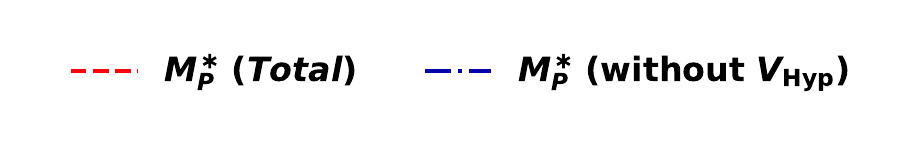}
    \end{minipage} \hspace{-4pt}
    \begin{minipage}{0.48\linewidth}
        \centering
        \includegraphics[width=\linewidth]{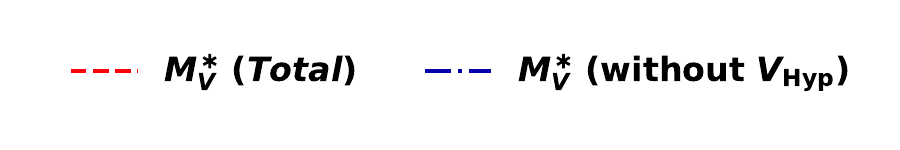}
    \end{minipage}
    \caption{In-medium masses of pseudoscalar \(D^0\) and \(D_s\) mesons [left panel] and the vector \(D^{0*}\) and \(D_s^*\) mesons [right panel] as functions of baryon density \(\rho_B\) (in units of $\rho_0$) at temperature \(T=0\) GeV and isospin asymmetry \(\eta=0\). Results are shown both with and without the contribution of the hyperfine potential.}
    \label{fig:9}
\end{figure}

 The in-medium masses of the vector mesons \(D^{0*}\), \(D^{+*}\), and \(D_s^*\) are illustrated in Fig.~\ref{fig:2} as a function of baryon density $\rho_B$ (in units of $\rho_0$) for temperatures \(T=0, 0.1,\) and \(0.15\) GeV. Each subplot presents results for isospin asymmetry values $\eta = 0, 0.3$, and $0.5$. At fixed isospin asymmetry and temperature of the medium, the effective masses of the \(D^{0*}\) and \(D^{+*}\) mesons are observed to decrease at lower densities and subsequently increase at higher $\rho_B$, whereas the \(D_s^*\) meson displays a gradual mass decrease with increasing density. This contrasting behaviour between pseudoscalar \(D\) and vector \(D^*\) mesons can be explained by the spin-spin interaction term in Eq.~\eqref{eq:6}, where the expectation value \(\langle \mathbf S_q \cdot \mathbf S_{\bar{q}}\rangle\) takes a value of \(+1/4\) for vector mesons and \(-3/4\) for pseudoscalar mesons. Consequently, the hyperfine potential \(V_\text{Hyp}\) acts to reduce the mass of pseudoscalar \(D\) mesons while increasing that of vector \(D^*\) mesons with increasing density. This effect is explicitly illustrated in Fig.~\ref{fig:9}, which shows the density dependence of effective masses of $D$ and $D^*$ mesons at \(T=0\) GeV and \(\eta=0\). Specifically, Figs.~\ref{fig:9}(a) and \ref{fig:9}(c), clearly demonstrate that $D^0$ mass decreases, while $D^{0*}$ mass increases in the presence of \(V_\text{Hyp}\) term at large baryon densities. Isospin asymmetry is observed to cause splitting in the effective masses of isospin doublet \(D^*=\begin{pmatrix} D^{+*} \\ D^{0*} \end{pmatrix}\), due to the mass difference between \(u\) and \(d\) quarks in an isospin asymmetric nuclear medium. As shown in Figs.~\ref{fig:2}(a)-\ref{fig:2}(c), the effective mass of the \(D^{0*}\) meson remains largely unaffected by isospin asymmetry for low baryon densities, specifically, up to \(\rho_B \sim \rho_0\) at \(T=0\) GeV, \(\sim 1.5\rho_0\) at \(T=0.1\) GeV, and \(\sim 2\rho_0\) at \(T=0.15\) GeV. At higher densities, a reduction in the effective masses due to isospin asymmetry is observed, although the overall magnitude of the mass splitting diminishes with rising temperature. At \(\rho_B=3\rho_0\), the observed mass shifts of $D^{0*}$, for $\eta=$0 (0.5), are \(0.028 (-0.003)\), \(0.008 (-0.008)\), and \(0.001 (-0.007)\) GeV at $T=0, 0.1,$ and $0.15$ GeV, respectively, relative to its vacuum mass of \(2.007\) GeV. For the $D^{+*}$ meson, the response of the isospin asymmetry differs. In Figs.~\ref{fig:2}(d), (e), and (f), the mass of \(D^{+*}\) is observed to rise as \(\eta\) goes from \(0\) to \(0.3\). However, for \(\eta=0.5\), the effective mass decreases in Fig.~\ref {fig:2}(d) and gradually shows an increase in Fig.~\ref{fig:2}(f).
% it decreases for $\eta=0.5$  as \(\rho_B\) increases. The effect of isospin is more profound for $D^{+*}$ as compared\textcolor{red}{In Fig.~\ref{fig:2}(e), the effective mass for \(D^{+*}\) at \(\eta=0.5\) gradually increases, subsequently revealing a contrasting behavior in Fig.~\ref{fig:2}(f), compared to \(D^{0*}\) with isospin asymmetry, as shown in Fig.~\ref{fig:2}(c)}. 
% Correspondingly, for \(\rho_B=\rho_0(3\rho_0)\), the change in the mass of \(D^{0*}\) meson from its vacuum value due to the isospin asymmetry \(\eta=0.5\) in the medium is noted to be \(0.019(0.003)\) GeV, \(0.018(0.008)\) GeV and \(0.016(0.007)\) GeV at temperatures \(T=0, 0.1\) and \(0.15\) GeV respectively. 
At \(\rho_B=3\rho_0\) and \(\eta=0(0.5)\), the mass shifts for \(D^{+*}\) meson are observed to be \(0.02 (0.019)\), \(0.02 (0.007)\) and \(0.02 (0.005)\) GeV from its vacuum value corresponding to \(T=0, 0.1\) and \(0.15\) GeV, respectively.

\par Expanding the analysis to include the pseudoscalar \(D_s\) and vector \(D_s^*\) mesons shows that both mesons demonstrate qualitatively similar behaviour as the baryon density rises, though with differing magnitudes. As shown in Figs.~\ref{fig:9}(b) and \ref{fig:9}(d), for given baryon density, the effective mass of the \(D_s\) meson decreases  while that of \(D_s^*\) meson increases when the hyperfine interaction term \(V_\text{Hyp}\)  is included in the calculations. This is attributed to the reduced influence of nuclear medium and the greater mass of \(s\) quark relative to the \(u/d\) quarks, leading to smaller fluctuations in the \(V_\text{Hyp}\) term as baryon density rises. An increase in the isospin asymmetry of the medium causes a rise in  the effective masses of \(D_s\) and \(D_s^*\) mesons which is more pronounced at lower temperatures. As shown in subplots (g) and (i) of Figs.~\ref{fig:1} and \ref{fig:2}, the effective mass increases slightly with $\eta$, but this trend becomes less distinct as temperature rises. Nonetheless, at \(T=0.15\) GeV and \(\rho_B<3\rho_0\), the effective masses are found to overlap for \(\eta=0\) and \(0.3\), while it shows a decrease for \(\eta=0.5\). Beyond this density, there is a slight increase in the masses for \(\eta=0.3\) and \(0.5\). For $\eta=0.5$, the mass shift of the \(D_s\) meson from its vacuum value of 2.008 GeV is \(-0.029(-0.048)\) GeV, \(-0.026(-0.045)\), and \(-0.023(-0.043)\) GeV at $T=0,0.1,$ and 0.15 GeV, respectively at \(\rho_B=\rho_0(3\rho_0)\). Similarly, the mass shift for the \(D_s^*\) meson from its vacuum value of \(2.111\) GeV is \(-0.024(-0.039)\), \(-0.021(-0.037)\), and \(-0.02(-0.035)\) GeV at the same temperatures. These results indicate that the \(s\) quark experiences minimal modification in an isospin asymmetric medium. Meanwhile, an increase in medium temperature leads to a decrease in mass shift for all pseudoscalar \(D\) mesons and vector \(D^*\) mesons. Additionally, at \(T=0.15\) GeV isospin doublet mesons reveal a contrasting behavior with isospin asymmetry. 

\par Several theoretical studies have reported significant modifications in the masses of pseudoscalar \(D\) and vector \(D^*\) mesons when placed in a nuclear medium. Using a chiral hadronic model, the mass of the \(D^0\) meson in an isospin asymmetric nuclear medium was found to shift by \(-0.058\) GeV, while the \(D^+\) meson exhibited a larger shift of \(-0.099\) GeV when \(\eta\) is set to 0.5 at nuclear saturation density \cite{Mishra:2008cd}. Under similar conditions employing a combined chiral SU(3) and QCD sum-rule framework, mass shifts of \(-0.040\), \(-0.081\), \(-0.068\) and \(-0.047\) GeV were reported for \(D_s\), \(D^{0*}\), \(D^{+*}\) and \(D_s^*\) mesons, respectively, in an isospin asymmetric strange hadronic medium \cite{Chhabra:2017rxz, Kumar:2015fca}. In symmetric nuclear medium, the quark-meson coupling model predicted a mass shift of \(-0.062\) GeV for both \(D\) and \(\bar{D}\) mesons at a density of \(\rho_B=0.15\) fm\(^{-3}\) \cite{Tsushima:1998ru}. A  mass splitting of \(0.060\) GeV for the \(D-\bar D\) doublet at \(\rho_B=0.15\) fm\(^{-3}\) was  found using QCD sum rules in cold nuclear medium \cite{Hilger:2008jg}. Additionally, only a mild change in the mass of the \(D\) meson, but a significant broadening of its width was observed using a self-consistent coupled channel method \cite{Tolos:2005ft}. In Ref.~~\cite{Kumar:2013tna} the effects of symmetric nuclear medium was analysed on the vector \(D^*\) meson, where mass shifts of \(-0.076\), \(-0.071\), \(-0.065\) and \(-0.056\) GeV were observed at \(T=0\), \(0.05\), \(0.1\) and \(0.15\) GeV, respectively, at \(\rho_B=\rho_0\).

% \par \textcolor{red}{A comparable mass shift is also anticipated in the nuclear medium using asymmetric hadronic model \cite{Mishra:2008cd}, where an increase in mass of \(0.023\) GeV was noted for the \(D^0\) meson, while the \(D^+\) meson exhibited a mass decrease of \(0.018\) GeV at baryon density \(\rho_B=\rho_0\). The medium modification of \(D\) mesons has been explored in an asymmetric strange hadronic medium utilizing the chiral SU(3) approach and QCD sum rules, revealing a mass shift in the pseudoscalar meson \(D_s\) meson at \(\rho_B=\rho_0(4\rho_0)\) of \(-40(-61)\) in Ref.~\cite{Chhabra:2017rxz}, while for vector mesons the observed shift is \(-81(-137)\) for \(D^{*0}\), and \(-68.4(-104.2)\) has been noted for \(D^{*+}\) at \(\rho_B=\rho_0(4\rho_0)\), with \(-47.4(-65.7)\) recorded for \(D^*_s\) at \(\rho_B=\rho_0(2\rho_0)\), respectively \cite{Kumar:2015fca}. Additionally, the mass differences of \(D\) and \(\bar {D} \) mesons were studied using the quark meson coupling model in Ref.~\cite{Tsushima:1998ru}, while Ref.~\cite{Hilger:2008jg} employed QCD sum rules to investigate mass changes of pseudoscalar mesons in a symmetric nuclear medium at zero temperature. Furthermore, the medium effects on vector \(D\) mesons in a symmetric nuclear medium were discussed in Ref.~\cite{Kumar:2013tna}.}

\subsection{In-medium weak decay constant of \(D\) and \(D^*\) meson}
% The weak decay constants are associated with normalizing the leading twist quark DAs of the relevant mesons—determining the probability of valence-quark distributions, and can be obtained from hard exclusive reaction processes \cite{Chernyak:1983ej, Lepage:1980fj}. The decay constants of a pseudoscalar meson $P$ and a vector meson $V$ in terms of four-momentum $P^\mu$ and a mass $\textit{M}$ are defined by
The decay constant is  defined through the matrix elements connecting the vacuum to a meson state with four-vector momenta \(\textit{P}\) for pseudoscalar mesons  and vector mesons through relations \cite{Arifi:2023jfe, Choi:2007se},
\begin{eqnarray}
\langle 0 |\bar q \gamma^\mu \gamma_5 q | P(\textit{P}) \rangle & = & i f_P \textit{P}^\mu ,
\end{eqnarray}
and
\begin{eqnarray}
	\langle 0 |\bar q \gamma^\mu q | V(\textit{P},  \textit{J}_z) \rangle & = & f_V \textit{M}_V \epsilon^\mu (\textit{J}_z),
\end{eqnarray}
respectively. Here, $f_P$ and $f_V$ are the decay constants for pseudoscalar and vector mesons, respectively. Also, \(\epsilon^\mu (\textit{J}_z)\) and \(\textit{M}_V\) denote the polarization vector and the mass of the vector meson. The explicit forms of the medium-modified decay constants of pseudoscalar and vector mesons are given by \cite{Arifi:2023jfe}
\begin{eqnarray}
\label{eq:2}
f^*_P & = & 2\sqrt{6} \int_0^1 dx \int \frac{d^2 \mathbf{k}_\perp}{2(2\pi)^3} \frac{\mathbf{\Phi}^*(x, \mathbf{k}_\perp) }{\sqrt{\mathcal{A}^{*2} + \mathbf{k}_\perp^2} }\ \mathcal{A^*},\\
f^*_V & = & 2\sqrt{6} \int_0^1 dx \int \frac{d^2 \mathbf{k}_\perp}{2(2\pi)^3} \frac{\mathbf{\Phi}^*(x, \mathbf{k}_\perp) }{\sqrt{\mathcal{A}^{*2} + \mathbf{k}_\perp^2} }\left(\ \mathcal{A^*}+\frac{2\textbf{k}^2_\perp}{\mathcal{M}^*}\right).
\label{eq:3}
\end{eqnarray}
% where operators $\mathcal O_M$ are given by
% \begin{eqnarray}
% \mathcal O_P=\mathcal A,
% \label{eq:2}\\
% \mathcal O_V=\mathcal{A}+\frac{2\textbf{k}_\perp}{D_0},
% \label{eq:3}
% \end{eqnarray} for pseudoscalar and vector mesons, respectively. 
% where $\mathcal A^* = (1 - x) m^*_q + x m^*_{\bar{q}}
% $ and $\mathcal{M}_0^* =\textit{ M}_0^* + m^*_q + m^*_{\bar{q}}
% $.
\par
\begin{figure}[h]
    \centering
    % First Row (T = 0)
    \begin{minipage}{0.33\textwidth}
        \centering
        \includegraphics[width=\textwidth,height=1.05\textwidth]{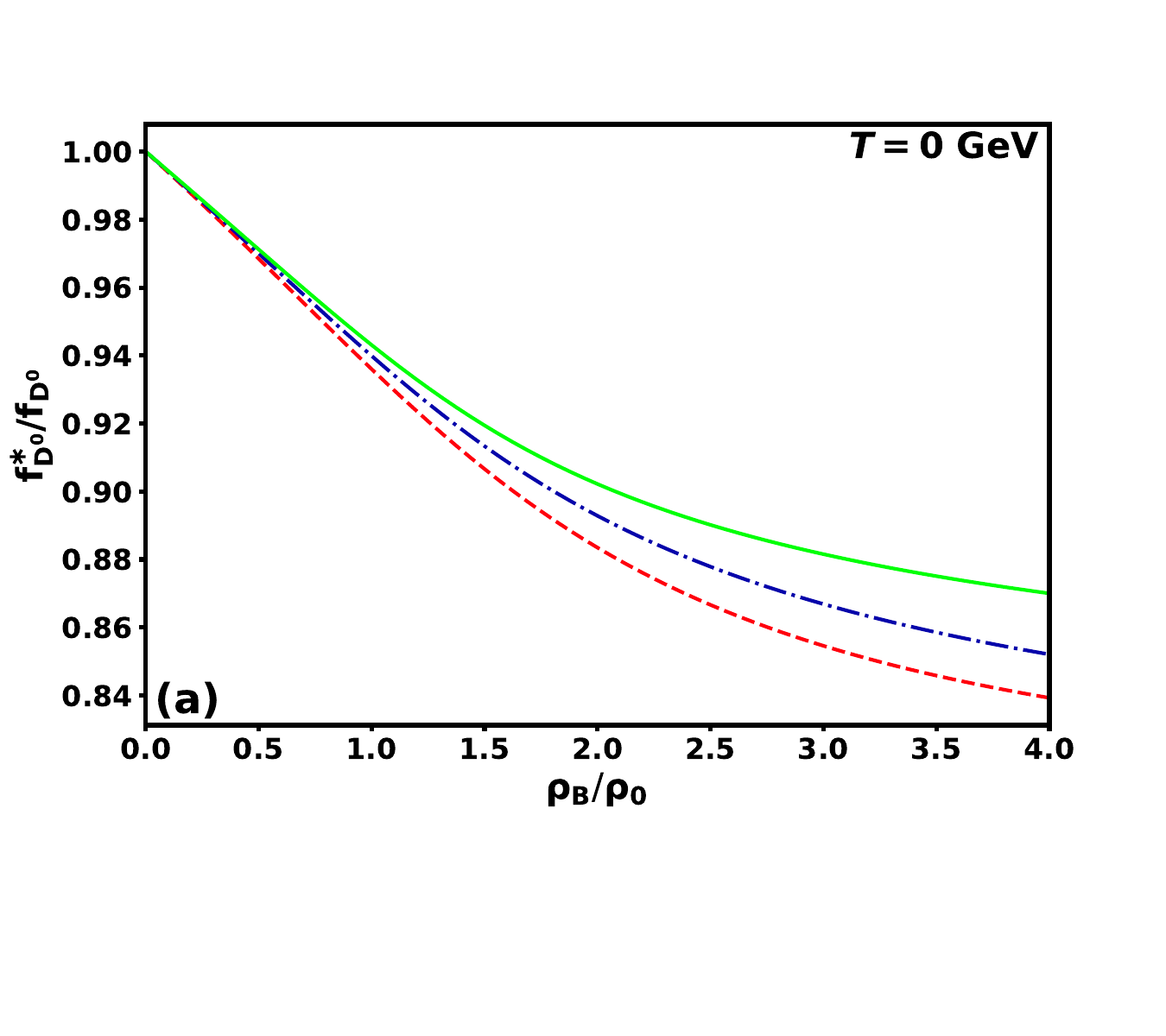}
    \end{minipage}\hspace{-1mm}
    \begin{minipage}{0.33\textwidth}
        \centering
        \includegraphics[width=\textwidth,height=1.05\textwidth]{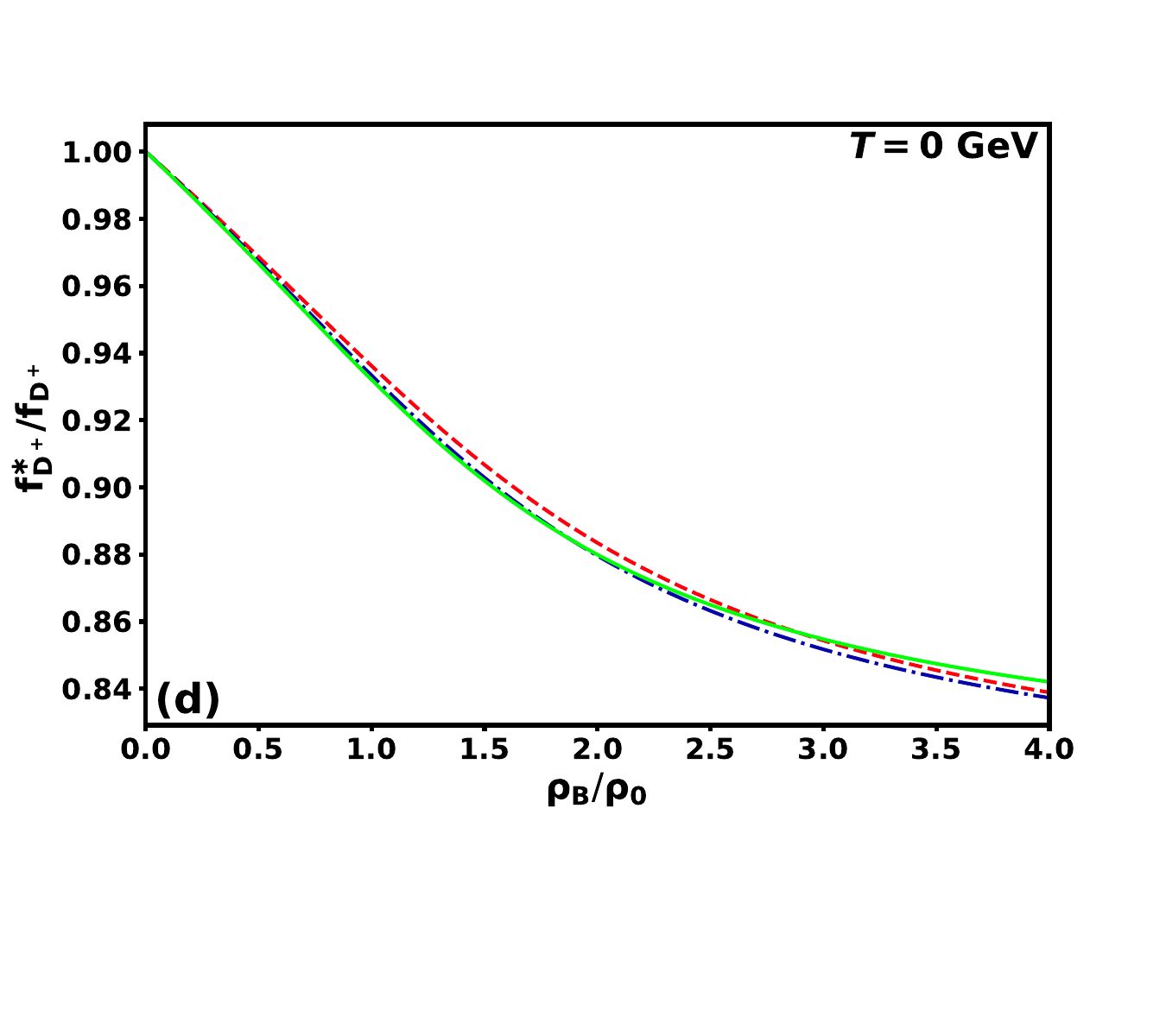}
    \end{minipage}\hspace{-1mm}
    \begin{minipage}{0.33\textwidth}
        \centering
        \includegraphics[width=\textwidth,height=1.05\textwidth]{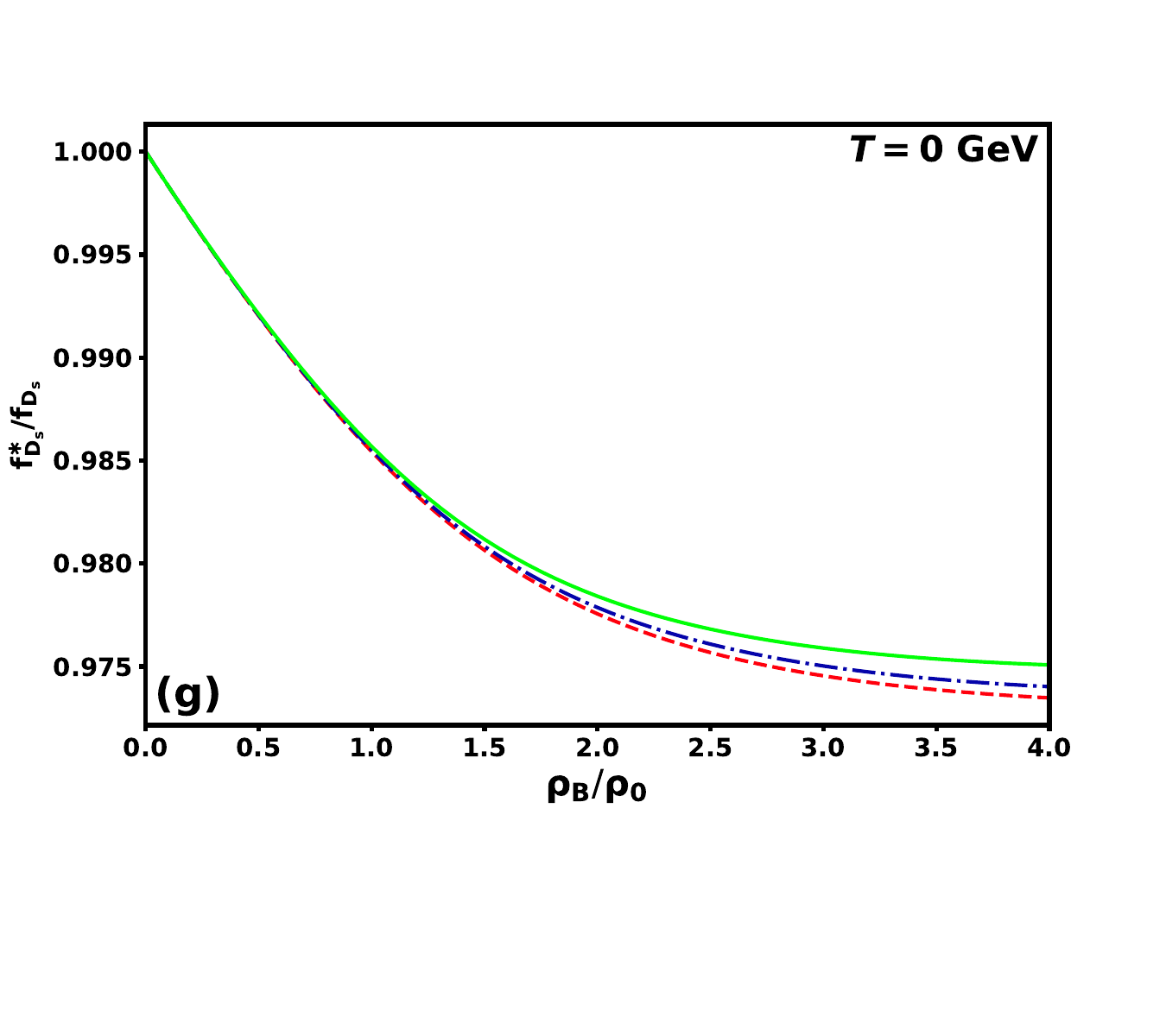}
    \end{minipage}
    
    \vspace{-10mm} % Reduce vertical spacing

    % Second Row (T = 100)
    \begin{minipage}{0.33\textwidth}
        \centering
        \includegraphics[width=\textwidth,height=1.05\textwidth]{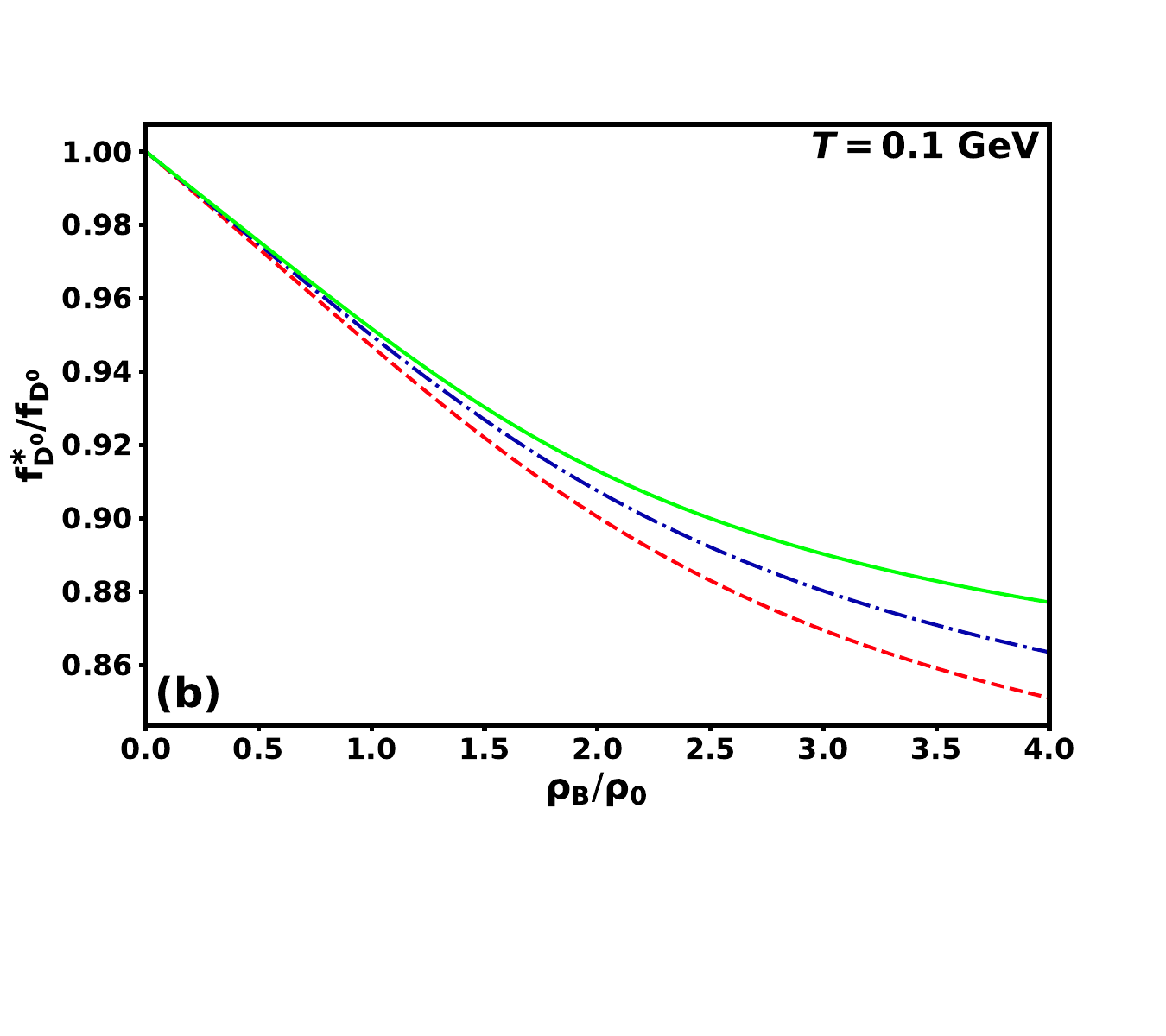}
    \end{minipage}\hspace{-1mm}
    \begin{minipage}{0.33\textwidth}
        \centering
        \includegraphics[width=\textwidth,height=1.05\textwidth]{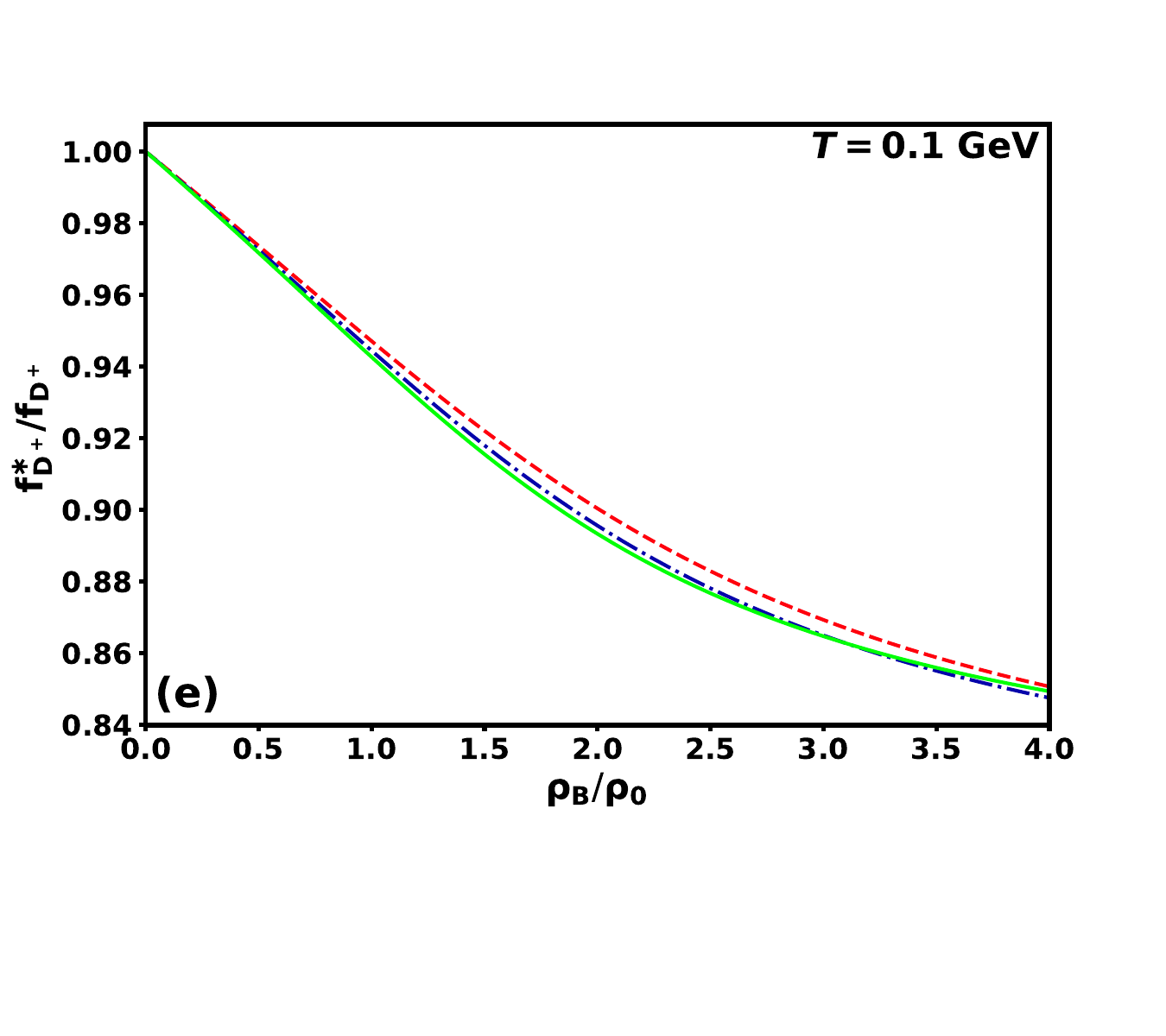}
    \end{minipage}\hspace{-1mm}
    \begin{minipage}{0.33\textwidth}
        \centering
        \includegraphics[width=\textwidth,height=1.05\textwidth]{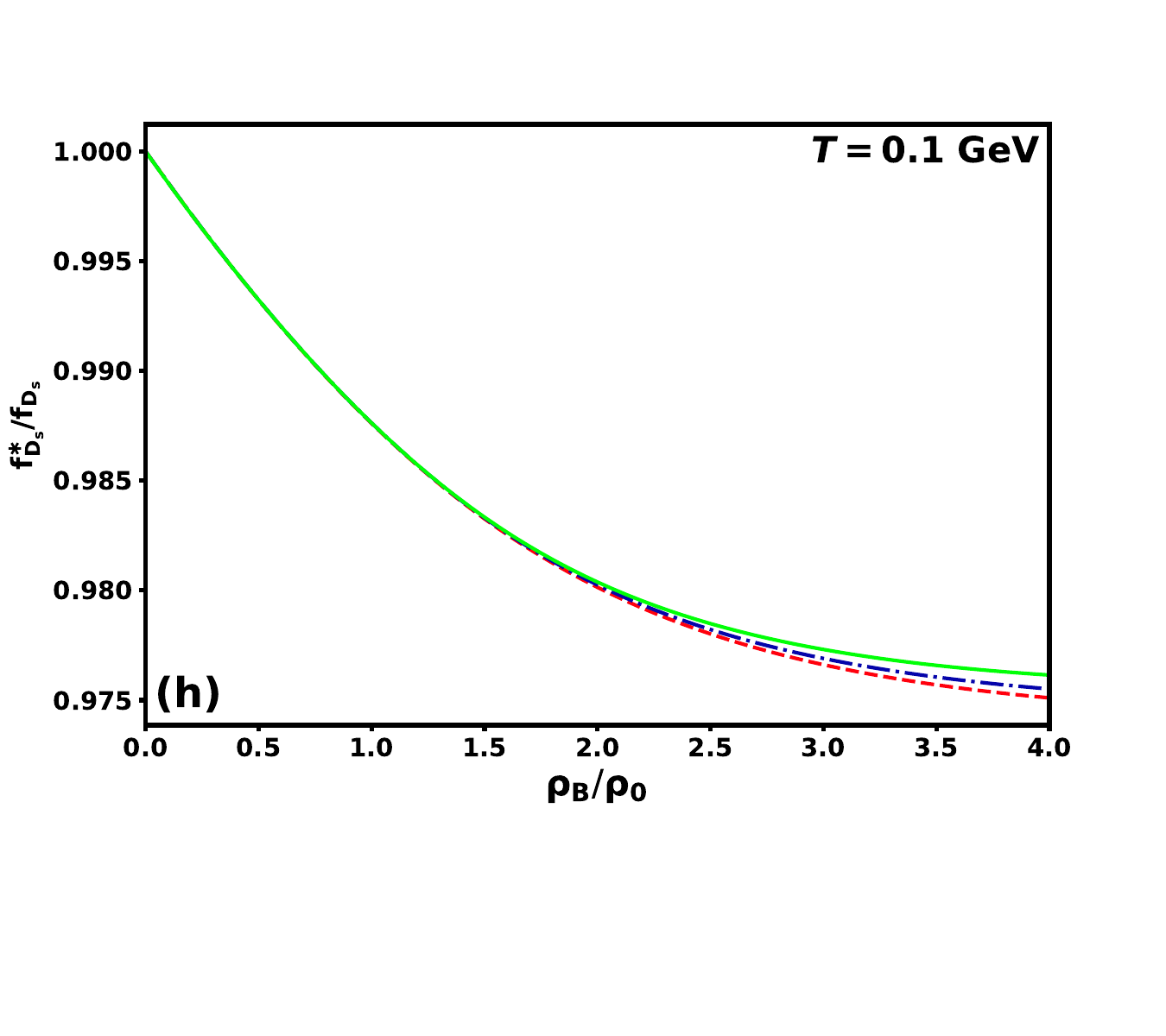}
    \end{minipage}

    \vspace{-10mm} % Reduce vertical spacing

    % Third Row (T = 150)
    \begin{minipage}{0.33\textwidth}
        \centering
        \includegraphics[width=\textwidth,height=1.05\textwidth]{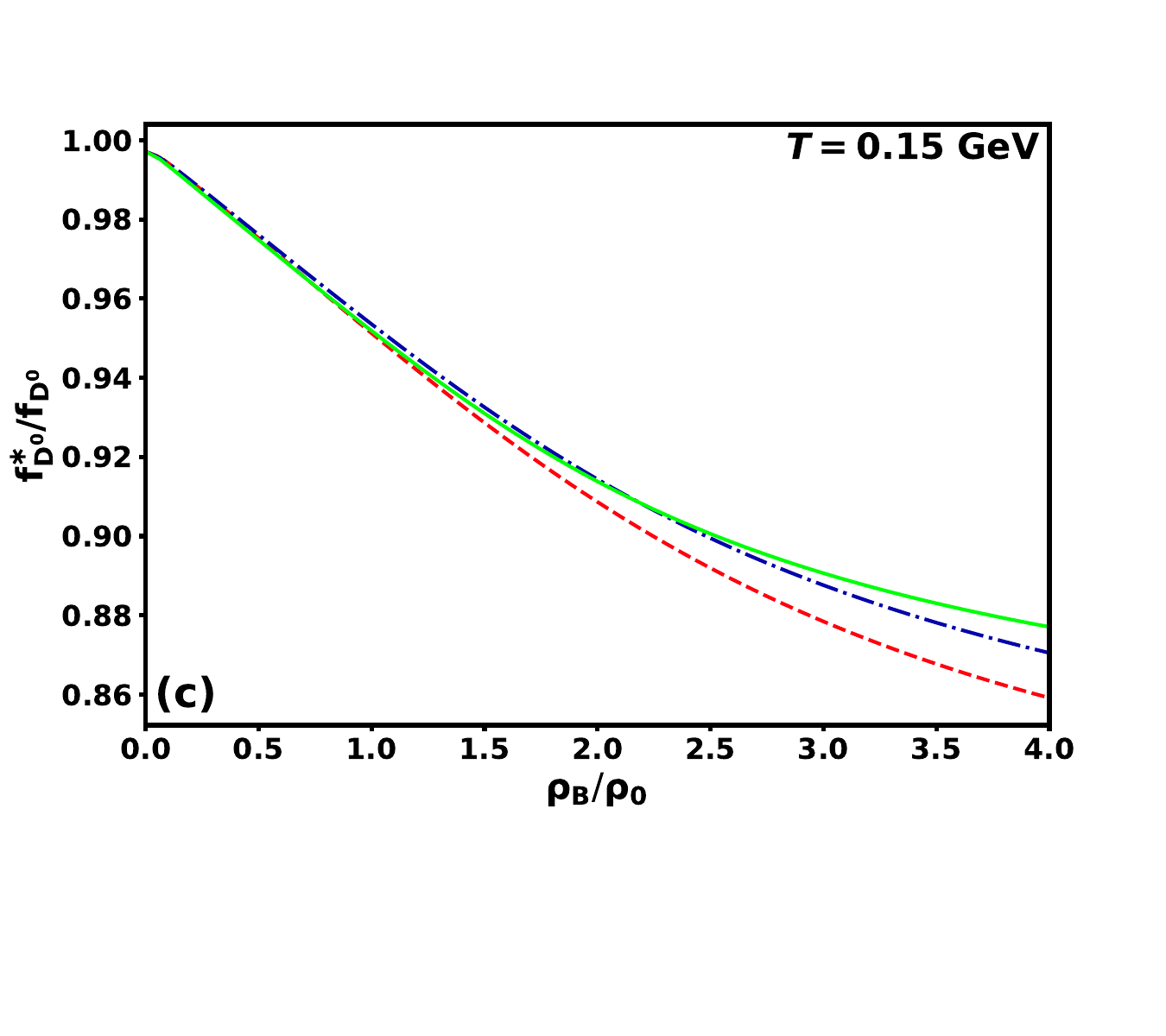}
    \end{minipage}\hspace{-1mm}
    \begin{minipage}{0.33\textwidth}
        \centering
        \includegraphics[width=\textwidth,height=1.05\textwidth]{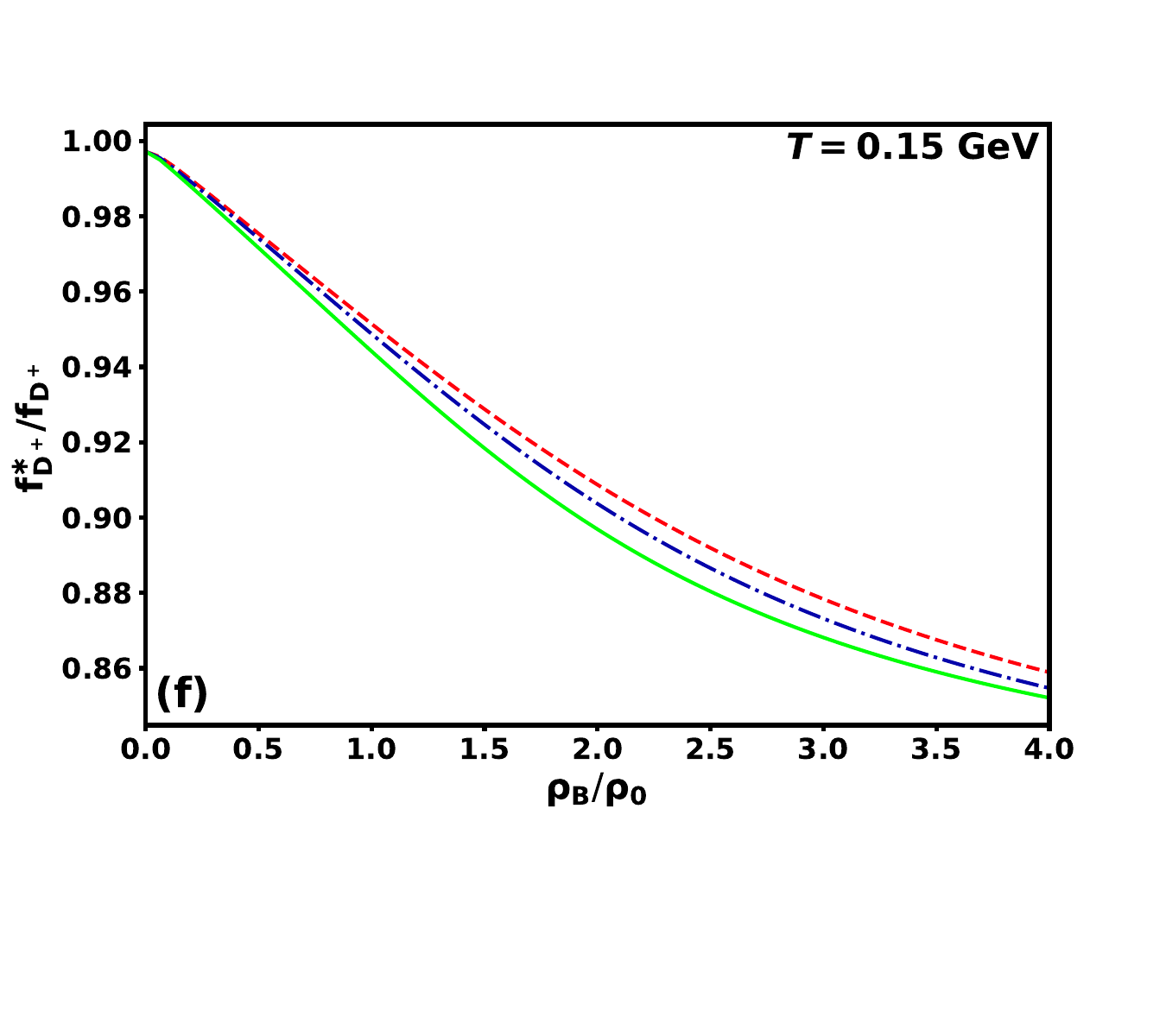}
    \end{minipage}\hspace{-1mm}
    \begin{minipage}{0.33\textwidth}
        \centering
        \includegraphics[width=\textwidth,height=1.05\textwidth]{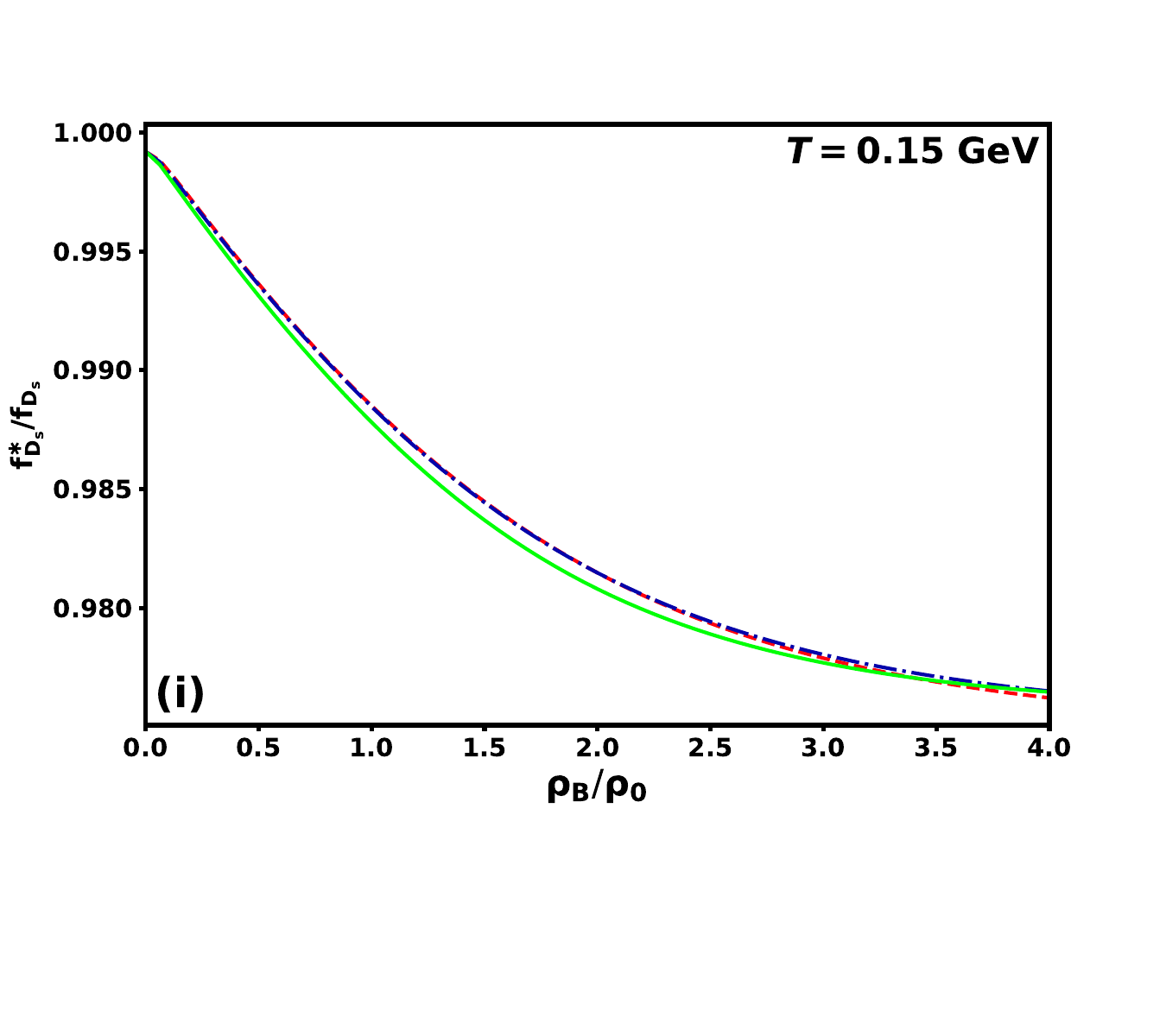}
    \end{minipage}

    \vspace{-10mm}
    
    \begin{minipage}{0.67\textwidth} 
        \centering
        \includegraphics[width=\textwidth,height=0.15\textwidth]{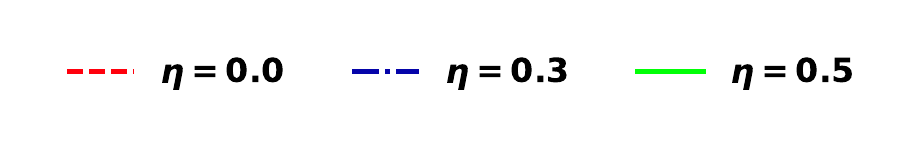} % Extra image that spans across
    \end{minipage}
    \caption{The ratio of the in-medium weak decay constant to the free-space value for pseudoscalar \(D^0\), \(D^+\), and \(D_s\) mesons plotted as a function of baryon density (in units of $\rho_0$). The results are presented for temperature \(T=0\) [in subplots (a), (d) and (g)], \(0.1\) [in subplots (b), (e) and (h)] and \(0.15\) [in subplots (c), (f) and (i)] GeV at isospin asymmetry $\eta= 0, 0.3$ and \(0.5\).}
    \label{fig:3}
\end{figure}

\begin{figure}[h]
    \centering
    % First Row (T = 0)
    \begin{minipage}{0.33\textwidth}
        \centering
        \includegraphics[width=\textwidth,height=1.05\textwidth]{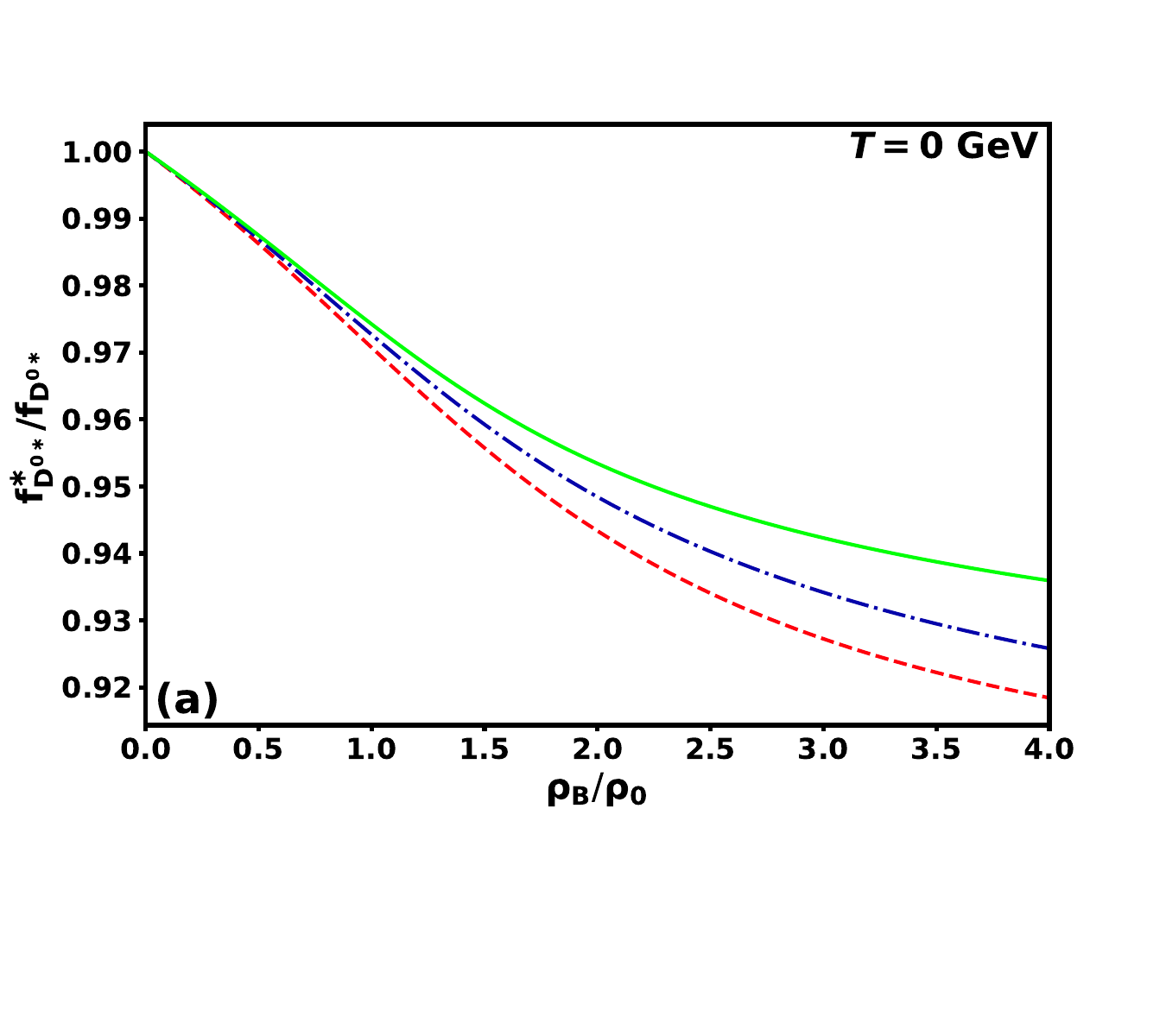}
    \end{minipage}\hspace{-1mm}
    \begin{minipage}{0.33\textwidth}
        \centering
        \includegraphics[width=\textwidth,height=1.05\textwidth]{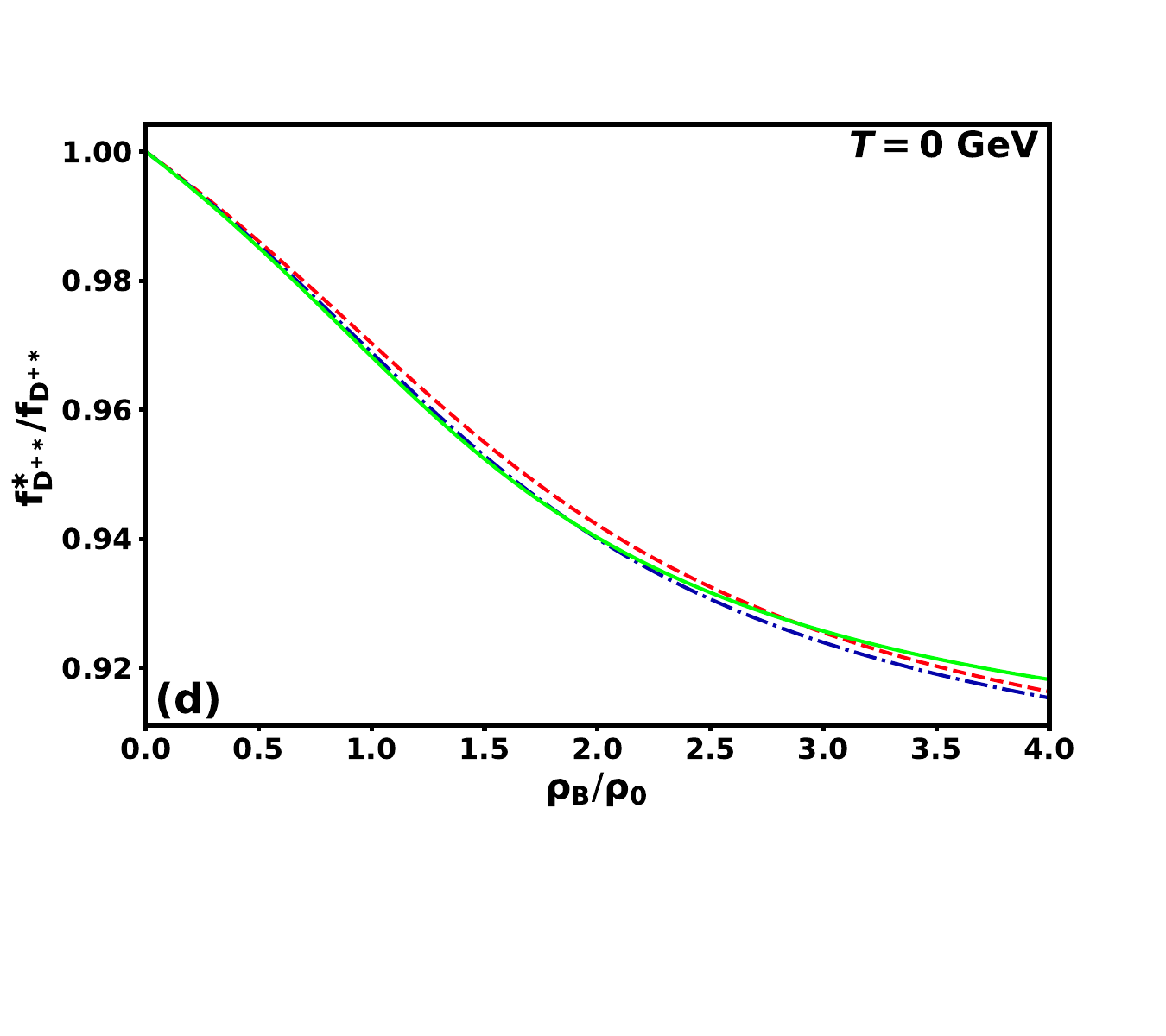}
    \end{minipage}\hspace{-1mm}
    \begin{minipage}{0.33\textwidth}
        \centering
        \includegraphics[width=\textwidth,height=1.05\textwidth]{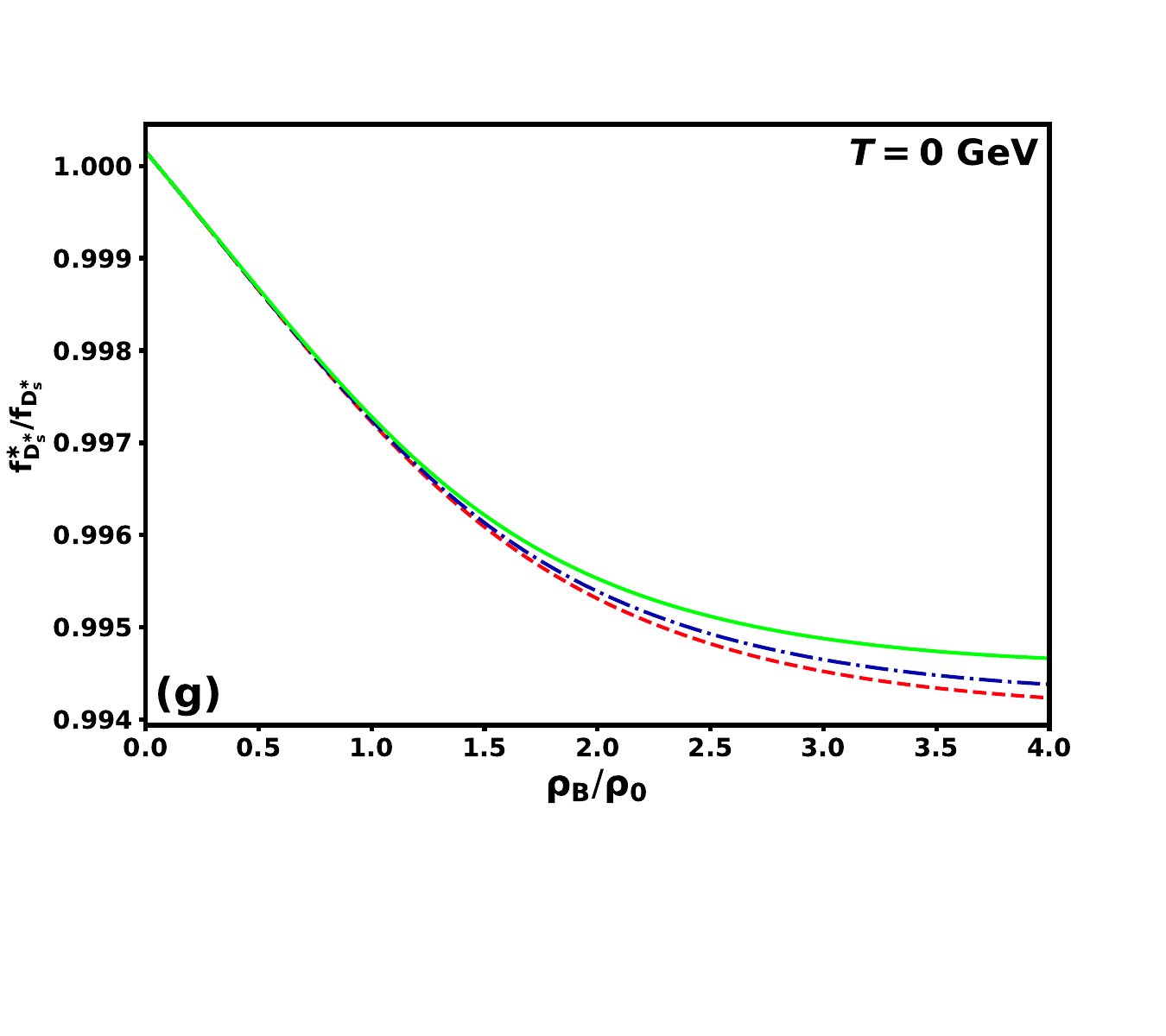}
    \end{minipage}
    
    \vspace{-10mm} % Reduce vertical spacing

    % Second Row (T = 100)
    \begin{minipage}{0.33\textwidth}
        \centering
        \includegraphics[width=\textwidth,height=1.05\textwidth]{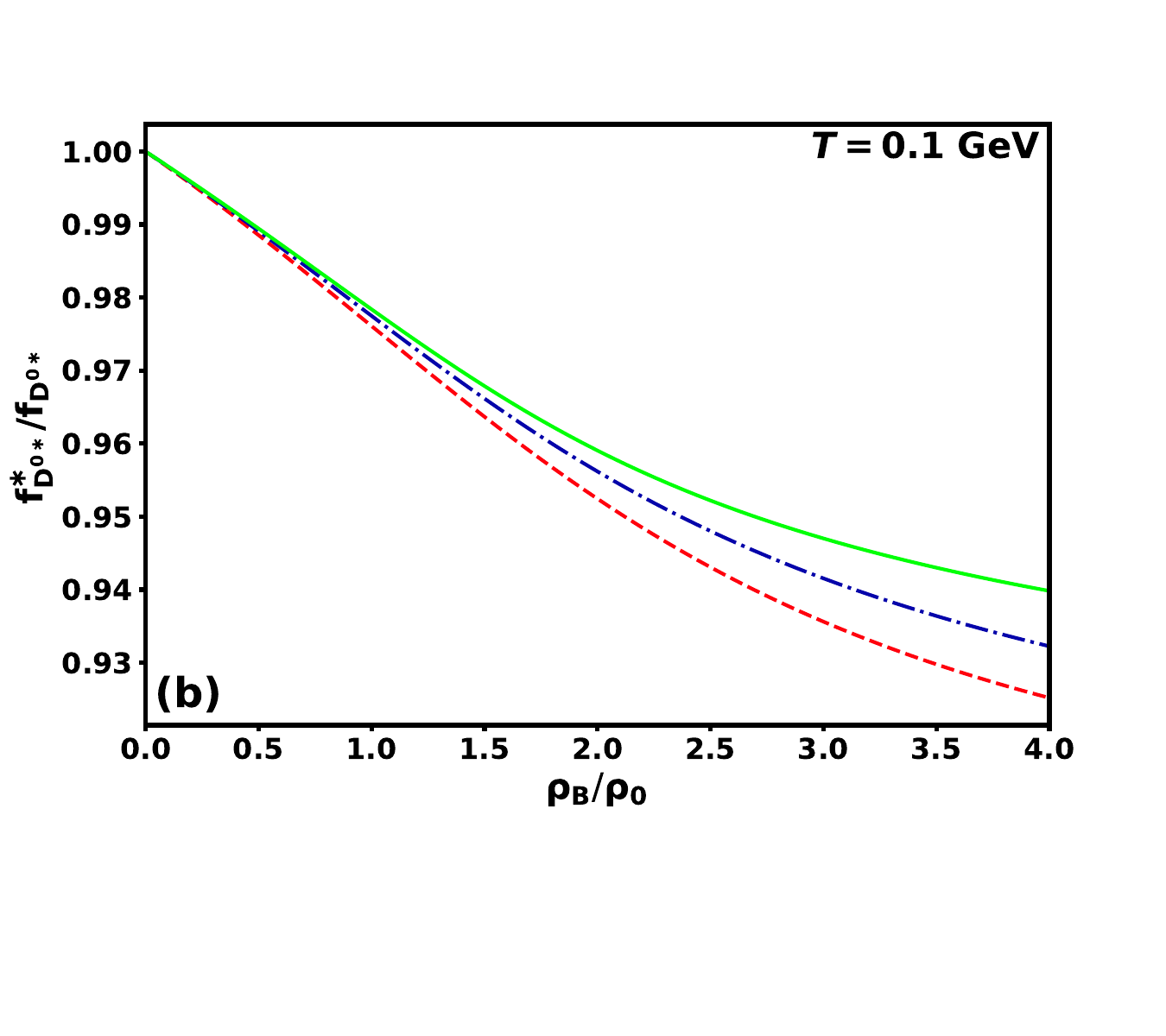}
    \end{minipage}\hspace{-1mm}
    \begin{minipage}{0.33\textwidth}
        \centering
        \includegraphics[width=\textwidth,height=1.05\textwidth]{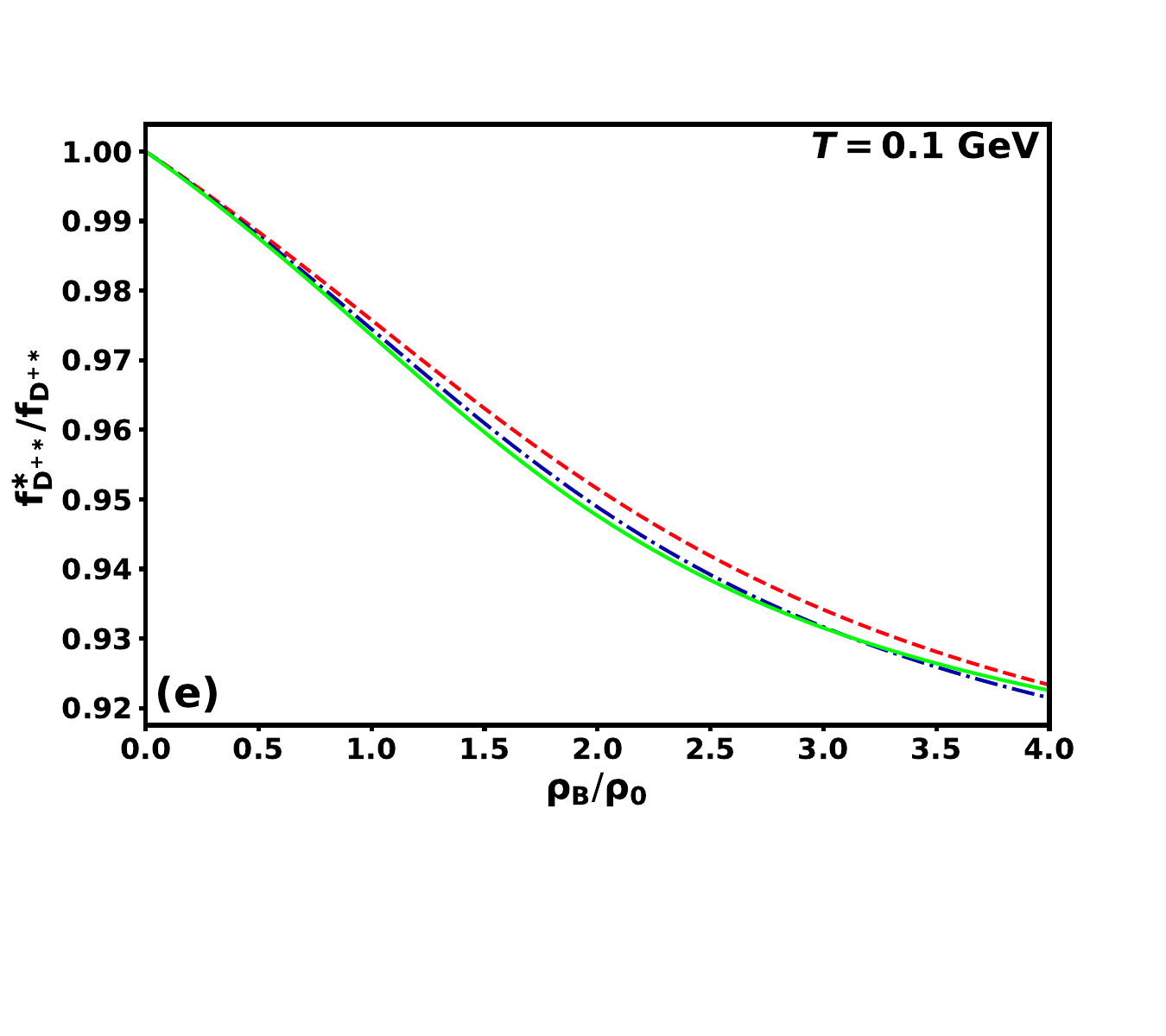}
    \end{minipage}\hspace{-1mm}
    \begin{minipage}{0.33\textwidth}
        \centering
        \includegraphics[width=\textwidth,height=1.05\textwidth]{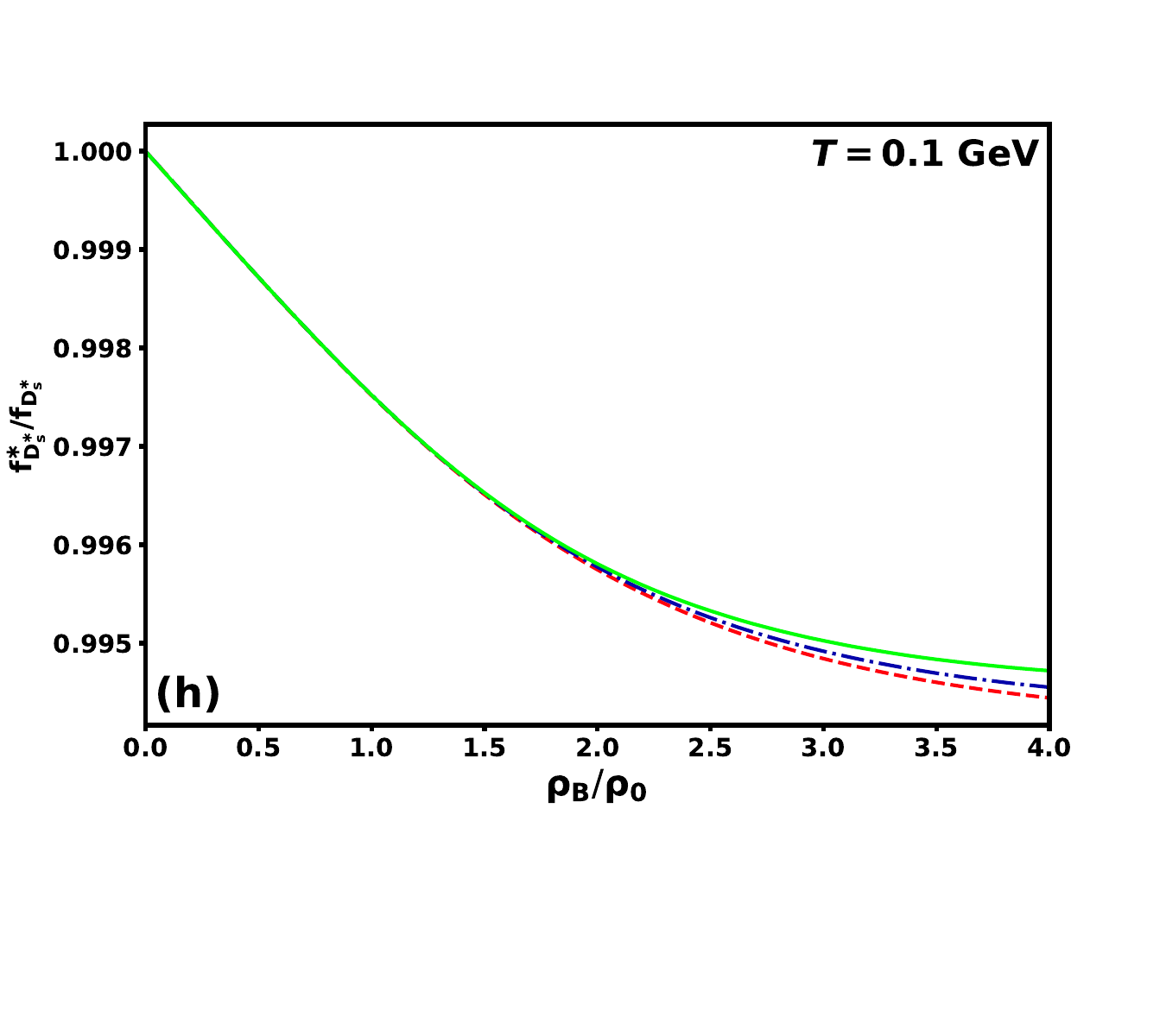}
    \end{minipage}

    \vspace{-10mm} % Reduce vertical spacing

    % Third Row (T = 150)
    \begin{minipage}{0.33\textwidth}
        \centering
        \includegraphics[width=\textwidth,height=1.05\textwidth]{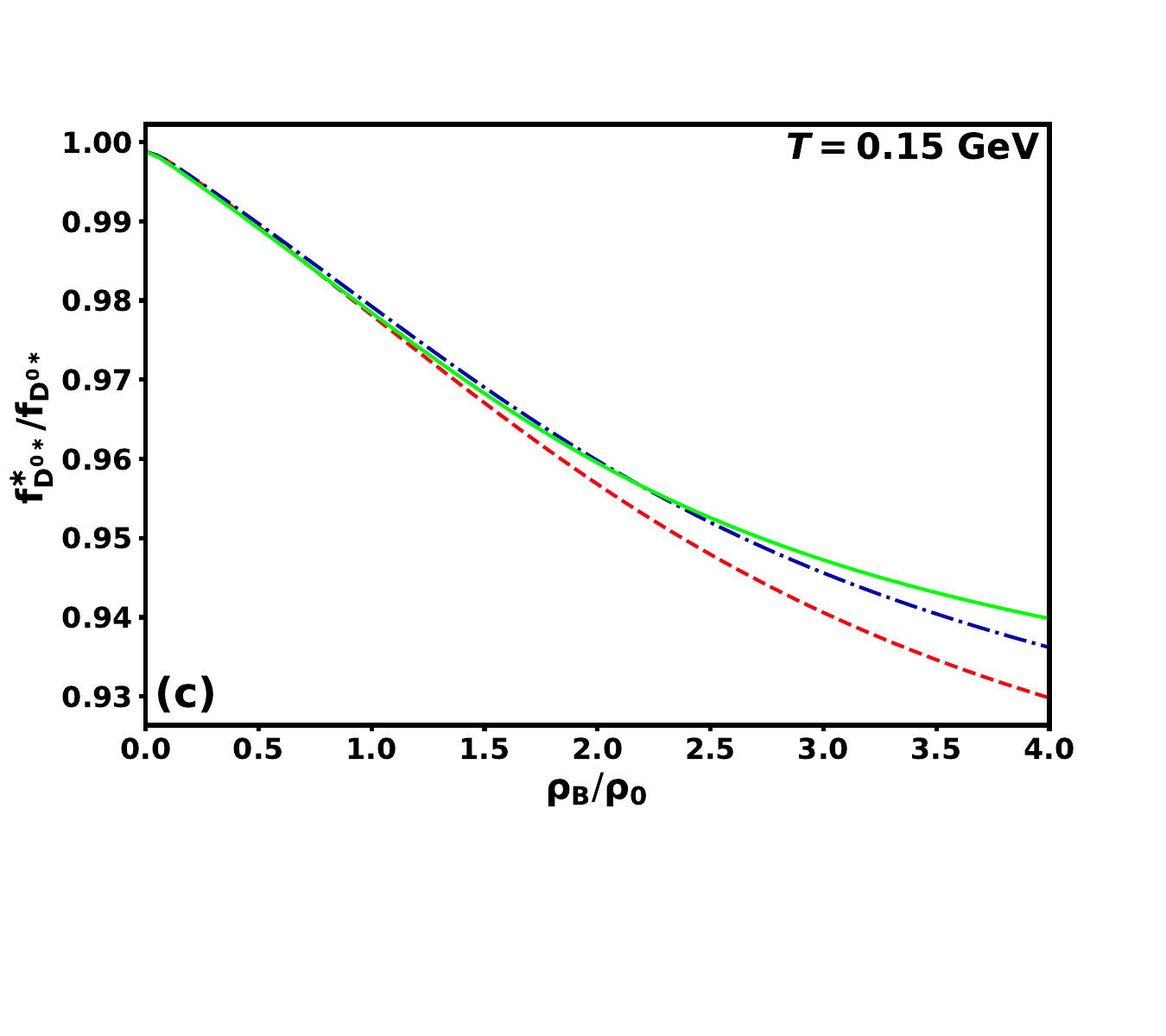}
    \end{minipage}\hspace{-1mm}
    \begin{minipage}{0.33\textwidth}
        \centering
        \includegraphics[width=\textwidth,height=1.05\textwidth]{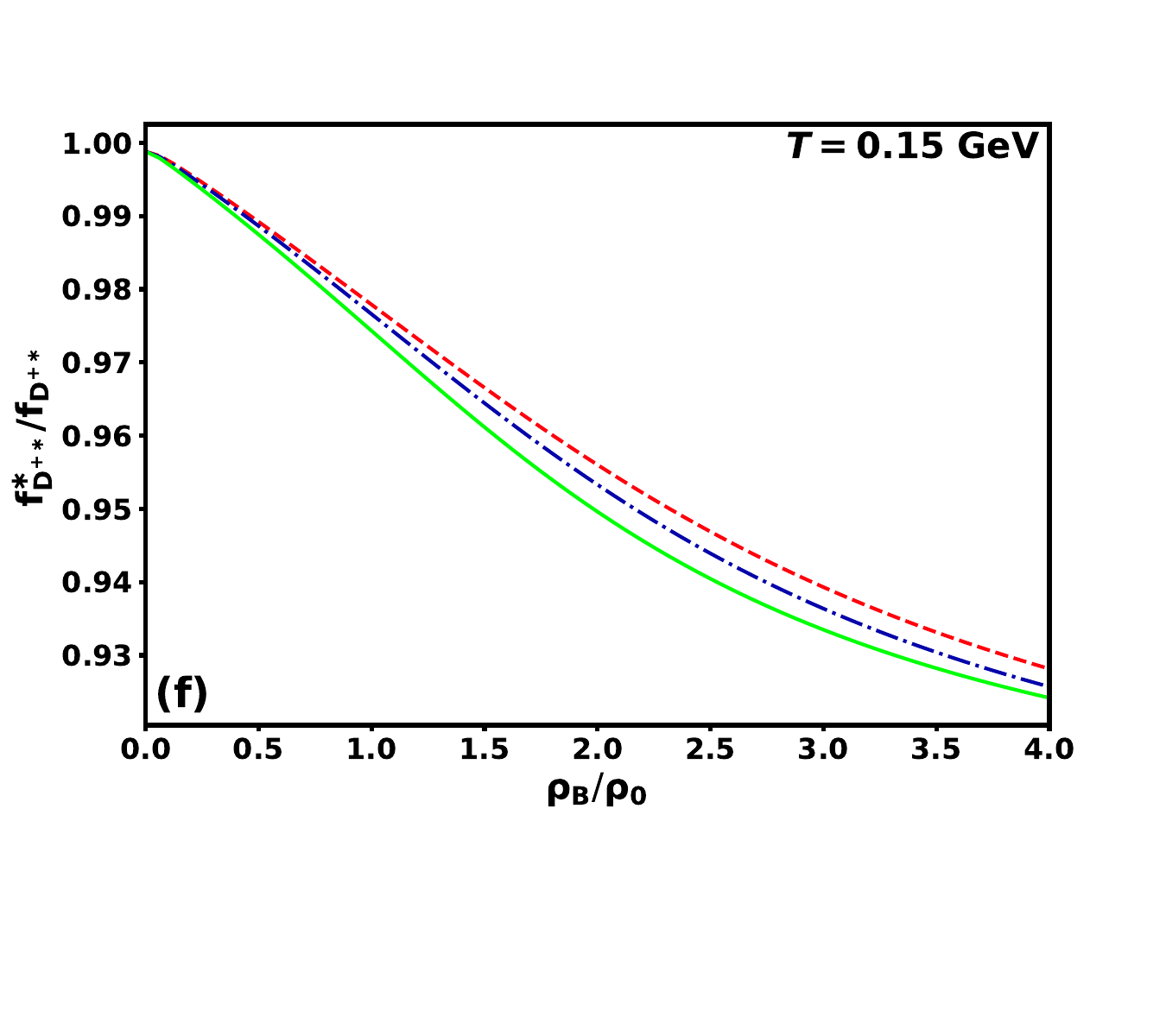}
    \end{minipage}\hspace{-1mm}
    \begin{minipage}{0.33\textwidth}
        \centering
        \includegraphics[width=\textwidth,height=1.05\textwidth]{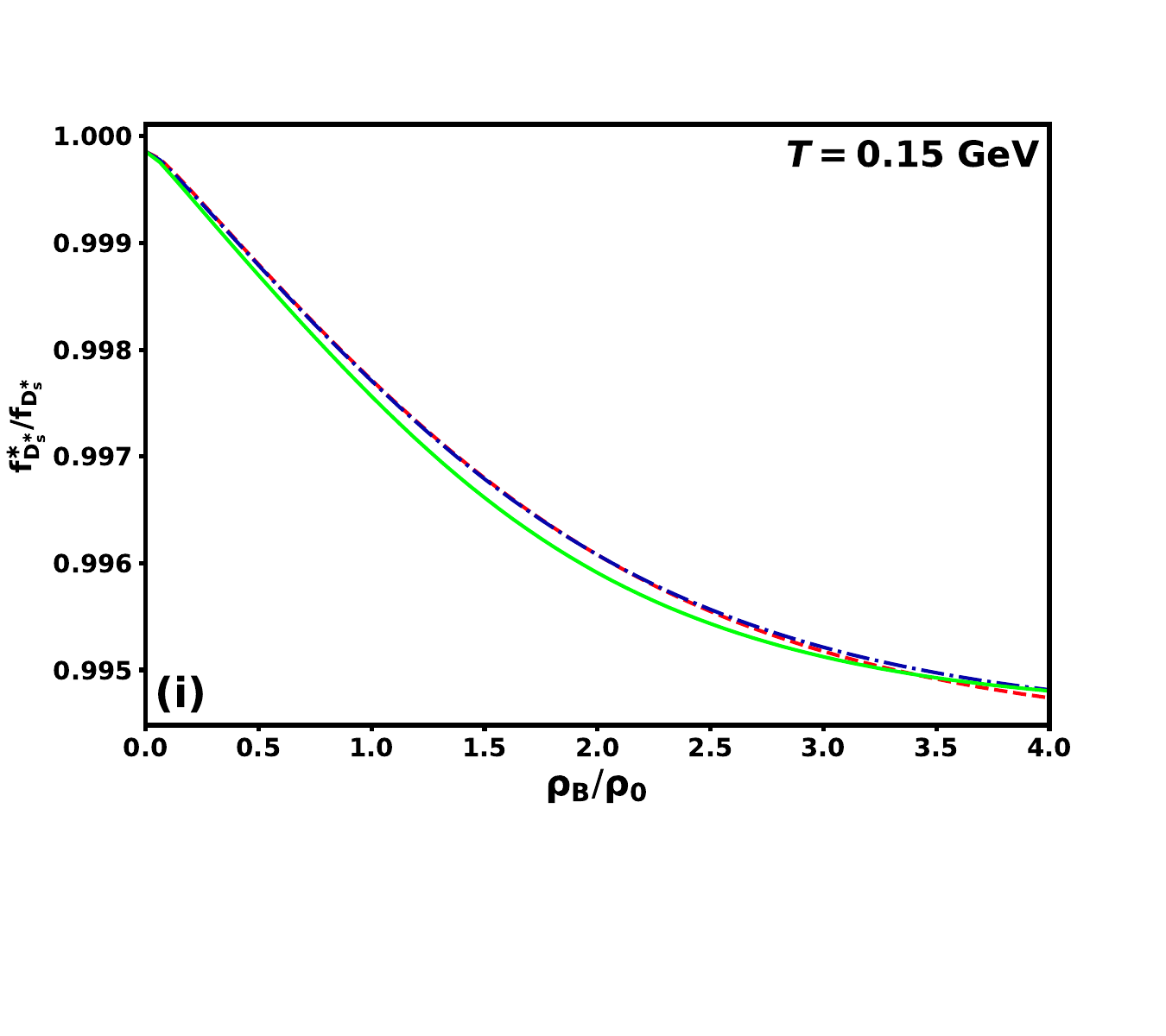}
    \end{minipage}

    \vspace{-10mm}
    
    \begin{minipage}{0.67\textwidth} 
        \centering
        \includegraphics[width=\textwidth,height=0.15\textwidth]{new2/T0/DC_D0_T0_diff_eta_legend.pdf} % Extra image that spans across
    \end{minipage}
    \caption{The ratio of the in-medium weak decay constant to the free-space value for vector \(D^{0*}\), \(D^{+*}\) and \(D_s^*\) mesons plotted as a function of baryon density (in units of $\rho_0$). The results are presented for temperature \(T=0\) [in subplots (a), (d), and (g)], \(0.1\) [in subplots (b), (e), and (h)], and \(0.15\) [in subplots (c), (f) and (i)] GeV at isospin asymmetry $\eta= 0, 0.3$ and \(0.5\).}
    \label{fig:4}
\end{figure}

 The ratio of the weak decay constant in an isospin asymmetric nuclear medium to that in free space, $f_M^*/f_M$, is shown in Figs.~\ref{fig:3} and ~\ref{fig:4}. The variation of $f_M^*/f_M$ for pseudoscalar mesons \(D^0\), \(D^+\) and \(D_s\) and vector mesons \(D^{0*}\), \(D^{+*}\) and \(D_s^*\) is plotted as a function of the baryon density $\rho_B$ for different values of temperatures \(T=0, 0.1\) and \(0.15\) GeV. Each subplot presents results corresponding to the isospin asymmetry $\eta=0, 0.3 $ and $0.5$. The weak decay constant ratio is observed to decrease with increasing baryon density \(\rho_B\) for all pseudoscalar \(D\) and vector \(D^*\) mesons. Both pseudoscalar \(D\) and vector \(D^*\) mesons exhibit a similar trend in the weak decay constant ratio as the isospin asymmetry increases in the medium. However, the ratio for vector \(D^*\) mesons is higher in magnitude than that for pseudoscalar \(D\) mesons at any particular value of $\rho_B$. In an isospin symmetric medium at $\rho_B=\rho_0(3\rho_0)$ and \(T=0\) GeV, the observed magnitude of the ratio is \(0.936 (0.855)\) for \(D^0\), \(0.936 (0.854)\) for \(D^+\), \(0.986 (0.975)\) for the \(D_s\) meson. In contrast, the corresponding values of the ratio for the vector \(D^{0*}\), \(D^{+*}\) and \(D_s^*\) mesons are found to be \(0.970 (0.927)\), \(0.970 (0.925)\) and \(0.997 (0.994)\), respectively. This indicates that vector mesons experience reduced suppression in the nuclear medium and can be understood through Eqs.~\eqref{eq:2} and \eqref{eq:3}. The additional term ${2\textbf{k}^2_\perp}/{\mathcal{M}^*}$ in Eq.~\eqref{eq:3} compensates for the reduction caused by the term $\mathcal{A}^*$, resulting in a higher value of weak decay constant for vector mesons.
\par As shown in Figs.~\ref{fig:3}(a)-\ref{fig:3}(c), the ratio for \(D^0\) meson increases with increasing asymmetry, with a more pronounced rise at higher densities. The splitting in the ratios due to asymmetry decreases as the temperature increases from \(T=0\) to \(0.15\) GeV, and the influence of asymmetry emerges at higher \(\rho_B\). For the \(D^+\) meson, Fig.~\ref{fig:3}(d) shows a slight increase in the ratio at larger \(\rho_B\) when \(\eta=0.5\). However, when the temperature is raised to \(T=0.1\) GeV in  Fig.~\ref{fig:3}(e), the ratio gradually decreases for the same \(\eta\). In Fig.~\ref{fig:3}(f), a further increase in temperature yields a notable decline in the ratio as \(\eta\) increases. This contrasting behavior for the ratio of \(D^+\) and \(D^0\) mesons arises from the splitting of the constituent \(u\) and \(d\) quark masses in an isospin asymmetric medium.
% A similar trend is observed with the weak decay ratio of vector mesons \(D^{0*}\) and \(D^{+*}\) in Fig~\ref{fig:4}. The subplots (a), (b), (c), (d), (e), and (f) mirror the corresponding subplots for the pseudoscalar \(D^0\) and \(D^+\) mesons in Fig~\ref{eq:3}, differing only in the magnitude of the ratio.
As shown in Figs.~\ref{fig:3}(g) and (h), the ratio for the $D_s$ meson increases as \(\eta\) increases at higher densities, while the effect of asymmetry diminishes for rising temperature. In Fig.~\ref{fig:3}(i), the ratio for \(D_s\)  decreases at lower densities as $\eta$ changes from $0$ to  $0.5$, before rising at higher \(\rho_B\). In contrast, the ratios for \(\eta=0\) and \(0.3\) remain identical. A similar pattern appears for the weak decay ratios of the vector mesons \(D^{0*}\), \(D^{+*}\), and \(D_s^*\) in Fig.~\ref{fig:4}. Each subplot mirrors that of the corresponding pseudoscalar mesons in Fig.~\ref{fig:3}, differing only in magnitude. Moreover, the pseudoscalar \(D_s\) and vector \(D_s^*\) meson exhibit minimal fluctuations across the parameter space, as indicated by the nearly constant values of \(f^*_{D_s}/f_{D_s}\) and \(f^*_{D_s^*}/f_{D_s^*}\) with the increasing \(\eta\). This uniformity arises from the lower sensitivity of medium effects on the \(s\) quark. In contrast, the lighter \(u\) and \(d\) quarks undergo more substantial mass reduction in the nuclear medium, due to the partial restoration of chiral symmetry \cite{Kumar:2023owb}. Table~\ref{tab:5} illustrates these trends at $\rho_B=3\rho_0$. To conclude, rising temperature generally enhances the decay constant ratios for both pseudoscalar and vector \(D\) mesons. 
%To the best of our knowledge, no theoretical framework has yet explored the in-medium decay constants of pseudoscalar \(D\) and vector \(D^*\) mesons in an isospin asymmetric nuclear medium.
 Using a hybrid chiral SU(3) model combined with QCD sum rules, shifts in the weak decay constants have been studied in a strange hadronic medium \cite{Chhabra:2016vhp, Chhabra:2017rxz, Kumar:2015fca}. Similarly, QCD sum rules were used in Ref.~\cite{Wang:2015uya} to investigate   decay constants of heavy mesons.

% it has been observed that within a strange hadronic medium at \(rho_B=\rho_0\), the decay constant experiences a drop from their vacuum value of about \(-0.035(-0.003\) GeV, \(-0.008(-0.007)\) GeV, \(-0.022(-0.021\) GeV, \(-0.02(-0.021\) GeV and \(-0.0.13(-0.012)\) GeV corresponding to \(D^+\), \(D^0\), \(D^{*0}\), \(D^{*+}\) and \(D^*_s\) mesons, respectively, due to isospin asymmetry \(\eta=0.5\).

% \textcolor{red}{In comparison to our result, in Ref.~\cite{Chhabra:2016vhp}, a change in the decay constant of -3.5(-3) and -8(-7) at temperatures \(T=0(0.1)\) and baryon density \(\rho_B=\rho_0\) has been observed, attributed to the influence of isospin asymmetry \(\eta=0.5\) within a strange hadronic medium for the pseudoscalar \(D^+\) and \(D^0\) mesons, utilizing a hybrid approach that combines the chiral SU(3) model and QCD sum rules. Under similar conditions, the values of shifts in decay constant of vector \(D\) mesons have been investigated in Ref.~\cite{Kumar:2015fca} and is found to be \(-22(-20.7)\), \(-20(-20.6)\), and \(-12.5(-12.2)\) for \(D^{*0}\), \(D^{*+}\) and \(D_s^*\) meson respectively. All above values for the shift of decay constant are presented at strangeness fraction \(\textit{f}_s=0\). Compared to our findings in isospin asymmetric nuclear medium, the modifications noted in Ref.~\cite{Kumar:2009xc, Hayashigaki:2000es, Chhabra:2017rxz} contribute to the study of medium properties, such as temperature, baryon density, and isospin asymmetry, on the behavior of \(D\) mesons.}

\begin{table}[htbp]
\centering
\small % Reduce overall text size
\renewcommand{\arraystretch}{1.5}
\setlength{\tabcolsep}{2pt}
\begin{tabular}{|c|c|c|c|c|c|c|c|}
\hline
\multicolumn{1}{|c|}{\textbf{Temperature}} &\multicolumn{1}{|c|}{\makecell{\textbf{Isospin}\\\textbf{asymmetry}}} & \multicolumn{6}{c|}{\textbf{Ratio of Decay Constants $f^*_M/f_M$}} \\ 
\cline{3-8}
% \hline
\textbf{$T$ (GeV)} & $\boldsymbol{\eta}$ & $\boldsymbol{f_{D^0}^*/f_{D^0}}$ & $\boldsymbol{f_{D^+}^*/f_{D^+}}$ & $\boldsymbol{f_{D_s}^*/f_{D_s}}$ & $\boldsymbol{f_{D^{0*}}^*/f_{D^{0*}}}$ & $\boldsymbol{f_{D^{+*}}^*/f_{D^{+*}}}$ & $\boldsymbol{f_{D_s^*}^*/f_{D_s^*}}$ \\
\hline
\multirow{3}{*}{0.00} & 0.0 & 0.855 & 0.854 & 0.975 & 0.927 & 0.925 & 0.994 \\
                   & 0.3 & 0.867 & 0.852 & 0.975 & 0.934 & 0.924 & 0.994 \\
                   & 0.5 & 0.882 & 0.855 & 0.976 & 0.942 & 0.926 & 0.995 \\
\hline
\multirow{3}{*}{0.10} & 0.0 & 0.870 & 0.870 & 0.977 & 0.936 & 0.934 & 0.995 \\
                    & 0.3 & 0.880 & 0.865 & 0.977 & 0.942 & 0.932 & 0.995 \\
                    & 0.5 & 0.890 & 0.865 & 0.977 & 0.947 & 0.932 & 0.995 \\
\hline
\multirow{3}{*}{0.15} & 0.0 & 0.878 & 0.878 & 0.978 & 0.94 & 0.939 & 0.995 \\
                     & 0.3 & 0.888 & 0.873 & 0.978 & 0.946 & 0.936 & 0.995 \\
                     & 0.5 & 0.890 & 0.868 & 0.978 & 0.947 & 0.933 & 0.995 \\
\hline
\end{tabular}
\vspace{0.5cm}
\caption{Ratios of in-medium to vacuum weak decay constants $f^*_M/f_M$ for pseudoscalar  (\(D^0\),\(D^+\),\(D_s\)) and vector  (\(D^{0*}\),\(D^{+*}\),\(D_s^*\)) mesons, at baryon density $\rho_B=3\rho_0$ for different temperatures ($T$) and isospin asymmetry ($\eta$).}
\label{tab:5}
\end{table}

\subsection{In-medium distribution amplitude}
The leading twist quark DAs of the mesons describe the longitudinal momentum fraction carried by the valence-quarks and can be derived from hard exclusive reaction processes \cite{Chernyak:1983ej, Lepage:1980fj}. These DAs are formally defined through matrix elements that connect free space states to meson states, and can be interpreted as the probability amplitudes for the hadron to be in its lowest Fock state with small separation in transverse momentum. The correlators for defining the pseudoscalar and vector mesons can be expressed as \cite{Li:2017mlw, Choi:2007yu, Arifi:2023jfe}
\begin{eqnarray}
A_{P}^+ &=& \langle 0 | \bar{q}(z) \gamma^+ \gamma_5 q(-z) | P(\textit{P}) \rangle,\nonumber \\
&=& i f_P \textit{P}^+ \int_0^1 dx \, e^{i (2x-1) \mathbf{\textit{P}} \cdot z} \, \phi_P(x)\Big|_{z^+ = z_\perp = 0}, \\
A_{V}^+ &=& \langle 0 | \bar{q}(z) \gamma^+ q(-z) | V(\textit{P}, 0) \rangle, \nonumber \\
&=& f_V \textit{M}_V \epsilon^+(0) \int_0^1 dx \, e^{i (2x-1) \textit{P} \cdot z} \, \phi_V(x)\Big|_{z^+ = z_\perp = 0}.
\end{eqnarray}
The medium-modified $\phi^*_M(x)$ for pseudoscalar and vector mesons can be obtained by integrating the LFWF over transverse momentum as 
\begin{eqnarray}
\label{eq:7}
\phi^*_P(x) &=& \frac{2\sqrt{6}}{ f^*_P} \int \frac{d^2 \mathbf{k}_\perp}{2(2\pi)^3} \frac{\mathbf \Phi^*(x, \mathbf{k}_\perp)}{\sqrt{\mathcal A^{*2} + \mathbf{k}_\perp^2}}\mathcal{A^*},\\
\phi^*_V(x) &=& \frac{2\sqrt{6}}{ f^*_V} \int \frac{d^2 \mathbf{k}_\perp}{2(2\pi)^3} \frac{\mathbf \Phi^*(x, \mathbf{k}_\perp)}{\sqrt{\mathcal A^{*2} + \mathbf{k}_\perp^2}}\left(\mathcal{A^*}+\frac{2 \mathbf{k^2_\perp}}{\mathcal{M}^*}\right).
\label{eq:8}
\end{eqnarray}

\par
\begin{figure}[htbp]
    \centering

    % 2x2 Grid of Figures
    \begin{minipage}{0.48\linewidth}
        \centering
        \includegraphics[width=\linewidth]{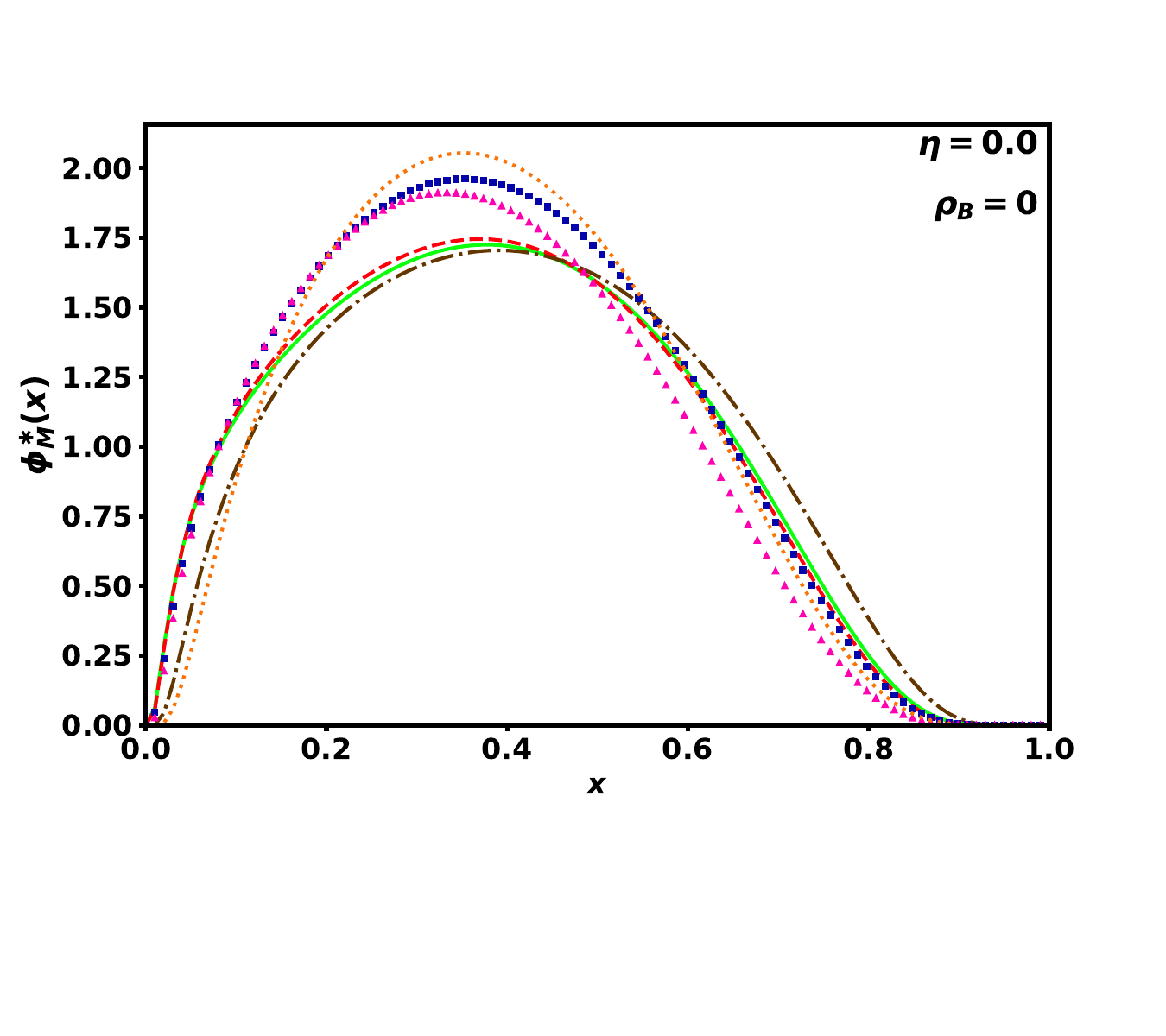}
    \end{minipage} \hspace{-6pt}
    % \begin{minipage}{0.48\linewidth}
    %     \centering
    %     \includegraphics[width=\linewidth]{final/DA_allD_T0_eta5_rho3.pdf}
    % \end{minipage}

    \vspace{-10mm}
      \begin{subfigure}{0.7\linewidth}
        \centering
        \includegraphics[width=0.7\linewidth]{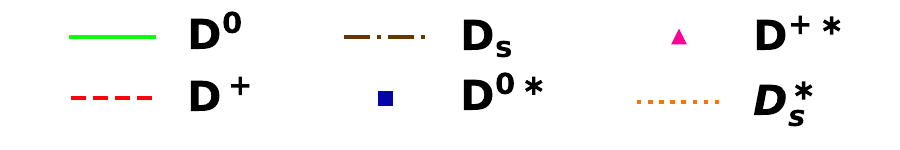}
     \end{subfigure}

    \caption{Distribution amplitude $\phi^*_M(x)$ plotted as a function of longitudinal momentum fraction $x$ for pseudoscalar (\(D^0\), \(D^+\) and \(D_s\)) and vector (\(D^{0*}\), \(D^{+*}\) and \(D_s^*\)) mesons in free space ($\rho_B=0$) at temperature \(T=0\) GeV.
    }
    \label{fig:5}
\end{figure}

 There has been limited investigation in the literature on the determination of DAs for \(D\) mesons in a nuclear medium \cite{Arifi:2023jfe}. The behavior of the \(D\) meson DA in a vacuum has been studied using the revised light-cone harmonic oscillator model in Ref.~\cite{Zhong:2020cqr}. Additionally, QCD sum rules have been employed within the background field theory framework to study the leading-twist light-cone DA of the \(D_s\) meson \cite{Zhang:2021wnv}. In this work, we present the DAs, $\phi^*_M (x)$, as a function of longitudinal momentum fraction $x$ for both pseudoscalar \(D\) and vector \(D^*\) mesons for free space ($\rho_B=0$) at temperature \(T=0\) GeV in Fig.~\ref{fig:5}.
% The results are shown for free space (baryon density $\rho_B=0$, isospin asymmetry $\eta=0.0$) in subplot (a) and in an isospin-asymmetric nuclear medium ($\rho_B=3\rho_0$, $\eta = 0.5$) (b). 
The DAs exhibit a convex-concave shape, indicating that the heavier charm quark carries the majority of the meson’s momentum \cite{Serna:2020txe}. Moreover, our results are in qualitative agreement with those in Refs.~\cite {Zhong:2020cqr, Zhang:2021wnv}. In addition, the DAs for pseudoscalar mesons having \(u\) or \(d\) quarks reach higher peaks than those with an \(s\) quark, due to the smaller mass difference between the strange and charm quarks. In contrast, the vector mesons exhibit an opposite trend, as the meson having \(s\) quark attains a higher peak. The difference arises from the additional term \({2\mathbf{k}^2_\perp}{/\mathcal{M}^*}\) in Eq.~\eqref{eq:8}, which increases with baryon density, whereas the \(\mathcal{A}^*\) term decreases. The combined effect of these terms lead to overall higher DAs for the vector \(D^*\) mesons.
% The distribution amplitudes for the pseudoscalar \(D\) meson and the vector \(D^*\) meson in a symmetric nuclear medium are examined in Ref.~\cite{Arifi:2023jfe}, and our findings are observed to be in qualitative agreement with their results.
% On introducing the isospin asymmetry in subplot (b), we observe changes in the DA maxima from its vacuum value \(1.724\) to \(1.748\) for \(D^0\), \(1.745\) to \(1.786\) for \(D^+\), \(1.705\) to \(1.675\) for \(D_s\), \(1.96\) to \(1.961\) for \(D^{0*}\), \(1.911\) to \(1.908\) for \(D^{+*}\) and \(2.078\) to \(2.037\) for the \(D_s^*\) meson. Additionally, we observe that the peaks for pseudoscalar mesons \(D^0\) and \(D^+\) significantly shift toward the right from \(x=0.374(0.414)\), while the peak for the \(D_s\) meson remains consistent at \(x=0.394\). In the case of vector mesons, the shifts in the peaks are relatively minor, moving from \(x=0.354(0.363)\) for \(D^{*0}\) and \(D^{*+}\) mesons, and for the \(D^*_s\) meson, a shift is observed toward the left from \(x=0.35\) to \(0.34\). The influence of asymmetry is seen to cause a reduction in the DAs of all pseudoscalar and vector mesons with lighter quark content up to around \(x<0.33\). Meanwhile, both the \(D_s\) and \(D^*_s\) mesons exhibit a slight decline in the DAs across the entire range of \(x\).

\par
\begin{figure}[htbp]
    \centering

    % 2x2 Grid of Figures
    \begin{minipage}{0.48\linewidth}
        \centering
        \includegraphics[width=\linewidth]{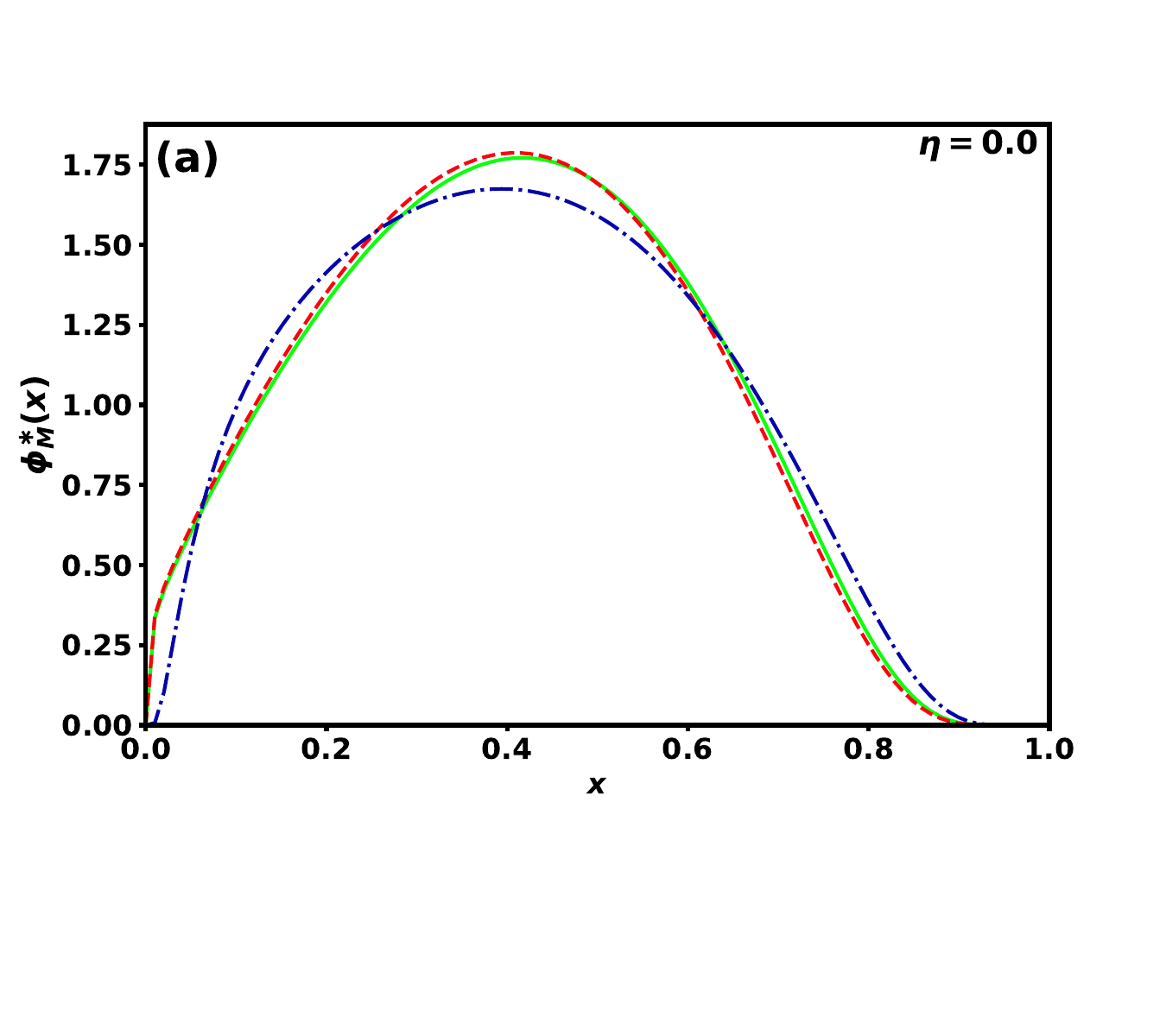}
    \end{minipage} \hspace{-6pt}
    \begin{minipage}{0.48\linewidth}
        \centering
        \includegraphics[width=\linewidth]{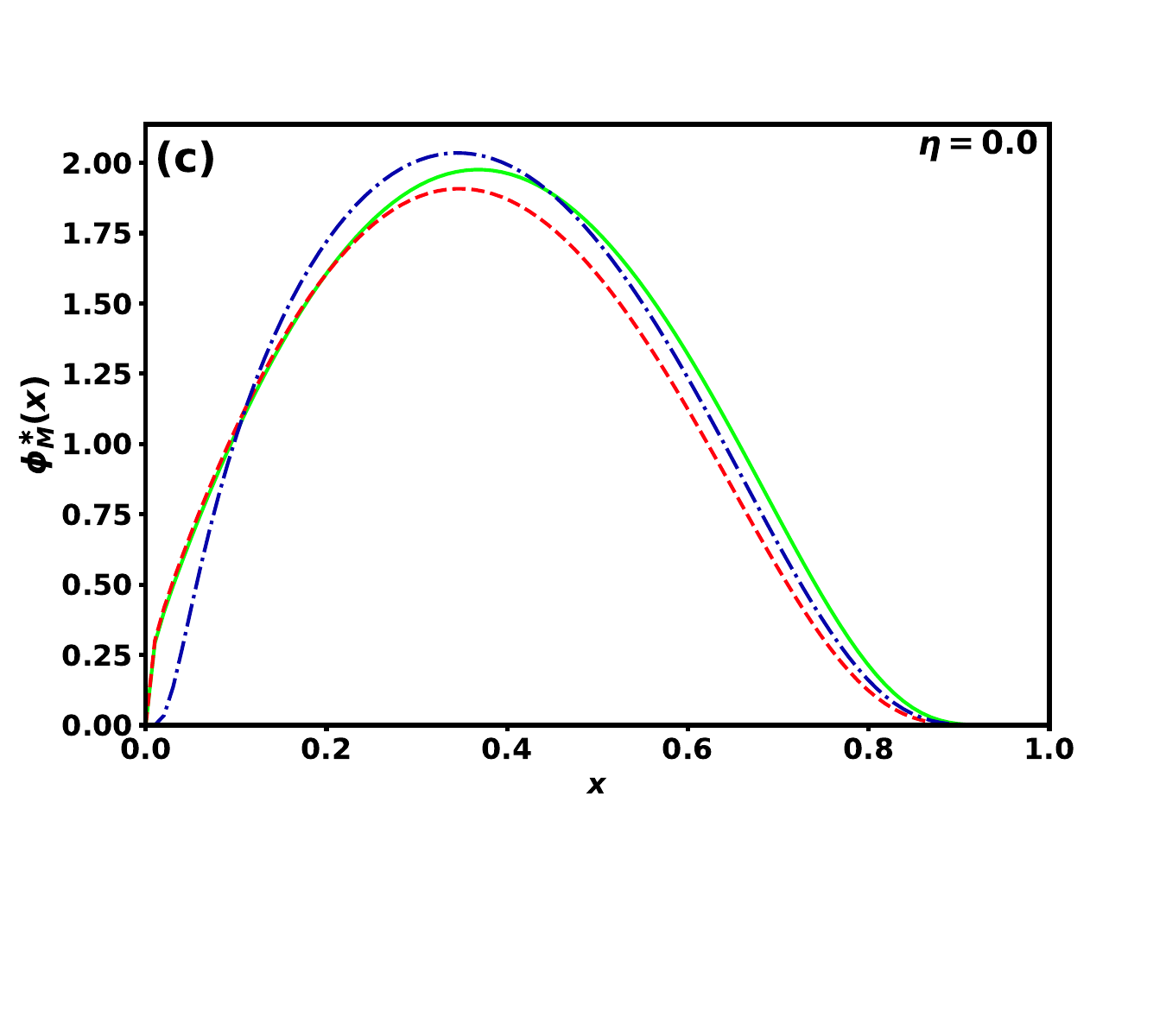}
    \end{minipage}

    \vspace{-15mm} % Reduce vertical space

    \begin{minipage}{0.48\linewidth}
        \centering
        \includegraphics[width=\linewidth]{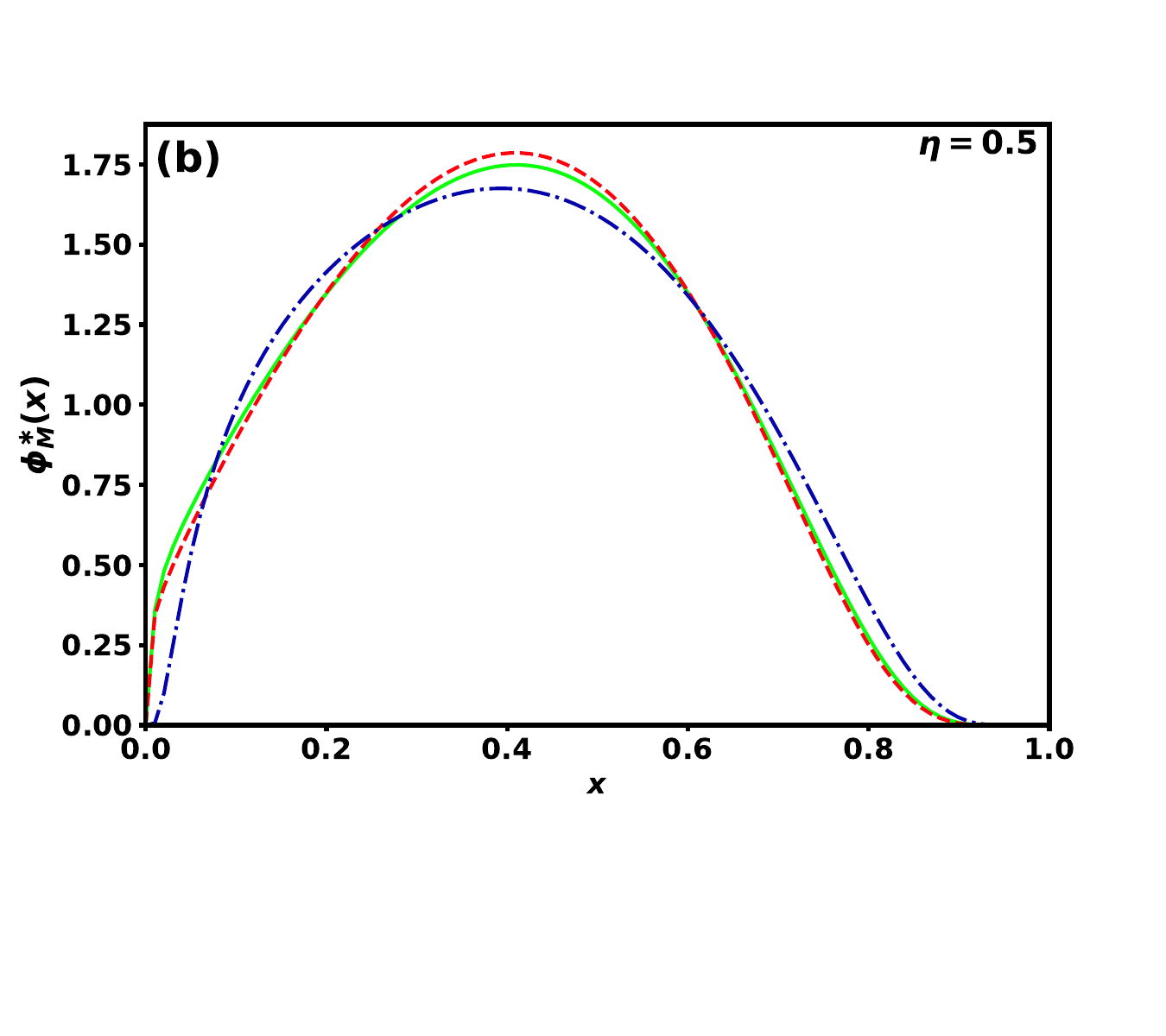}
    \end{minipage} \hspace{-6pt}
    \begin{minipage}{0.48\linewidth}
        \centering
        \includegraphics[width=\linewidth]{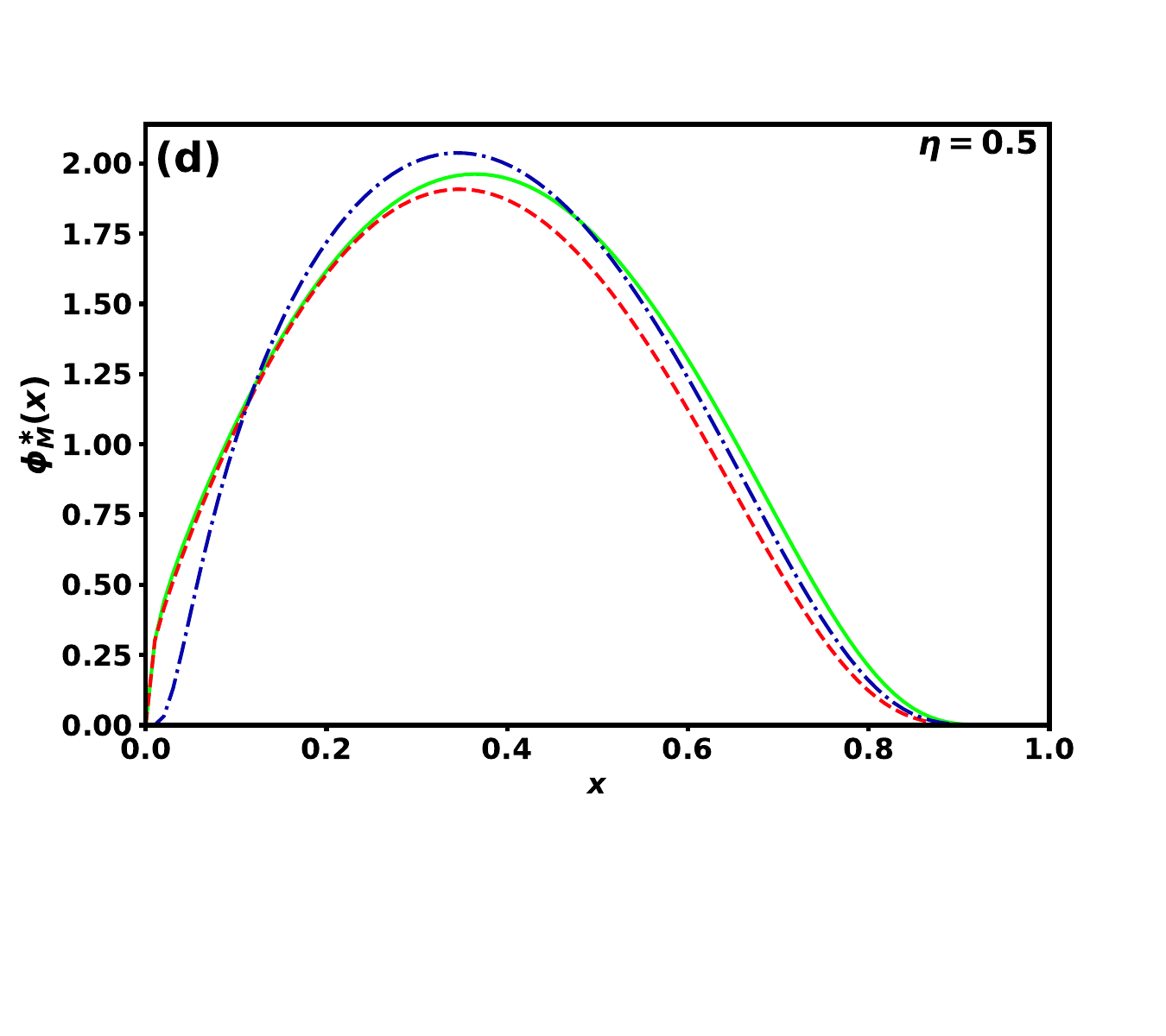}
    \end{minipage}

    \vspace{-15mm} % Reduce space before the last figure

    % Last Figure (Centered Below)
    \begin{minipage}{0.48\linewidth}
        \centering
        \includegraphics[width=\linewidth]{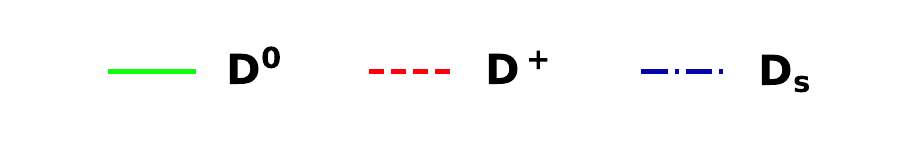}
    \end{minipage} \hspace{-6pt}
    \begin{minipage}{0.48\linewidth}
        \centering
        \includegraphics[width=\linewidth]{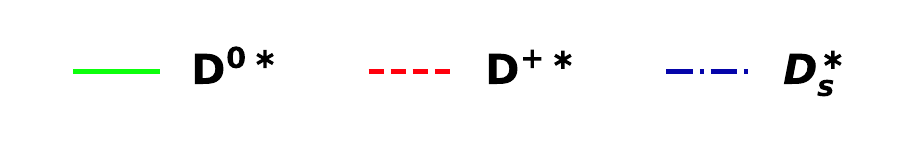}
    \end{minipage}

    \caption{Distribution amplitudes \(\phi^*_M(x)\) for pseudoscalar (\(D^0\), \(D^+\) and \(D_s\)) mesons [left panel] and  vector (\(D^{0*}\), \(D^{+*}\) and \(D_s^*\)) mesons [right panel] plotted as a function of longitudinal momentum fraction $x$ at baryon density $\rho_B=3\rho_0$ and temperature \(T=0\) GeV. The results are presented for symmetric nuclear medium ($\eta=0.0$) [in subplots (a) and (c)] and isospin asymmetric nuclear medium ($\eta=0.5$) [in subplots (b) and (d)].}
    
    \label{fig:6}
\end{figure}

 Fig.~\ref{fig:6} presents the DAs for pseudoscalar \(D\) and vector \(D^*\) mesons as a function of longitudinal momentum fraction \(x\), evaluated at baryon density  \(\rho_B=3\rho_0\) and temperature \(T=0\) GeV, at $\eta = 0$ and $0.5$.
% The results are shown for both isospin symmetric (\(\eta=0\)) and isospin asymmetric (\(\eta=0.5\)) nuclear medium. 
In the left panel, Fig.~\ref{fig:6}(a) indicates that the DAs for the \(D^0\) and \(D^+\) mesons are nearly identical in a symmetric medium. However, Fig.~\ref{fig:6}(b) shows that under isospin asymmetry (\(\eta=0.5\)), a noticeable separation emerges between their DAs. The right panel displays the DAs for vector \(D^*\) mesons. In Fig.~\ref{fig:6}(c) corresponding to the symmetric case, the DAs for \(D^{0*}\) and \(D^{+*}\) are observed to overlap for \(x<0.24\), followed by a noticable distinction beyond this region. When isospin asymmetry is introduced as shown, in Fig.~\ref{fig:6}(d), the DA for the \(D^{0*}\) meson shifts slightly upward compared to that of the \(D^{+*}\) meson for \(x<0.24\). Beyond this point, both DAs remain nearly unchanged from those in Fig.~\ref{fig:6}(c). In contrast, the DAs for the pseudoscalar \(D_s\) and vector \(D_s^*\) meson remain largely unaffected by isospin asymmetry. This stability is attributed to the presence of the \(s\) quark, which is less sensitive to medium effects. A similar behavior for the \(D\) and \(D^*\) mesons in symmetric nuclear medium was previously reported in Ref.~\cite{Arifi:2023jfe}.

\par
\begin{figure}[htbp]
    \centering

    % 2x2 Grid of Figures
    \begin{minipage}{0.48\linewidth}
        \centering
        \includegraphics[width=\linewidth]{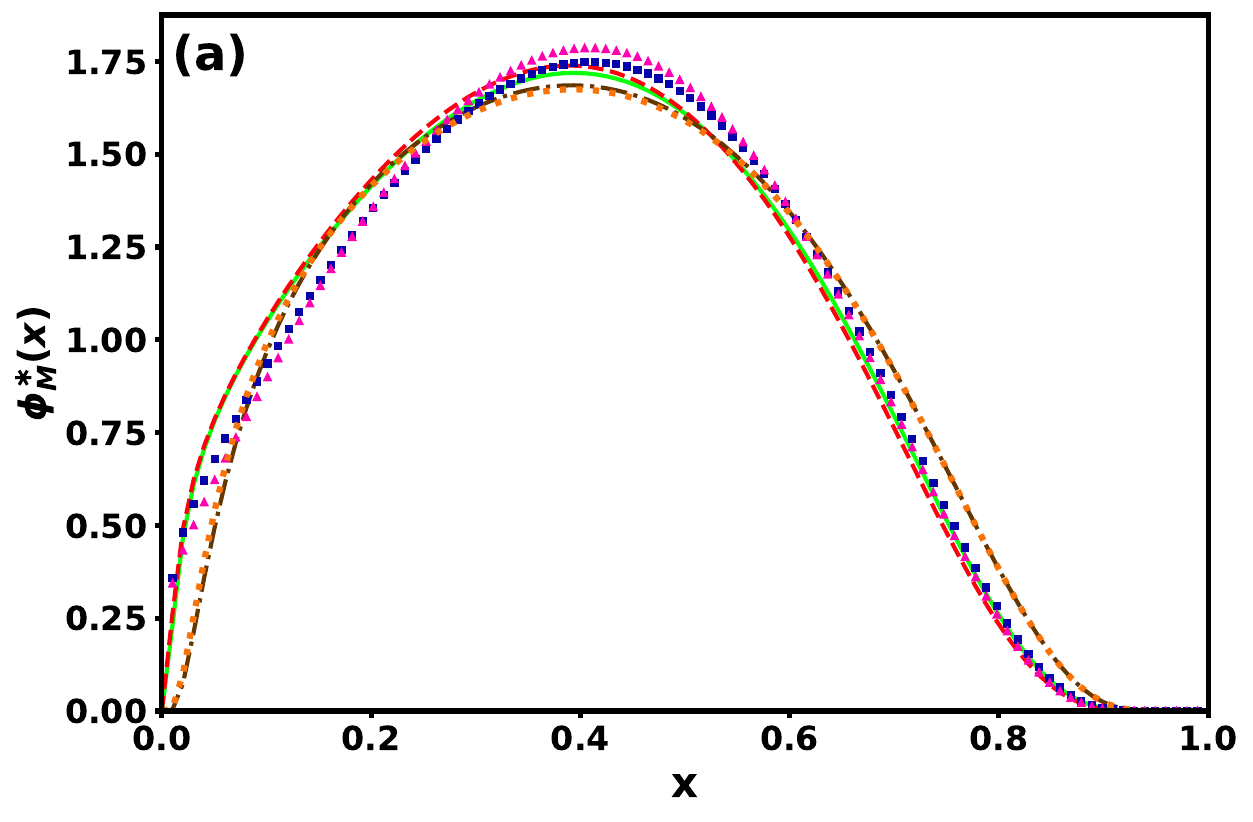}
    \end{minipage} \hspace{-6pt}
    \begin{minipage}{0.48\linewidth}
        \centering
        \includegraphics[width=\linewidth]{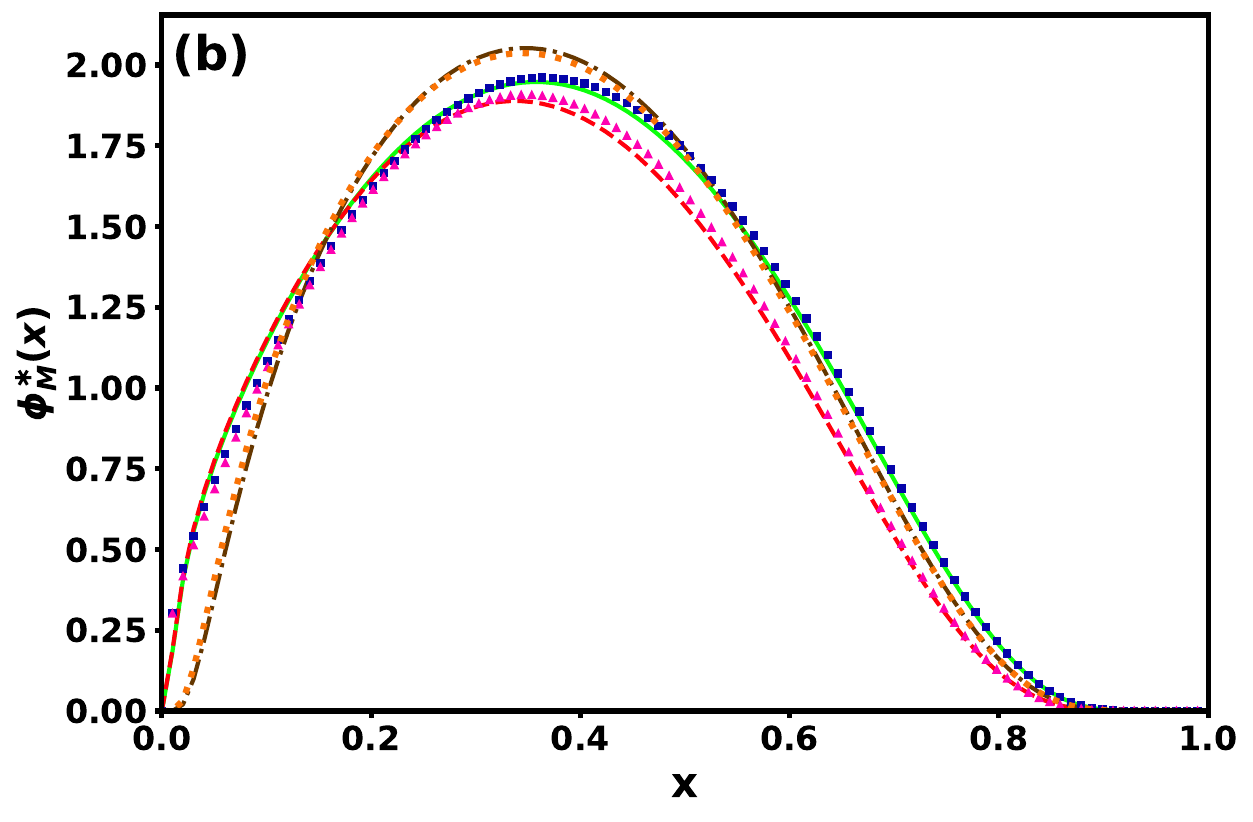}
    \end{minipage}

    \vspace{-2mm}
    \begin{minipage}{0.48\linewidth}
        \centering
        \includegraphics[width=\linewidth]{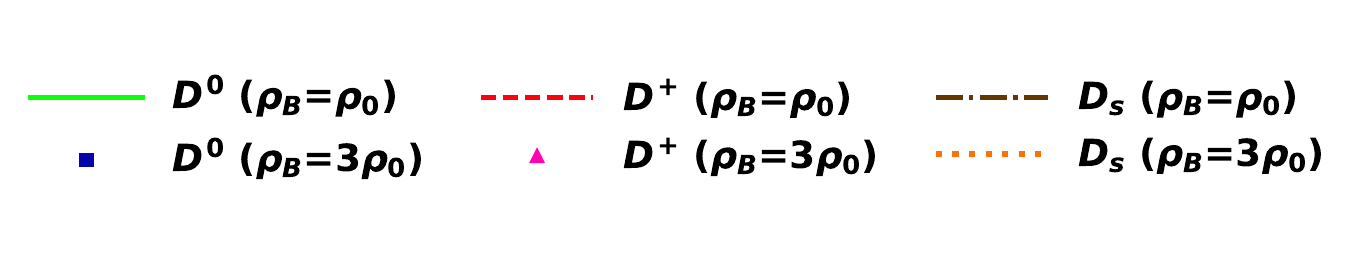}
    \end{minipage} \hspace{-6pt}
    \begin{minipage}{0.48\linewidth}
        \centering
        \includegraphics[width=\linewidth]{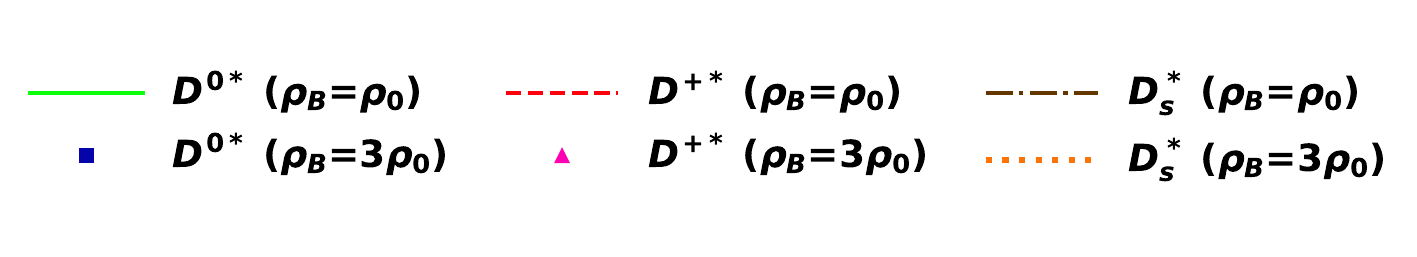}
    \end{minipage}

    \caption{Distribution amplitudes \(\phi^*_M(x)\) plotted as a function of longitudinal momentum fraction $x$ for pseudoscalar mesons (\(D^0\), \(D^+\) and \(D_s\)) [left panel] and  vector mesons (\(D^{0*}\), \(D^{+*}\) and \(D_s^*\)) [right panel] at isopin asymmetry $\eta=0.5$ and temperature \(T=0\) GeV. Each subplot displays results for baryon densities \(\rho_B=\rho_0\) and \(3\rho_0\).
    }
    
    \label{fig:7}
\end{figure}

 The density variation of in-medium DAs of  \(D\) and  \(D^*\) mesons in an isospin asymmetric nuclear medium (\(\eta=0.5\)) as a function of longitudinal momentum fraction $x$ at temperature \(T=0\) GeV is represented in Fig.~\ref{fig:7}. 
As illustrated in Fig.~\ref{fig:7}(a), for $\rho_B=\rho_0$, the DAs of \(D^0\) and \(D^+\) mesons overlaps at lower $x$, but separate noticeably at higher values of $x$. However, for $\rho_B=3\rho_0$, this separation becomes significant across the entire range of $x$. This behavior is attributed to the difference between the \(u\) and \(d\) quark masses induced by isospin asymmetry in the medium. The DA curves for \(D^0\) and \(D^+\) corresponding to higher density show a reduction in magnitude compared to the curves at $\rho_B=\rho_0$ for \(x<0.3\). In contrast, for larger values of $x$, there is a noticeable rise in magnitude, resulting in both curves converging around $x>0.8$. The variation in DAs of vector \(D^*\) mesons is presented in Fig.~\ref{fig:7}(b). Here, DAs for \(D^{0*}\) and \(D^{+*}\) mesons also coincide at low \(x\), while a clear splitting appears at higher $x$. However, the overall change in DAs for these mesons remains minor as the baryon density increases from \(\rho_B=\rho_0\) to $3\rho_0$. A similar pattern in the behavior of DAs with baryon density  for pseudoscalar and vector mesons is observed  in Ref.~\cite{Arifi:2023jfe}. For the pseudoscalar \(D_s\) and vector \(D_s^*\) mesons, the DAs remain largely unchanged near the endpoint $x\rightarrow0$ and $x\rightarrow1$ with increasing \(\rho_B\). There is, however, a slight decrease in the DAs for $\rho_B=3\rho_0$ within the interval $0.3<x<0.5$.

\par
\begin{figure}[htbp]
    \centering
    % 2x2 Grid of Figures
    \begin{minipage}{0.48\linewidth}
        \centering
        \includegraphics[width=\linewidth]{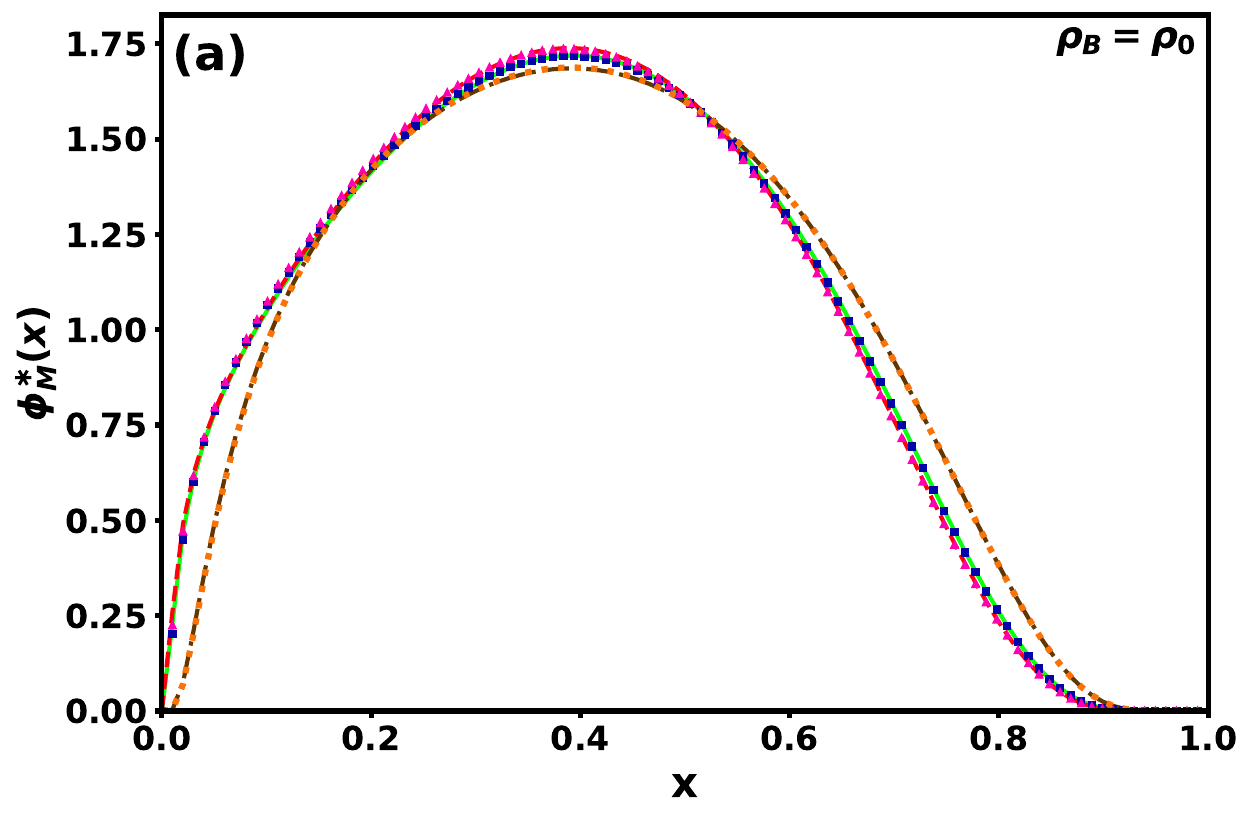}
    \end{minipage} \hspace{-6pt}
    \begin{minipage}{0.48\linewidth}
        \centering
        \includegraphics[width=\linewidth]{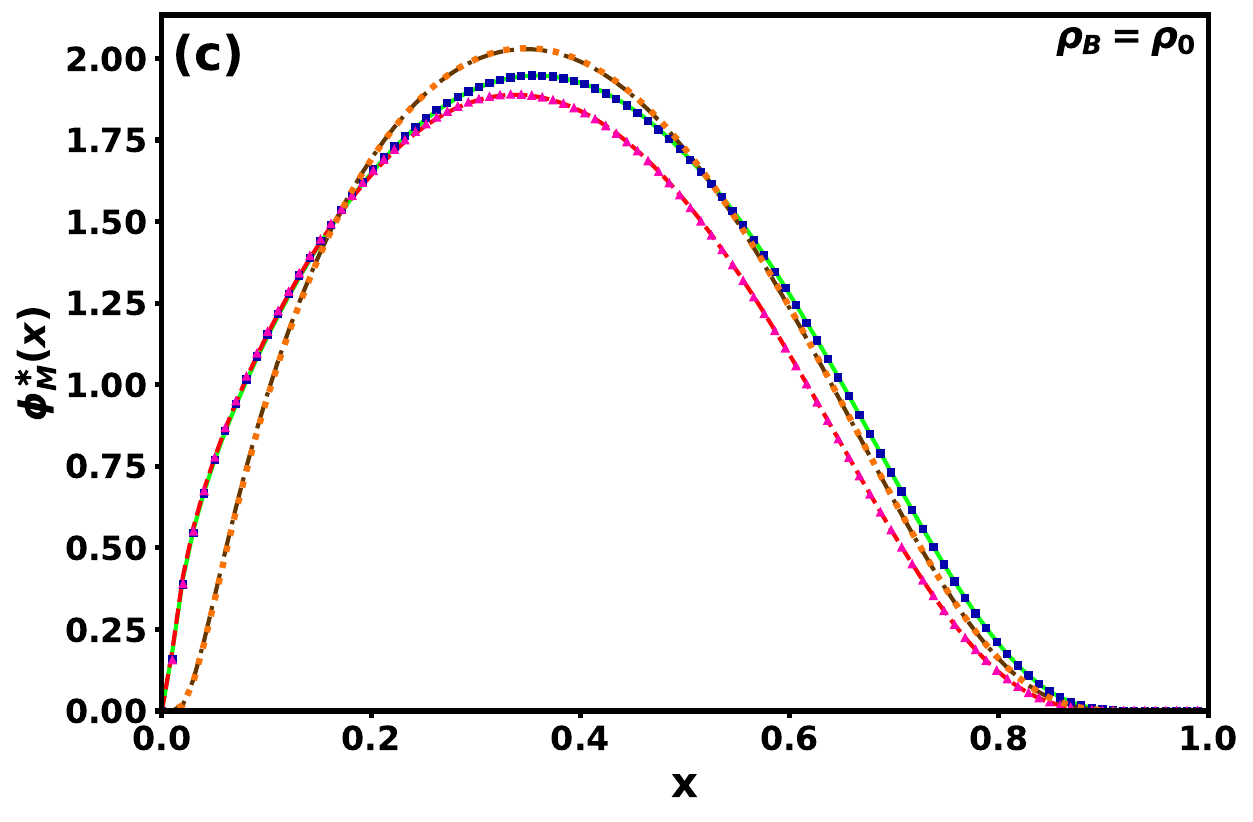}
    \end{minipage}

    \vspace{-3pt} % Reduce vertical space

    \begin{minipage}{0.48\linewidth}
        \centering
        \includegraphics[width=\linewidth]{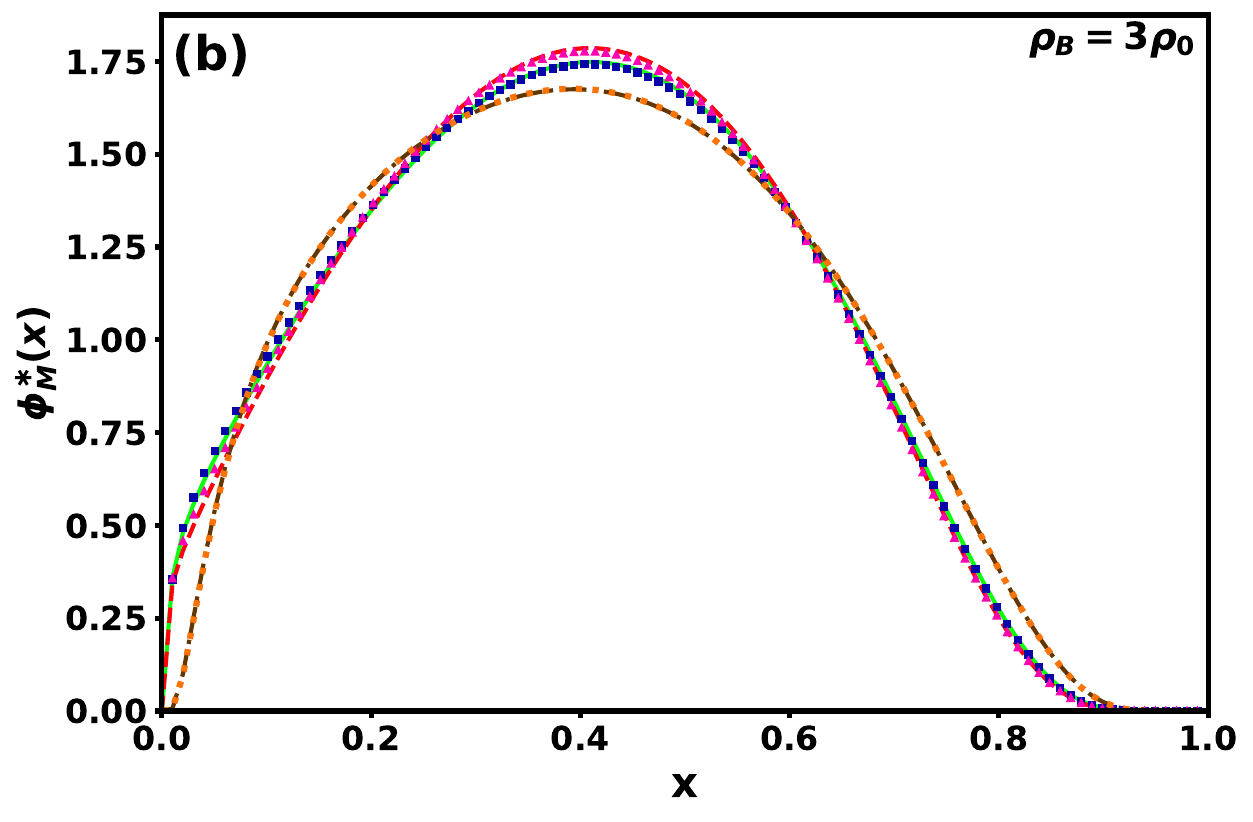}
    \end{minipage} \hspace{-6pt}
    \begin{minipage}{0.48\linewidth}
        \centering
        \includegraphics[width=\linewidth]{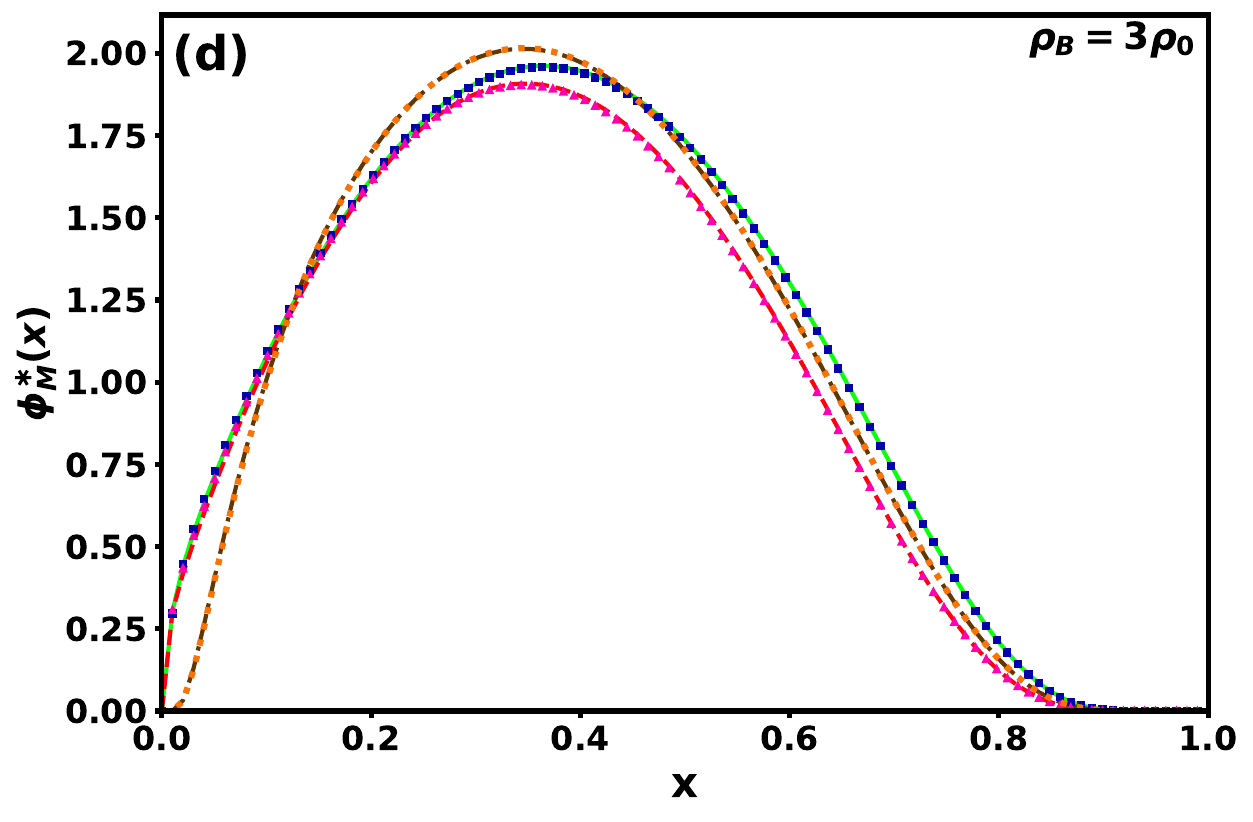}
    \end{minipage}

    \vspace{-3pt} % Reduce space before the last figure

    % Last Figure (Centered Below)
    \begin{minipage}{0.48\linewidth}
        \centering
        \includegraphics[width=\linewidth]{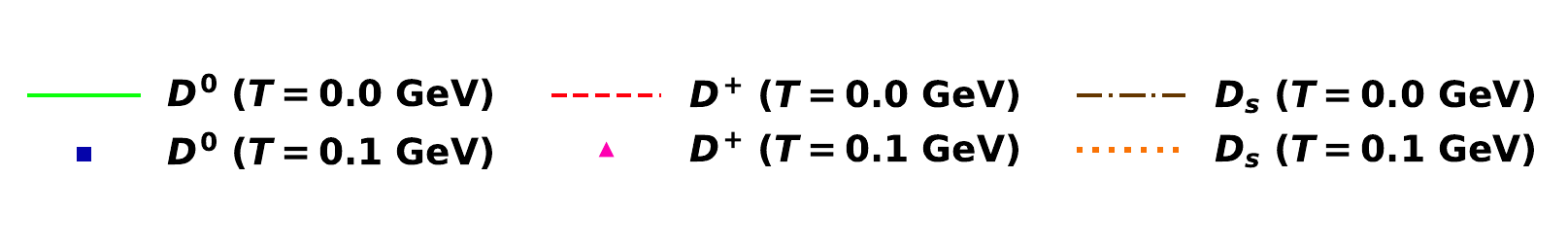}
    \end{minipage} \hspace{-6pt}
    \begin{minipage}{0.48\linewidth}
        \centering
        \includegraphics[width=\linewidth]{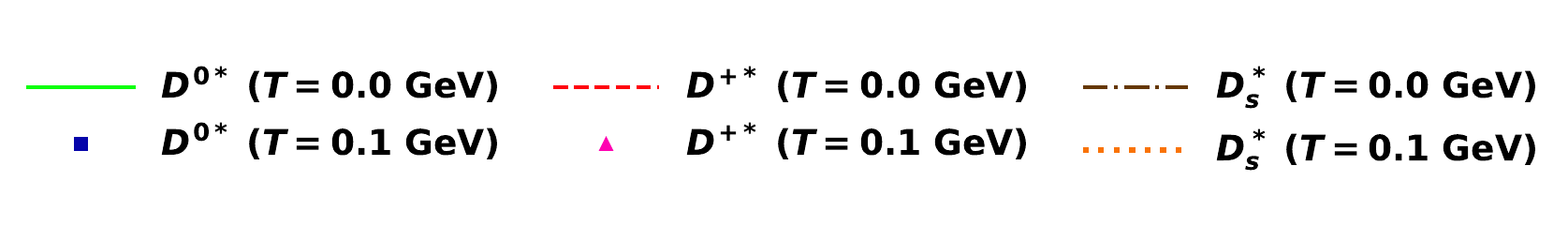}
    \end{minipage}

    \caption{Distribution amplitudes \(\phi^*_M(x)\) for pseudoscalar mesons (\(D^0\), \(D^+\) and \(D_s\)) [left panel] and vector mesons (\(D^{0*}\), \(D^{+*}\) and \(D_s^*\)) [right panel] plotted as a function of longitudinal momentum fraction $x$ at isospin asymmetry $\eta=0.5$. The results are presented for baryon densities $\rho_B=\rho_0$ [in subplots (a) and (c)] and  $3\rho_0$ [in subplots (b) and (d)] at temperature \(T=0\) and \(T=0.1\) GeV.}
    
    \label{fig:8}
\end{figure}

 To investigate the influence of temperature on the DAs of pseudoscalar \(D\) and vector \(D^*\) mesons, we plot $\phi^*_M(x)$ as a function of longitudinal momentum fraction $x$, as shown in Fig.~\ref{fig:8}. The analysis is carried out for an isospin asymmetry $\eta = 0.5$ at two baryon densities $\rho_B = \rho_0$ and $ 3\rho_0$. Each subplot presents results at temperatures \(T = 0\) and \( 0.1\) GeV. For both densities, the DAs for the pseudoscalar \(D^0\), \(D^+\), and \(D_s\) mesons as well as the vector \(D^{0*}\), \(D^{+*}\) and \(D^*_s\) mesons are noted to remain almost unchanged when temperatures increases from \(T=0\) to \(T=0.1\) GeV. However, a closer inspection of the peak values reveals small but noteworthy shifts with increasing baryon density. In particular, the maxima of the DA for the \(D^0\) meson changes from \(1.719 (1.718)\) to \(1.748(1.742)\), for the \(D^+\) meson from \(1.739(1.737)\) to \(1.786(1.777)\), and for the \(D_s\) meson from \(1.686(1.688)\) to \(1.675(1.677)\) as \(\rho_B\) increases from \(\rho_0\) to \(3\rho_0\) at \(T=0(0.1)\) GeV. Similarly, for vector mesons, the peak of the DA for the \(D^{0*}\) meson shifts from \(1.948(1.947)\) to \(1.962(1.958)\), the \(D^{+*}\) meson from \(1.888(1.89)\) to \(1.908(1.903)\), and the \(D^*_s\) meson from \(2.029(2.032)\) to \(2.012(2.016)\), under the same conditions. These changes highlight the sensitivity of DAs to the baryon density, even though their overall temperature dependence remains minimal in the range considered.
% \par In literature, there has not been much research done on determining the DAs for the \(D\) mesons in the nuclear medium. However, the behavior of \(D\) mesons DA has been studied through a revised light-cone harmonic oscillator model in Ref.~\cite{Zhong:2020cqr}. Moreover, QCD sum rules under the background theory have been applied to investigate the leading-twist light-cone DA of the \(D_s\) meson \cite{Zhang:2021wnv}.

% Second figure (Grid of images)

\section{Summary}
\label{sec:4}
In this study, we investigated the medium modifications of pseudoscalar (\(D^0\), \(D^+\), \(D_s\)) and vector (\(D^{0*}\), \(D^{+*}\), \(D_s^*\)) mesons using a hybrid framework that combines the LFQM with the CQMF model. The in-medium quark masses, determined using the CQMF model were used as input in the LFQM to evaluate meson properties in an isospin asymmetric nuclear medium at zero and finite temperature.
\par We first determined the meson masses and weak decay constants in vacuum using the variational principle to calculate the Gaussian parameter \(\beta\). These results were found to be comparable with the available experimental data and theoretical predictions. In-medium effects were analyzed over a range of baryon density (\(\rho_B\)) and isospin asymmetries (\(\eta\)). The masses of pseudoscalar \(D^0\) and \(D^+\) mesons were observed to decrease with increasing \(\rho_B\), while that of vector \(D^*\) mesons, containing \(u\) or \(d\) quarks, initially decreased but increased at higher densities. The isosppin asymmetry induced mass splitting within the isospin doublet \(D=\begin{pmatrix} D^+ \\ D^0 \end{pmatrix}\) and \(D^*=\begin{pmatrix} D^{+*} \\ D^{0*} \end{pmatrix}\). The pseudoscalar \(D_s\) and vector \(D_s^*\) mesons exhibited minimal sensitivity to isospin asymmetry due to the presence of the \(s\) quark. Furthermore, the effect of finite temperature was found to suppress medium modifications, leading to a reduced mass shift of pseudoscalar \(D\) and vector \(D^*\) mesons from their vacuum values. The weak decay constant ratios \(f^*_{M}/f_{M}\) decreased with increasing baryon density, especially for mesons containing light (\(u/d\)) quarks, while mesons having heavier \(s\) and \(c\) quark pair showed minimal change with rising temperature and isospin asymmetry \(\eta\). We also analyzed the medium-dependent DAs, \(\phi^*_M(x)\), of \(D\) and \(D^*\) mesons as a function of the longitudinal momentum fraction \(x\). In vacuum, the DAs showed the expected convex-concave structure, indicating that the charm quark carries majority of the meson's momentum. The DAs of vector \(D^*\) mesons display higher peaks in comparison to the pseudoscalar \(D\) mesons due to contributions from the \(2\textbf{k}^2_\perp/\mathcal{M^*}\) term in the wave function. The DA profile for the \(D_s\) meson had the lowest peak due to the small mass difference between the \(c\) and \(s\) quarks. In contrast, for vector mesons, the trend was reversed, with \(D_s^*\) displaying a maxima  compared to \(D^{0*}\) and \(D^{+*}\) mesons. At finite $\rho_B$ and $\eta$, significant modifications in the DAs were observed. Increasing $\eta$ from \(0\) to \(0.5\) led to a noticeable DA splitting between the \(D^0\) and \(D^+\) mesons for \(x<0.7\), and between \(D^{0*}\) and \(D^{+*}\) mesons for \(x<0.24\). A slight shift in the maxima towards the left was observed for both pseudoscalar and vector \(D\) mesons. Increasing \(\rho_B\) from \(\rho_0\) to \(3\rho_0\) resulted in DAs of \(D^0\), \(D^+\), and their vector counterparts to increase in the region \(x>0.3\), while a suppression was observed at lower \(x\). In contrast, \(D_s\) and \(D_s^*\) mesons exhibit only a slight decrease in the intermediate \(0.3<x<0.5\). A change in temperature  from \(T=0\) to \(0.1\) GeV produced no change in the DAs at finite \(\rho_B\), further confirming weak thermal sensitivity. 
\par 
% By bridging the gap between effective hadronic models and fundamental quark-level dynamics for the nuclear matter, our work offers valuable insights towards the development of more refined models based on the nuclear matter grounded in QCD. 
This work strengthens the link between quark-level dynamics and phenomenological hadronic models probing the internal structure of mesons in dense nuclear environments. It lays a theoretical foundation for understanding the behavior of medium effects on open charm mesons under extreme conditions. The results are particularly relevant for ongoing and upcoming experiments, such as CBM and PANDA at FAIR, where in-medium modifications of charm hadrons may be probed. The framework presented here can be extended to study other non-perturbative QCD phenomena, including in-medium electromagnetic form factors, Mellin moments, and related hadronic observables of heavy mesons in the dense matter.

\section{Acknowledgement}
H.D. would like to thank  the Science and Engineering Research Board, Anusandhan-National Research Foundation, Government of India under the scheme SERB-POWER Fellowship (Ref No. SPF/2023/000116) for financial support. A. K. sincerely acknowledge Anusandhan National
Research Foundation (ANRF), Government of India for
funding of the research project under the Science and
Engineering Research Board-Core Research Grant
(SERB-CRG) scheme (File No. CRG/2023/000557).


\section{Reference}

\begin{thebibliography}{3}
\bibitem{EuropeanMuon:1983wih}
J.~J.~Aubert \textit{et al.} [European Muon],Phys. Lett. B \textbf{123}, 275 (1983).

\bibitem{Suzuki:2002ae}
K.~Suzuki, M.~Fujita, H.~Geissel, H.~Gilg, A.~Gillitzer, R.~S.~Hayano, S.~Hirenzaki, K.~Itahashi, M.~Iwasaki and P.~Kienle, \textit{et al.}
Phys. Rev. Lett. \textbf{92}, 072302 (2004).

\bibitem{Friedman:2004jh}
E.~Friedman, M.~Bauer, J.~Breitschopf, H.~Clement, H.~Denz, E.~Doroshkevich, A.~Erhardt, G.~J.~Hofman, R.~Meier and G.~J.~Wagner, \textit{et al.}, Phys. Rev. Lett. \textbf{93}, 122302 (2004).

\bibitem{CHAOS:1996nql}
F.~Bonutti \textit{et al.} [CHAOS],
Phys. Rev. Lett. \textbf{77}, 603 (1996).

\bibitem{CHAOS:2004rhl}
P.~Camerini \textit{et al.} [CHAOS],
Nucl. Phys. A \textbf{735}, 89 (2004).

% \bibitem{8}
% See https://www.gsi.de/en/work/research/pandahadrons/experiments/panda

\bibitem{PANDA:2009yku}
M.~F.~M.~Lutz \textit{et al.} [PANDA],
%``Physics Performance Report for PANDA: Strong Interaction Studies with Antiprotons,''
[arXiv:0903.3905 [hep-ex]].

\bibitem{Ohnishi:2019cif}
H.~Ohnishi, F.~Sakuma and T.~Takahashi,
%``Hadron Physics at J-PARC,''
Prog. Part. Nucl. Phys. \textbf{113}, 103773 (2020).
% doi:10.1016/j.ppnp.2020.103773
% [arXiv:1912.02380 [nucl-ex]].

\bibitem{Sawada:2007gy}
S.~Sawada,
%``J-PARC, Japan Proton Accelerator Research Complex,''
Nucl. Phys. A \textbf{782}, 434 (2007).
% doi:10.1016/j.nuclphysa.2006.10.070

\bibitem{Sakaguchi:2019xjv}
T.~Sakaguchi [J-PARC-HI],
%``High density matter physics at J-PARC-HI,''
PoS \textbf{CORFU2018}, 189 (2019).
% doi:10.22323/1.347.0189
% [arXiv:1904.12821 [nucl-ex]].

% \bibitem{7}
% See https://www.gsi.de/en/researchaccelerators/fair

\bibitem{Senger:2012wr}
P.~Senger,
%``The compressed baryonic matter experiment at FAIR,''
Central Eur. J. Phys. \textbf{10}, 1289 (2012).
% doi:10.2478/s11534-012-0048-5

% \bibitem{8}
% See https://www.gsi.de/en/work/research/pandahadrons/experiments/panda

% \bibitem{PANDA:2009yku}
% M.~F.~M.~Lutz \textit{et al.} [PANDA],
% %``Physics Performance Report for PANDA: Strong Interaction Studies with Antiprotons,''
% [arXiv:0903.3905 [hep-ex]].

\bibitem{Kekelidze:2017ghu}
V.~D.~Kekelidze, \textit{et al.},
%``Project Nuclotron-based Ion Collider fAcility at JINR,''
Phys. Part. Nucl. \textbf{48}, 727 (2017).
% doi:10.1134/S1063779617050239

\bibitem{Kekelidze:2017tgp}
V.~Kekelidze, \textit{et al.},
%``Feasibility study of heavy-ion collision physics at NICA JINR,''
Nucl. Phys. A \textbf{967}, 884 (2017).
% doi:10.1016/j.nuclphysa.2017.06.031

\bibitem{Rapp:2008tf}
R.~Rapp, D.~Blaschke and P.~Crochet,
Prog. Part. Nucl. Phys. \textbf{65}, 209 (2010).

% \bibitem{Aoki:2023qgl}
% K.~Aoki, D.~Arimizu, S.~Ashikaga, W.~C.~Chang, T.~Chujo, K.~Ebata, H.~En\textquoteright{}yo, S.~Esumi, H.~Hamagaki and R.~Honda, \textit{et al.}
% %``Experimental Study of In-medium Spectral Change of Vector Mesons at J-PARC,''
% Few Body Syst. \textbf{64}, 63 (2023)
% % doi:10.1007/s00601-023-01828-7

% \bibitem{CBELSATAPS:2005iwc}
% D.~Trnka \textit{et al.} [CBELSA/TAPS],
% %``First observation of in-medium modifications of the omega meson,''
% Phys. Rev. Lett. \textbf{94}, 192303 (2005).
% % doi:10.1103/PhysRevLett.94.192303
% % [arXiv:nucl-ex/0504010 [nucl-ex]].

% \bibitem{KEK-PS-E325:2005wbm}
% R.~Muto \textit{et al.} [KEK-PS-E325],
% %``Evidence for in-medium modification of the phi meson at normal nuclear density,''
% Phys. Rev. Lett. \textbf{98}, 042501 (2007).
% % doi:10.1103/PhysRevLett.98.042501
% % [arXiv:nucl-ex/0511019 [nucl-ex]].

\bibitem{Kim:2022lng}
D.~N.~Kim and G.~A.~Miller,
Phys. Rev. C \textbf{106}, 055202 (2022).

\bibitem{Hutauruk:2018qku}
P.~T.~P.~Hutauruk, Y.~Oh and K.~Tsushima,
%``Electroweak properties of pions in a nuclear medium,''
Phys. Rev. C \textbf{99}, 015202 (2019)
% doi:10.1103/PhysRevC.99.015202
% [arXiv:1810.08874 [nucl-th]].

\bibitem{Roberts:2000aa}
C.~D.~Roberts and S.~M.~Schmidt,
%``Dyson-Schwinger equations: Density, temperature and continuum strong QCD,''
Prog. Part. Nucl. Phys. \textbf{45}, S1-S103 (2000).
% doi:10.1016/S0146-6410(00)90011-5
% [arXiv:nucl-th/0005064 [nucl-th]].

\bibitem{deMelo:2016uwj}
J.~P.~B.~C.~de Melo, K.~Tsushima and I.~Ahmed,
Phys. Lett. B \textbf{766}, 125 (2017).

\bibitem{deMelo:2018hfw}
J.~P.~B.~C.~de Melo and K.~Tsushima,
Phys. Lett. B \textbf{788}, 137 (2019).

%\cite{Kaur:2024wze}
\bibitem{Kaur:2024wze}
N.~Kaur, S.~Puhan, R.~Pandey, A.~Kumar, S.~Dutt and H.~Dahiya,
%``Does nuclear medium affect the transverse momentum-dependent parton distributions of valence quark of pions?,''
Phys. Lett. B \textbf{859}, 139114 (2024).
%doi:10.1016/j.physletb.2024.139114
%[arXiv:2409.05394 [hep-ph]].
%6 citations counted in INSPIRE as of 09 Jun 2025

\bibitem{Singh:2024lra}
D.~Singh, S.~Puhan, N.~Kaur, M.~Kaur, A.~Kumar, S.~Dutt and H.~Dahiya,
%``Effect of an asymmetric nuclear medium on the valence quark structure of the kaons,''
Phys. Rev. D \textbf{111}, 054001 (2025).
% doi:10.1103/PhysRevD.111.054001
% [arXiv:2410.20181 [hep-ph]].

\bibitem{Puhan:2024xdq}
S.~Puhan, N.~Kaur, A.~Kumar, S.~Dutt and H.~Dahiya,
%``Pion valence quark distributions in asymmetric nuclear matter at finite temperature,''
Phys. Rev. D \textbf{110}, 054042 (2024).
% doi:10.1103/PhysRevD.110.054042
% [arXiv:2408.07334 [nucl-th]].

%\cite{Puhan:2025ibn}
\bibitem{Puhan:2025ibn}
S.~Puhan, N.~Kaur, A.~Kumar, S.~Dutt and H.~Dahiya,
%``Effect of nuclear medium on the spatial distribution of pions,''
Nucl. Phys. B \textbf{1017}, 116940 (2025).
%doi:10.1016/j.nuclphysb.2025.116940
%[arXiv:2501.16706 [hep-ph]].
%1 citations counted in INSPIRE as of 09 Jun 2025



%\cite{Tanisha:2025glu}
\bibitem{Tanisha:2025glu}
Tanisha, S.~Puhan, N.~Kaur, A.~Kumar and H.~Dahiya,
%``Impact of isospin asymmetric nuclear medium on pseudoscalar and vector $B$ mesons,''
[arXiv:2504.21392 [nucl-th]].
%0 citations counted in INSPIRE as of 09 Jun 2025

\bibitem{Tsushima:2016sep}
K.~Tsushima and J.~P.~B.~C.~de Melo,
%``In-Medium Pion Valence Distribution Amplitude,''
Few Body Syst. \textbf{58}, 85 (2017).

\bibitem{Zschiesche:2003qq}
D.~Zschiesche, A.~Mishra, S.~Schramm, H.~Stoecker and W.~Greiner,
%``In-medium vector meson masses in a chiral SU(3) model,''
Phys. Rev. C \textbf{70}, 045202 (2004).
% \bibitem{13}
% R.~Russo [ALICE], Nuovo Cim. C \textbf{037}, no.01, 315-317 (2014)

\bibitem{Matsui:1986dk}
T.~Matsui and H.~Satz,
%``$J/\psi$ Suppression by Quark-Gluon Plasma Formation,''
Phys. Lett. B \textbf{178}, 416 (1986).
% doi:10.1016/0370-2693(86)91404-8

\bibitem{NA50:1996lag}
M.~Gonin \textit{et al.} [NA50],
%``Anomalous J / psi suppression in Pb + Pb collisions at 158-A-GeV/c,''
Nucl. Phys. A \textbf{610}, 404C (1996)
% doi:10.1016/S0375-9474(96)00373-9

\bibitem{Garcia-Recio:2010fiq}
C.~Garcia-Recio, J.~Nieves and L.~Tolos,
%``D mesic nuclei,''
Phys. Lett. B \textbf{690}, 369 (2010).
% doi:10.1016/j.physletb.2010.05.056
% [arXiv:1004.2634 [nucl-th]].

\bibitem{Tolos:2013gta}
L.~Tolos,
%``Charming mesons with baryons and nuclei,''
Int. J. Mod. Phys. E \textbf{22}, 1330027 (2013).
% doi:10.1142/S0218301313300270
% [arXiv:1309.7305 [nucl-th]].

\bibitem{Tsushima:2011kh}
K.~Tsushima, D.~H.~Lu, G.~Krein and A.~W.~Thomas,
%``$J/\Psi$-nuclear bound states,''
Phys. Rev. C \textbf{83}, 065208 (2011).
% doi:10.1103/PhysRevC.83.065208
% [arXiv:1103.5516 [nucl-th]].

\bibitem{Tolos:2005ft}
L.~Tolos, J.~Schaffner-Bielich and H.~Stoecker,
%``D-mesons: In-medium effects at FAIR,''
Phys. Lett. B \textbf{635}, 85 (2006).
% doi:10.1016/j.physletb.2006.02.045
% [arXiv:nucl-th/0509054 [nucl-th]].

\bibitem{Tolos:2004yg}
L.~Tolos, J.~Schaffner-Bielich and A.~Mishra,
%``Properties of D-mesons in nuclear matter within a self-consistent coupled-channel approach,''
Phys. Rev. C \textbf{70}, 025203 (2004)
% doi:10.1103/PhysRevC.70.025203
% [arXiv:nucl-th/0404064 [nucl-th]].

\bibitem{Lutz:2005vx}
M.~F.~M.~Lutz and C.~L.~Korpa,
%``Open-charm systems in cold nuclear matter,''
Phys. Lett. B \textbf{633}, 43 (2006).
% doi:10.1016/j.physletb.2005.11.046
% [arXiv:nucl-th/0510006 [nucl-th]].

\bibitem{Cassing:2000vx}
W.~Cassing, E.~L.~Bratkovskaya and A.~Sibirtsev,
%``Open charm production in relativistic nucleus-nucleus collisions,''
Nucl. Phys. A \textbf{691}, 753 (2001).
% doi:10.1016/S0375-9474(01)00562-0
% [arXiv:nucl-th/0010071 [nucl-th]].

\bibitem{Hayashigaki:2000es}
A.~Hayashigaki,
%``Mass modification of D meson at finite density in QCD sum rule,''
Phys. Lett. B \textbf{487}, 96 (2000).
% doi:10.1016/S0370-2693(00)00760-7
% [arXiv:nucl-th/0001051 [nucl-th]].

\bibitem{Tsushima:1998ru}
K.~Tsushima, D.~H.~Lu, A.~W.~Thomas, K.~Saito and R.~H.~Landau,
%``Charmed mesic nuclei,''
Phys. Rev. C \textbf{59}, 2824 (1999).
% doi:10.1103/PhysRevC.59.2824
% [arXiv:nucl-th/9810016 [nucl-th]].



%\cite{Mishra:2008cd}
\bibitem{Mishra:2008cd}
A.~Mishra and A.~Mazumdar,
%``D-mesons in asymmetric nuclear matter,''
Phys. Rev. C \textbf{79}, 024908 (2009).
%doi:10.1103/PhysRevC.79.024908
%[arXiv:0810.3067 [nucl-th]].
%71 citations counted in INSPIRE as of 09 Jun 2025

%\cite{Kumar:2011ff}
\bibitem{Kumar:2011ff}
A.~Kumar and A.~Mishra,
%``D-mesons and charmonium states in hot isospin asymmetric strange hadronic matter,''
Eur. Phys. J. A \textbf{47}, 164 (2011).
%doi:10.1140/epja/i2011-11164-6
%[arXiv:1102.4792 [nucl-th]].
%75 citations counted in INSPIRE as of 09 Jun 2025


\bibitem{ReddyP:2017tgo}
S.~Reddy P., A.~Jahan C.~S., N.~Dhale, A.~Mishra and J.~Schaffner-Bielich,
%``D mesons in strongly magnetized asymmetric nuclear matter,''
Phys. Rev. C \textbf{97}, 065208 (2018).
% doi:10.1103/PhysRevC.97.065208
% [arXiv:1712.07997 [nucl-th]].

\bibitem{Kumar:2019axp}
R.~Kumar and A.~Kumar,
%``Analysis of pseudoscalar and scalar $D$ mesons and charmonium decay width in hot magnetized asymmetric nuclear matter,''
Phys. Rev. C \textbf{101}, 015202 (2020).
% doi:10.1103/PhysRevC.101.015202
% [arXiv:1908.09172 [hep-ph]].

\bibitem{Chhabra:2017emy}
R.~Chhabra and A.~Kumar,
%``Masses and decay widths of scalar $D_0$ and $D_{s0}$ mesons in a strange hadronic medium,''
Phys. Rev. C \textbf{98}, 025205 (2018).
% doi:10.1103/PhysRevC.98.025205
% [arXiv:1710.08320 [hep-ph]].

\bibitem{Chhabra:2016vhp}
R.~Chhabra and A.~Kumar,
%``In-medium pseudoscalar $D/B$ mesons and charmonium decay width,''
Eur. Phys. J. A \textbf{53}, 105 (2017).
% doi:10.1140/epja/i2017-12285-6

\bibitem{Kumar:2013tna}
A.~Kumar,
%``Heavy Scalar, Vector and Axial-Vector Mesons in Hot and Dense Nuclear Medium,''
Adv. High Energy Phys. \textbf{2014}, 549726 (2014).
% doi:10.1155/2014/549726
% [arXiv:1308.3768 [nucl-th]].

\bibitem{Kumar:2015fca}
A.~Kumar and R.~Chhabra,
%``Heavy Vector and Axial-Vector Mesons in Asymmetric Strange Hadronic Matter,''
Phys. Rev. C \textbf{92}, 035208 (2015).
% doi:10.1103/PhysRevC.92.035208
% [arXiv:1506.02115 [nucl-th]].

\bibitem{Hosaka:2016ypm}
A.~Hosaka, T.~Hyodo, K.~Sudoh, Y.~Yamaguchi and S.~Yasui,
%``Heavy Hadrons in Nuclear Matter,''
Prog. Part. Nucl. Phys. \textbf{96}, 88 (2017).
% doi:10.1016/j.ppnp.2017.04.003
% [arXiv:1606.08685 [hep-ph]].

\bibitem{Chhabra:2018oyl}
R.~Chhabra and A.~Kumar,
%``D and B Mesons in Hot and Dense Symmetric Nuclear Medium,''
Springer Proc. Phys. \textbf{203}, 657 (2018).
% doi:10.1007/978-3-319-73171-1\_155

\bibitem{Chhabra:2017rxz}
R.~Chhabra and A.~Kumar,
%``In-medium properties of pseudoscalar $D_s$ and $B_s$ mesons,''
Eur. Phys. J. C \textbf{77}, 726 (2017).
% doi:10.1140/epjc/s10052-017-5277-8
% [arXiv:1711.00631 [hep-ph]].

\bibitem{Arifi:2024mff}
A.~J.~Arifi, L.~Happ, S.~Ohno and M.~Oka,
%``Structure of heavy mesons in the light-front quark model,''
Phys. Rev. D \textbf{110}, 014020 (2024).
% doi:10.1103/PhysRevD.110.014020
% [arXiv:2401.07933 [hep-ph]].

\bibitem{Arifi:2023jfe}
A.~J.~Arifi, P.~T.~P.~Hutauruk and K.~Tsushima,
%``In-medium properties of the light and heavy-light mesons in a light-front quark model,''
Phys. Rev. D \textbf{107}, 114010 (2023).
% doi:10.1103/PhysRevD.107.114010
% [arXiv:2302.12382 [hep-ph]].

\bibitem{Brodsky:1997de}
S.~J.~Brodsky, H.~C.~Pauli and S.~S.~Pinsky,
%``Quantum chromodynamics and other field theories on the light cone,''
Phys. Rept. \textbf{301}, 299 (1998).
% doi:10.1016/S0370-1573(97)00089-6
% [arXiv:hep-ph/9705477 [hep-ph]].

\bibitem{Wilson:1994fk}
K.~G.~Wilson, T.~S.~Walhout, A.~Harindranath, W.~M.~Zhang, R.~J.~Perry and S.~D.~Glazek,
%``Nonperturbative QCD: A Weak coupling treatment on the light front,''
Phys. Rev. D \textbf{49}, 6720 (1994).
% doi:10.1103/PhysRevD.49.6720
% [arXiv:hep-th/9401153 [hep-th]].

%\bibitem{Liu:2001iz}
%B.~Liu, V.~Greco, V.~Baran, M.~Colonna and M.~Di Toro,
%%``Asymmetric nuclear matter: The Role of the isovector scalar channel,''
%Phys. Rev. C \textbf{65}, 045201 (2002).
%% doi:10.1103/PhysRevC.65.045201
%% [arXiv:nucl-th/0112034 [nucl-th]].
%
%\bibitem{Thakur:2022dxb}
%V.~Thakur, R.~Kumar, P.~Kumar, V.~Kumar, M.~Kumar, C.~Mondal, B.~K.~Agrawal and S.~K.~Dhiman,
%%``Effects of an isovector scalar meson on the equation of state of dense matter within a relativistic mean field model,''
%Phys. Rev. C \textbf{106}, 045806 (2022).
%% doi:10.1103/PhysRevC.106.045806
%% [arXiv:2210.02793 [nucl-th]].
%
%\bibitem{Li:2022okx}
%F.~Li, B.~J.~Cai, Y.~Zhou, W.~Z.~Jiang and L.~W.~Chen,
%%``Effects of Isoscalar- and Isovector-scalar Meson Mixing on Neutron Star Structure,''
%Astrophys. J. \textbf{929}, 183 (2022).
%% doi:10.3847/1538-4357/ac5e2a
%% [arXiv:2202.08705 [nucl-th]].



\bibitem{Wang:2001jw}
P.~Wang, Z.~Y.~Zhang and Y.~W.~Yu,
Commun. Theor. Phys. \textbf{36}, 71 (2001).

% \bibitem{Kumar:2010hs}
% A.~Kumar and A.~Mishra,
% %``$J/\psi$ and $\eta_{c}$ masses in isospin asymmetric hot nuclear matter: A QCD sum rule approach,''
% Phys. Rev. C \textbf{82}, 045207 (2010).
% % doi:10.1103/PhysRevC.82.045207
% % [arXiv:1005.2748 [nucl-th]].

\bibitem{Kumar:2018ujk}
R.~Kumar and A.~Kumar,
%``$J/\psi $ and $\eta _c$ in asymmetric hot magnetized nuclear matter: a unified approach of Chiral SU(3) model and QCD sum rules,''
Eur. Phys. J. C \textbf{79}, 403 (2019).
% doi:10.1140/epjc/s10052-019-6913-2
% [arXiv:1810.09185 [hep-ph]].

\bibitem{Papazoglou:1998vr}
P.~Papazoglou, D.~Zschiesche, S.~Schramm, J.~Schaffner-Bielich, H.~Stoecker and W.~Greiner,
Phys. Rev. C \textbf{59}, 411 (1999).

\bibitem{Kumar:2010hs}
A.~Kumar and A.~Mishra,
%``$J/\psi$ and $\eta_{c}$ masses in isospin asymmetric hot nuclear matter: A QCD sum rule approach,''
Phys. Rev. C \textbf{82}, 045207 (2010).
% doi:10.1103/PhysRevC.82.045207
% [arXiv:1005.2748 [nucl-th]].

\bibitem{Singh:2016hiw}
H.~Singh, A.~Kumar and H.~Dahiya,
Chin. Phys. C \textbf{41}, 094104 (2017).

\bibitem{Wang:2001hw}
P.~Wang, Z.~Y.~Zhang, Y.~W.~Yu, R.~K.~Su and Q.~Song,
Nucl. Phys. A \textbf{688}, 791 (2001).

\bibitem{Wang:2002aq}
P.~Wang, Z.~Y.~Zhang, Y.~W.~Yu, H.~Guo, R.~K.~Su and H.~Q.~Song,
Nucl. Phys. A \textbf{705}, 455 (2002).

\bibitem{Kumar:2023owb}
A.~Kumar, S.~Dutt and H.~Dahiya,
Eur. Phys. J. A \textbf{60}, 4 (2024).

\bibitem{Barik:1985rm}
N.~Barik, B.~K.~Dash and M.~Das,
%``STATIC PROPERTIES OF THE NUCLEON OCTET IN A RELATIVISTIC POTENTIAL MODEL WITH CENTER-OF-MASS CORRECTION,''
Phys. Rev. D \textbf{31}, 1652 (1985).
% doi:10.1103/PhysRevD.31.1652

\bibitem{Barik:2013lna}
N.~Barik, R.~N.~Mishra, D.~K.~Mohanty, P.~K.~Panda and T.~Frederico,
%``Nuclear equation of state in a relativistic independent quark model with chiral symmetry and dependence on quark masses,''
Phys. Rev. C \textbf{88}, 015206 (2013).
% doi:10.1103/PhysRevC.88.015206
% [arXiv:1307.0934 [nucl-th]].

% \bibitem{Bakamjian:1953kh}
% B.~Bakamjian and L.~H.~Thomas,
% %``Relativistic particle dynamics. 2,''
% Phys. Rev. \textbf{92}, 1300 (1953).
% % doi:10.1103/PhysRev.92.1300

\bibitem{Keister:1991sb}
B.~D.~Keister and W.~N.~Polyzou,
%``Relativistic Hamiltonian dynamics in nuclear and particle physics,''
Adv. Nucl. Phys. \textbf{20}, 225 (1991).

\bibitem{Choi:1997iq}
H.~M.~Choi and C.~R.~Ji,
%``Mixing angles and electromagnetic properties of ground state pseudoscalar and vector meson nonets in the light cone quark model,''
Phys. Rev. D \textbf{59}, 074015 (1999).
% doi:10.1103/PhysRevD.59.074015
% [arXiv:hep-ph/9711450 [hep-ph]].

% \bibitem{Arifi:2024mff}
% A.~J.~Arifi, L.~Happ, S.~Ohno and M.~Oka,
% %``Structure of heavy mesons in the light-front quark model,''
% Phys. Rev. D \textbf{110}, 014020 (2024).
% % doi:10.1103/PhysRevD.110.014020
% % [arXiv:2401.07933 [hep-ph]].

\bibitem{Choi:2009ai}
H.~M.~Choi and C.~R.~Ji,
%``Semileptonic and radiative decays of the B(c) meson in light-front quark model,''
Phys. Rev. D \textbf{80}, 054016 (2009).
% doi:10.1103/PhysRevD.80.054016
% [arXiv:0903.0455 [hep-ph]].
\bibitem{Choi:1999nu}
H.~M.~Choi and C.~R.~Ji,
%``Light front quark model analysis of exclusive 0- ---\ensuremath{>} 0- semileptonic heavy meson decays,''
Phys. Lett. B \textbf{460}, 461 (1999).
% doi:10.1016/S0370-2693(99)00817-5
% [arXiv:hep-ph/9903496 [hep-ph]].

\bibitem{Arifi:2022pal}
A.~J.~Arifi, H.~M.~Choi, C.~R.~ji and Y.~Oh,
%``Mixing effects on 1S and 2S state heavy mesons in the light-front quark model,''
Phys. Rev. D \textbf{106}, 014009 (2022).
% doi:10.1103/PhysRevD.106.014009
% [arXiv:2205.04075 [hep-ph]].

% \bibitem{Wu:2025rto}
% Q.~Wu, Z.~F.~Cui and J.~Segovia,
% %``Parton distribution functions of ground state mesons composed of $c$ or $b$ quarks,''
% [arXiv:2503.07055 [hep-ph]].

% \bibitem{Jaus:1989au}
% W.~Jaus,
% %``Semileptonic Decays of B and d Mesons in the Light Front Formalism,''
% Phys. Rev. D \textbf{41}, 3394 (1990).
% % doi:10.1103/PhysRevD.41.3394

% \bibitem{Chung:1988mu}
% P.~L.~Chung, F.~Coester and W.~N.~Polyzou,
% %``Charge Form-Factors of Quark Model Pions,''
% Phys. Lett. B \textbf{205}, 545 (1988).
% % doi:10.1016/0370-2693(88)90995-1


% \bibitem{Melosh:1974cu}
% H.~J.~Melosh,
% %``Quarks: Currents and constituents,''
% Phys. Rev. D \textbf{9}, 1095 (1974).
% % doi:10.1103/PhysRevD.9.1095

% \bibitem{Jaus:1991cy}
% W.~Jaus,
% %``Relativistic constituent quark model of electroweak properties of light mesons,''
% Phys. Rev. D \textbf{44}, 2851 (1991).
% % doi:10.1103/PhysRevD.44.2851

% \bibitem{Dhiman:2019ddr}
% N.~Dhiman, H.~Dahiya, C.~R.~Ji and H.~M.~Choi,
% %``Twist-2 Pseudoscalar and Vector Meson Distribution Amplitudes in Light-Front Quark Model with Exponential-type Confining Potential,''
% Phys. Rev. D \textbf{100}, 014026 (2019).
% % doi:10.1103/PhysRevD.100.014026
% % [arXiv:1902.09160 [hep-ph]]. 

% \bibitem{Arifi:2023jfe}
% A.~J.~Arifi, P.~T.~P.~Hutauruk and K.~Tsushima,
% %``In-medium properties of the light and heavy-light mesons in a light-front quark model,''
% Phys. Rev. D \textbf{107}, 114010 (2023).
% % doi:10.1103/PhysRevD.107.114010
% % [arXiv:2302.12382 [hep-ph]].

\bibitem{Choi:2015ywa}
H.~M.~Choi, C.~R.~Ji, Z.~Li and H.~Y.~Ryu,
%``Variational analysis of mass spectra and decay constants for ground state pseudoscalar and vector mesons in the light-front quark model,''
Phys. Rev. C \textbf{92}, 055203 (2015).
% doi:10.1103/PhysRevC.92.055203
% [arXiv:1502.03078 [hep-ph]].

\bibitem{ParticleDataGroup:2024cfk}
S.~Navas \textit{et al.} [Particle Data Group],
Phys. Rev. D \textbf{110}, 030001 (2024).

\bibitem{Dhiman:2019ddr}
N.~Dhiman, H.~Dahiya, C.~R.~Ji and H.~M.~Choi,
%``Twist-2 Pseudoscalar and Vector Meson Distribution Amplitudes in Light-Front Quark Model with Exponential-type Confining Potential,''
Phys. Rev. D \textbf{100}, 014026 (2019).
% doi:10.1103/PhysRevD.100.014026
% [arXiv:1902.09160 [hep-ph]]. 

\bibitem{Wang:2015mxa}
Z.~G.~Wang,
%``Analysis of the masses and decay constants of the heavy-light mesons with QCD sum rules,''
Eur. Phys. J. C \textbf{75}, 427 (2015).
% doi:10.1140/epjc/s10052-015-3653-9
% [arXiv:1506.01993 [hep-ph]].

\bibitem{Becirevic:1998ua}
D.~Becirevic, P.~Boucaud, J.~P.~Leroy, V.~Lubicz, G.~Martinelli, F.~Mescia and F.~Rapuano,
%``Nonperturbatively improved heavy - light mesons: Masses and decay constants,''
Phys. Rev. D \textbf{60}, 074501 (1999).
% doi:10.1103/PhysRevD.60.074501
% [arXiv:hep-lat/9811003 [hep-lat]].

\bibitem{Cvetic:2004qg}
G.~Cvetic, C.~S.~Kim, G.~L.~Wang and W.~Namgung,
%``Decay constants of heavy meson of 0- state in relativistic Salpeter method,''
Phys. Lett. B \textbf{596}, 84 (2004).
% doi:10.1016/j.physletb.2004.06.092
% [arXiv:hep-ph/0405112 [hep-ph]].

\bibitem{Wang:2005qx}
G.~L.~Wang,
%``Decay constants of heavy vector mesons in relativistic Bethe-Salpeter method,''
Phys. Lett. B \textbf{633}, 492 (2006).
% doi:10.1016/j.physletb.2005.12.005
% [arXiv:math-ph/0512009 [math-ph]].

% \bibitem{Wang:2015mxa}
% Z.~G.~Wang,
% %``Analysis of the masses and decay constants of the heavy-light mesons with QCD sum rules,''
% Eur. Phys. J. C \textbf{75}, 427 (2015).
% % doi:10.1140/epjc/s10052-015-3653-9
% % [arXiv:1506.01993 [hep-ph]].

% \bibitem{Dhiman:2019ddr}
% N.~Dhiman, H.~Dahiya, C.~R.~Ji and H.~M.~Choi,
% %``Twist-2 Pseudoscalar and Vector Meson Distribution Amplitudes in Light-Front Quark Model with Exponential-type Confining Potential,''
% Phys. Rev. D \textbf{100}, 014026 (2019).
% % doi:10.1103/PhysRevD.100.014026
% % [arXiv:1902.09160 [hep-ph]].  

%\bibitem{Mishra:2008cd}
%A.~Mishra and A.~Mazumdar,
%%``D-mesons in asymmetric nuclear matter,''
%Phys. Rev. C \textbf{79}, 024908 (2009).
%% doi:10.1103/PhysRevC.79.024908
%% [arXiv:0810.3067 [nucl-th]].

% \bibitem{Chhabra:2017rxz}
% R.~Chhabra and A.~Kumar,
% %``In-medium properties of pseudoscalar $D_s$ and $B_s$ mesons,''
% Eur. Phys. J. C \textbf{77}, 726 (2017).
% % doi:10.1140/epjc/s10052-017-5277-8
% % [arXiv:1711.00631 [hep-ph]].



\bibitem{Hilger:2008jg}
T.~Hilger, R.~Thomas and B.~Kampfer,
%``QCD sum rules for D and B mesons in nuclear matter,''
Phys. Rev. C \textbf{79}, 025202 (2009).
% doi:10.1103/PhysRevC.79.025202
% [arXiv:0809.4996 [nucl-th]].

\bibitem{Choi:2007se}
H.~M.~Choi,
%``Decay constants and radiative decays of heavy mesons in light-front quark model,''
Phys. Rev. D \textbf{75}, 073016 (2007).
% doi:10.1103/PhysRevD.75.073016
% [arXiv:hep-ph/0701263 [hep-ph]].

% \bibitem{Kumar:2009xc}
% A.~Kumar and A.~Mishra,
% %``D mesons in asymmetric nuclear matter at finite temperatures,''
% [arXiv:0912.2477 [nucl-th]].

\bibitem{Wang:2015uya}
Z.~G.~Wang,
%``Analysis of heavy mesons in nuclear matter with a QCD sum rule approach,''
Phys. Rev. C \textbf{92}, 065205 (2015).
% doi:10.1103/PhysRevC.92.065205
% [arXiv:1501.05093 [hep-ph]].

\bibitem{Chernyak:1983ej}
V.~L.~Chernyak and A.~R.~Zhitnitsky,
%``Asymptotic Behavior of Exclusive Processes in QCD,''
Phys. Rept. \textbf{112}, 173 (1984).
% doi:10.1016/0370-1573(84)90126-1

\bibitem{Lepage:1980fj}
G.~P.~Lepage and S.~J.~Brodsky,
%``Exclusive Processes in Perturbative Quantum Chromodynamics,''
Phys. Rev. D \textbf{22}, 2157 (1980).
% doi:10.1103/PhysRevD.22.2157


\bibitem{Choi:2007yu}
H.~M.~Choi and C.~R.~Ji,
%``Distribution amplitudes and decay constants for (pi, K, rho, K*) mesons in light-front quark model,''
Phys. Rev. D \textbf{75}, 034019 (2007).
% doi:10.1103/PhysRevD.75.034019
% [arXiv:hep-ph/0701177 [hep-ph]].

\bibitem{Li:2017mlw}
Y.~Li, P.~Maris and J.~P.~Vary,
%``Quarkonium as a relativistic bound state on the light front,''
Phys. Rev. D \textbf{96}, 016022 (2017).
% doi:10.1103/PhysRevD.96.016022
% [arXiv:1704.06968 [hep-ph]].

% \bibitem{Serna:2020txe}
% F.~E.~Serna, R.~C.~da Silveira, J.~J.~Cobos-Mart\'\i{}nez, B.~El-Bennich and E.~Rojas,
% %``Distribution amplitudes of heavy mesons and quarkonia on the light front,''
% Eur. Phys. J. C \textbf{80}, 955 (2020).

\bibitem{Zhong:2020cqr}
T.~Zhong, K.~Li, Y.~Zhang and H.~B.~Fu,
%``$D$ Meson Leading-Twist Distribution Amplitude from $B\to Dl\bar{\nu }_{l}$ Semi-Leptonic Decay,''
Int. J. Theor. Phys. \textbf{59}, 2562-2571 (2020).
% doi:10.1007/s10773-020-04525-x

\bibitem{Zhang:2021wnv}
Y.~Zhang, T.~Zhong, H.~B.~Fu, W.~Cheng and X.~G.~Wu,
%``Ds-meson leading-twist distribution amplitude within the QCD sum rules and its application to the Bs\textrightarrow{}Ds transition form factor,''
Phys. Rev. D \textbf{103}, 114024 (2021).
% doi:10.1103/PhysRevD.103.114024
% [arXiv:2104.00180 [hep-ph]].

\bibitem{Serna:2020txe}
F.~E.~Serna, R.~C.~da Silveira, J.~J.~Cobos-Mart\'\i{}nez, B.~El-Bennich and E.~Rojas,
%``Distribution amplitudes of heavy mesons and quarkonia on the light front,''
Eur. Phys. J. C \textbf{80}, 955 (2020).

\end{thebibliography}
\end{document}